\documentclass[preprint]{aastex62}
\usepackage{float}
\usepackage{graphicx, graphics}
\usepackage{multirow}
\usepackage{natbib}

\newcommand{\Ms}{M$_\odot$} 

\shorttitle{VANDAM Class 0 and I Candidate Disks}
\shortauthors{Segura-Cox et al.}

\begin{document}


\title{The VLA Nascent Disk and Multiplicity Survey of Perseus Protostars (VANDAM).  V.  18 Candidate Disks around Class 0 and I Protostars in the Perseus Molecular Cloud}

\correspondingauthor{Dominique M. Segura-Cox}
\email{dom@mpe.mpg.de}

\author{Dominique M. Segura-Cox}
\affiliation{Department of Astronomy, University of Illinois, 1002 W. Green St., Urbana, IL 61801, USA}
\affiliation{Max-Planck-Institut f{\"u}r extraterrestrische Physik, Giessenbachstrasse 1, D-85748 Garching, Germany}

\author{Leslie W. Looney}
\affiliation{Department of Astronomy, University of Illinois, 1002 W. Green St., Urbana, IL 61801, USA}

\author{John J. Tobin}
\affiliation{Homer L. Dodge Department of Physics and Astronomy, University of Oklahoma, 440 W. Brooks St., Norman, OK 73019, USA}
\affiliation{Leiden Observatory, Leiden University, P.O. Box 9513, NL-2300RA Leiden, The Netherlands}

\author{Zhi-Yun Li}
\affiliation{Department of Astronomy, University of Virginia, 530 McCormick Rd., Charlottesville, VA 22903, USA}

\author{Robert J. Harris}
\affiliation{Department of Astronomy, University of Illinois, 1002 W. Green St., Urbana, IL 61801, USA}
\affiliation{National Center for Supercomputing Applications, 1205 W. Clark St., Urbana, IL 61801, USA}

\author{Sarah Sadavoy}
\affiliation{Harvard-Smithsonian Center for Astrophysics, 60 Garden St., Cambridge, MA 02138, USA}
\affiliation{Max-Planck-Institut f{\"u}r Astronomie, K{\"o}nigstuhl 17, D-69117 Heidelberg, Germany}

\author{Michael M. Dunham}
\affiliation{Department of Physics, State University of New York at Fredonia, 280 Central Ave., Fredonia, NY 14063, USA}
\affiliation{Harvard-Smithsonian Center for Astrophysics, 60 Garden St., Cambridge, MA 02138, USA}

\author{Claire Chandler}
\affiliation{National Radio Astronomy Observatory, P.O. Box O, 1003 Lopezville Rd., Socorro, NM 87801, USA}

\author{Kaitlin Kratter}
\affiliation{Department of Astronomy and Steward Observatory, University of Arizona, 933 N. Cherry Ave., Tucson, AZ 85721, USA}

\author{Laura P{\'e}rez}
\affiliation{Departamento de Astronom{\'i}a, Universidad de Chile, Camino El Observatorio 1515, Las Condes, Casilla 36-D, Santiago, Chile}

\author{Carl Melis}
\affiliation{Center for Astrophysics and Space Sciences, University of California, San Diego, 9500 Gilman Dr.,
La Jolla, CA 92093, USA}

\begin{abstract}
We present the full disk-fit results VANDAM survey of all Class 0 and I protostars in the Perseus molecular cloud.  We have 18 new protostellar disk candidates around Class 0 and I sources, which are well described by a simple, parametrized disk model fit to the 8 mm VLA dust-continuum observations.  33\% of Class 0 protostars and just 11\% of Class I protostars have candidate disks, while  78\% of Class 0 and I protostars do not have signs of disks within our 12 AU disk diameter resolution limit, indicating that at 8 mm most disks in the Class 0 and I phases are $<$10 AU in radius.  These small radii may be a result of surface brightness sensitivity limits.  Modeled 8 mm radii are similar to the radii of known Class 0 disks with detected Keplerian rotation.   Since our 8 mm data trace a population of larger dust grains which radially drift towards the protostar and are lower limits on true disk sizes, large disks at early times do not seem to be particularly rare.  We find statistical evidence that Class 0 and I disks are likely drawn from the same distribution, meaning disk properties may be defined early in the Class 0 phase and do not undergo large changes through the Class I phase.  By combining our candidate disk properties with previous polarization observations, we find a qualitative indication that misalignment between inferred envelope-scale magnetic fields and outflows may indicate disks on smaller scales in Class 0 sources.
\end{abstract}

\section{INTRODUCTION}

Disks of gas and dust around young protostars are fundamental to protostellar mass 
accretion and act as the mass reservoir from which stars and planetesimals form \citep{Armitage2011,Williams2011}.  
Circumstellar disks are expected to form around even the youngest Class 0 protostars which are embedded in their dense natal dust and gas envelope.  Class I protostars  are less embedded, having cleared a portion of their envelopes  \citep{Mckee2007}.  Until recently, disks around Class 0 and Class I protostars have remained elusive because $\sim$millimeter wavelengths are required to penetrate through the dense envelope \citep{Looney2000}, and sub-arcsecond resolution is required to spatially resolve the disk. 
Keplerian rotation is a tell-tale sign of true, rotationally supported disks that exist for long enough timescales to form long-lived disk structures and eventually planets; flattened structures without rotation quickly collapse inward \citep[e.g.,][]{Terebey1984}.
 So few young Keplerian disks are known and as a consequence, questions concerning disk frequency, disk radii, dust populations, disk evolution, and the presence of planetesimals in the youngest protostellar disks are only beginning to be addressed.

Keplerian rotation has been detected in disks around only 4 total low-mass Class 0 protostars to date with  $R$ $>$30 AU \citep{Ohashi2014,Tobin2012,Murillo2013,Codella2014,Yen2017,Lee2017}; however they are bright sources and may not represent typical disks at this stage of evolution.
Class I protostars have longer lifetimes and have cleared enough of their mass reservoir that more low-mass Class I disks have been detected \citep[$\sim$10 total, to date;][]{Harsono2014} than in Class 0 systems though not nearly as many as the $\sim$100 total in more-evolved Class II sources \citep[e.g.,][]{Andrews2009,Andrews2010}.  By the Class II stage the envelope has mostly dispersed, clearly revealing the circumstellar disk and allowing geometrical constraints to be found from the spectral energy distribution (SED) of the dust emission. The dense envelopes in Class 0 and I systems prevent disk parameters to be constrained from examining the SED alone. Recent observations of a disk around a Class II protostar have revealed the earliest known evidence of planet formation \citep{ALMA2015}.  The Class 0 and I protostellar stages have the largest mass reservoirs available to form disks and planetesimals; therefore understanding the properties of disks at the earliest possible epochs is crucial to determine the formation mechanism behind  circumstellar disks and the initial pathway to planet formation.

The morphology and strength of the magnetic fields in protostellar systems
also play an important role in star and disk formation \citep[e.g.,][]{Crutcher2012}. 
 Magnetic field effects on the small size scale of circumstellar disks ($\sim$0.5$^{\prime\prime}$ or $\sim$100 AU) of young stellar objects have started to be theoretically and observationally quantified in individual systems \citep{Mellon2008,Hennebelle2008,Stephens2014,SeguraCox2015,Cox2015}. Magnetic field morphology 
can be inferred from dust emission; spinning dust grains align their long axes perpendicular to the magnetic field, polarizing the dust emission \citep[e.g.,][]{Lazarian2007}.  When a strong magnetic field and the rotation axis of the circumstellar disk are aligned, magnetic braking can have a significant effect on disk formation by transporting away angular momentum, limiting forming disks to $R$ $<$10 AU \citep[e.g.,][]{Mellon2008,Dapp2010,Machida2011,Li2011,Dapp2012}.   In this scenario, disks only reach $R$ $\sim$100 AU at the end of the Class I phase
 when the envelope is less massive and magnetic braking becomes 
inefficient \citep[e.g.,][]{Dapp2012,Mellon2009,Machida2011}.  Conversely, recent works have highlighted the critical importance of the magnetic field direction relative to the rotation axis: when the field and rotation axis are misaligned, magnetic braking becomes less efficient, and $\sim$100 AU disks can form \citep{Joos2012,Li2013,Krumholz2013,SeguraCox2015,SeguraCox2016}.  Disks can form if the coupling of the magnetic field to the disk material can be lessened by non-ideal magnetohydrodynamic effects, allowing material to accrete from the envelope to the disk without dragging in a flux-frozen magnetic field. Because so few young embedded disks are currently known via observations, expanding the number of known Class 0 and I disks is critical to determining the role of magnetic braking in disk growth at early times.

To determine the properties of the youngest disks, we used  the Karl G. Jansky Very Large Array (VLA) to make continuum observations for the VLA Nascent Disk and Multiplicity (VANDAM) survey toward all known protostars in the
Perseus molecular cloud \citep{Tobin2015a}. The continuum observations trace dust emission (at $\lambda\sim8$ and $10$ mm)  and free-free emission from jets near the central protostar (at $\lambda\sim4$ and $6.4$ cm). The VANDAM observations form an unbiased survey of young protostellar disks down to $\sim$10 AU size scales, giving us the opportunity to potentially double the number of known disks in Class 0 and I protostars from $\sim$15 total to over 30.  We use the term ``candidate
disks" because we do not have kinematic data on small scales to determine whether these structures are rotationally supported.  The VANDAM sample contains all currently known Class 0 and I protostellar systems in  Perseus, with 37 Class 0 systems, 8 Class 0/I systems, and 37 Class I systems.   The 21 resolved sources in the VANDAM survey we examine in this paper (Table \ref{diskfulltab1}, Figure \ref{Per44_2x4}, and Figures in Appendix \ref{vandam:extended_images}) are the most complete sample of embedded sources in Perseus to-date \citep[see][for discussion of target selection]{Tobin2016a}. Per-emb-XX designations originate from \citet{Enoch2009}. We define resolved or extended sources as having spatial extents at least 1.1$\times$ the size of the FWHM of the beam, meaning we include marginally resolved sources in this study. We fit disk models to all protostars with relatively axisymmetric resolved emission (17 of 21 sources, see Table \ref{diskfulltab2} and Section \ref{vandam:modeling}) roughly perpendicular to known outflows; however, only sources with either axisymmetric resolved emission and a modeled disk-like profile or non-symmetric emission and other indirect evidence of a disk are considered candidate disks (see Appendix \ref{vandam:gallery}).

In this paper, we present the full results toward the protostellar disk candidates 
around the Class 0 and 
Class I protostars from the VANDAM survey.  This paper expands on the work done in \citet{SeguraCox2016}, which reported a subset of the candidate disks studied here.   The observations,  VLA set up, and data reduction are described in Section \ref{obssec}, with estimated masses from observed fluxes of extended sources presented in Section \ref{vandam:masses}.  Section \ref{vandam:modeling} describes our {\it u,v}-plane disk modeling procedure, and we describe the modeling results in Section \ref{vandam:results}.  The results of our study are discussed in Section \ref{vandam:discussion}, and the summary is given in Section \ref{vandam:summary}.  Appendix \ref{vandam:extended_images} shows 8 mm images of protostars with extended emission, Appendix \ref{vandam:gallery} lists previously known information on each candidate disk studied here, and Appendix \ref{vandam:bestfits} presents images of candidate-disk modeling results.

\section{OBSERVATIONS} \label{obssec}

Our VANDAM survey obtained VLA Ka-band
 lower-resolution ($\sim$0.28$^{\prime\prime}$, $\sim$65 AU) B-array data
 and high-resolution ($\sim$0.05$^{\prime\prime}$, $\sim$12 AU) 
A-array data. 
We detected 21 protostars with extended emission larger than the size of the beam in Perseus, with data collected in 2013, 2014, and 2015
 (Table \ref{diskfulltab1}, Figure \ref{Per44_2x4}, and Figures in Appendix \ref{vandam:extended_images}).
 The observations used the three-bit correlator mode and a bandwidth of 8 GHz divided into 64 sub-bands, each with 128 MHz bandwidth, 2 MHz channels, and full polarization products.  
Two 4 GHz basebands were centered at 36.9 GHz ($\sim$8.1 mm) and 29.0 GHz ($\sim$10.5 mm).  
 In each 3.5 hour block, three source were observed, and in each 2.75 hour block, two sources were observed. Some sources were observed in 1.5 hour blocks.  The flux calibrator was 3C48, and the bandpass calibrator is 3C84
 The observations were made in fast-switching mode to take into account rapid atmospheric phase variations, 
having a 2.5 minute total cycle time to switch between the target source and J0336+3218, the complex gain calibrator.  The total on-source integration time for each source was  $\sim$30 minutes in both A-array and B-array.  The data was reduced 
 with CASA 4.1.0 and the VLA pipeline (version 1.2.2). 
 We applied additional flagging beyond pipeline flagging by examining the phase, gain and bandpass
  calibration solutions.  It was not necessary to re-calibrate the data after additional flagging.  Only statistical uncertainties are considered in our study, though VLA Ka-band data sets have an estimated amplitude  calibration uncertainty of $\sim$10\%.

\begin{table*}[b]
\caption{Observed Positions of Resolved Sources and Beam Sizes}
  \begin{center}
\begin{tabular}{lcccc}
\hline
Source & $\alpha$ & $\delta$ & Combined Beam & Beam P.A.\\
& (J2000) & (J2000)  & (mas$\times$mas) & ($^{\circ}$)\\
\hline
SVS13B & 03:29:03.078 & +31:15:51.740 & 105 $\times$ 83  & -74.8\\
Per-emb-50 & 03:29:07.768 & +31:21:57.125  &  97 $\times$ 94 & 54.0 \\
Per-emb-14 & 03:29:13.548 & +31:13:58.153  & 91  $\times$ 75 & 82.4 \\
Per-emb-30 & 03:33:27.303 & +31:07:10.161   & 99   $\times$ 92 & -71.5\\
HH211-mms & 03:43:56.805 & +32:00:50.202 &  96 $\times$ 77 & 85.4\\
IC348 MMS & 03:43:57.064 & +32:03:04.789  &  89 $\times$ 80 & -57.7 \\
Per-emb-8 & 03:44:43.982 & +32:01:35.209 & 82$\times$ 72 & -71.6 \\
Per-emb-25 &03:26:37.511 &+30:15:27.813  & 70$\times$50 & -88.3  \\
NGC 1333 IRAS1 A &03:28:37.090 & +31:13:30.788  & 83$\times$74 & 99.9 \\
Per-emb-62 & 03:44:12.977&+32:01:35.419  & 61$\times$52 & -69.7 \\
Per-emb-63 & 03:28:43.271& +31:17:32.931 & 61$\times$52 & -69.7\\
SVS13C & 03:29:01.970& +31:15:38.053 & 83$\times$74 & -79.2 \\
NGC 1333 IRAS4A &03:29:10.537 &+31:13:30.933  & 74$\times$53 & 78.2 \\
NGC 1333 IRAS2A &03:28:55.569 &+31:14:37.025  & 91$\times$82 & -75.5 \\
IRAS 03292+3039 &03:32:17.928 &+30:49:47.825  & 107$\times$87 & 117.1 \\
IRAS 03282+3035  &03:31:20.939 & +30:45:30.273 & 93$\times$89 & -85.4 \\
Per-emb-18 & 03:29:11.258& +31:18:31.073 & 106$\times$94 & 84.2 \\
L1448 IRS3B &03:25:36.379 & +30:45:14.728  & 102$\times$89 & -63.2 \\
NGC 1333 IRAS4B & 03:29:12.010& +31:13:08.010 & 111$\times$92 & -85.3 \\
NGC 1333 IRAS1 B & 03:28:37.090 & +31:13:30.788  & 83$\times$74 & 99.9\\
B5-IRS1 & 03:47:41.591&+32:51:43.672  & 73$\times$64 & -76.5 \\
\hline
\end{tabular}\\[5.0pt]
 \end{center}
Positions reflect measured source center.  Combined beam sizes reflect robust = 0.25 weighting of A+B array data.  Position angle is measured counterclockwise from north.
\label{diskfulltab1} 
\end{table*}

  \begin{figure}[t]
        \centering
                \includegraphics[width=0.8\textwidth]{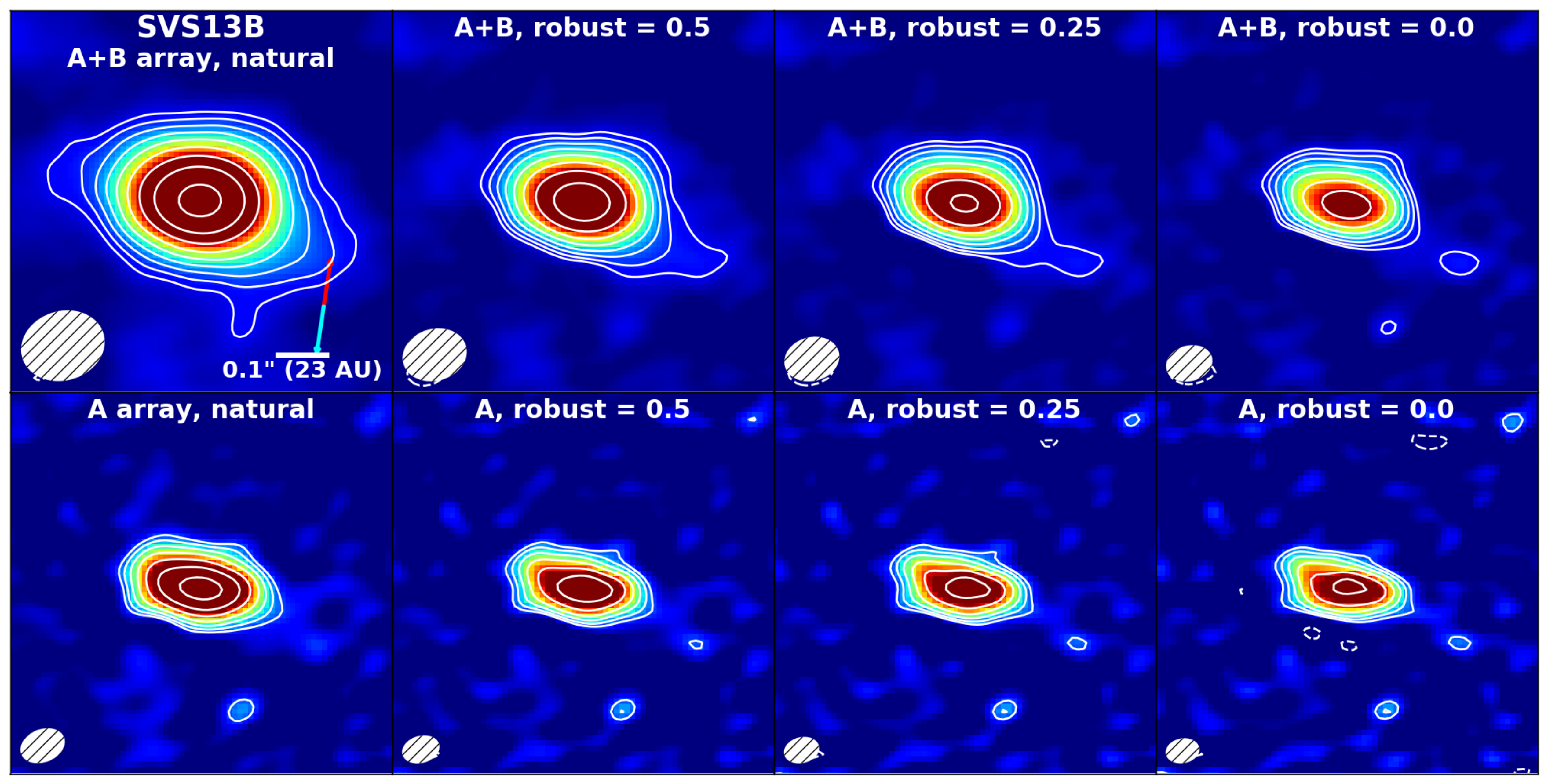}
        \caption{VLA A+B array data (top row), and A-array only data (bottom row) of SVS13B, all plotted with the same physical size scale. Images were produced with varying robust weighting values, labeled at the top of each panel. Contours start at 3$\sigma$ ($\sigma$ $\sim$15$\mu$Jy) with a factor of $\sqrt{2}$ spacing. The synthesized beam is in the lower left.  Outflow orientations are indicated by the red and blue arrows in the lower right corner of the upper left-most panel.}
        \label{Per44_2x4}
\end{figure} 
 
 \section{ESTIMATED MASSES FOR ALL EXTENDED SOURCES IN THE VANDAM SURVEY} \label{vandam:masses}
  
Protostellar disks are typically quantified by their radii and masses.  We must model the continuum emission to determine the radii accurately (Section \ref{vandam:modeling}), yet we can straightforwardly estimate masses from flux measurements, assuming no free-free contribution from the jets near the central protostar.
We estimate disk masses from the 8 mm dust continuum flux, assuming optically-thin emission, with the relation \citep{Hildebrand1983}:
\begin{equation}
M_{d}=\frac{d^{2}F_{\nu}}{B(T_{d})\kappa_{\nu}},
\label{massfull}
\end{equation}
where  $F_{\nu}$, $d$, $\kappa_{\nu}$, and $B_{\nu}(T_d)$, are the total observed flux, distance, grain opacity,
and blackbody intensity at dust temperature $T_d$ respectively.
To estimate $\kappa_{\nu}$ at 8.1 mm, we normalize to \citet{Ossenkopf1994} at 1.3 mm using a dust to gas ratio of $1/100$: 
$\kappa_{\nu}$ = $(1/100)(\nu/231\mathrm{GHz})^{\beta}$ cm$^{2}$~g$^{-1}$. $\beta$ = 1 is oftened assumed for protostellar disks \citep{Andrews2009}, yielding $\kappa_{\nu}=0.00146$ cm$^{2}$~g$^{-1}$.
 Mass estimates are ambiguous within an order of magnitude because of uncertainties in the dust-to-gas ratio, $T_d$, and $\beta$.  By varying $T_d$ we calculate upper and lower boundaries for each source rather than compute a single mass estimate due to the inherent uncertainty.  Upper bound masses were calculated using $T_d$ = 20 K, and lower bound masses were found assuming $T_d$ = 40 K.  The extended sources range in flux from 95.6 $\mu$Jy to 14836.1 $\mu$Jy, providing 
 estimated masses of 0.01--3.2 \Ms~(Table \ref{diskfulltab2}). The mass of L1448 IRS3B is highly underestimated because it is a triple system embedded in a larger disk \citep{Tobin2016b} which is marginally detected at 8 mm.  If we instead adopt  the \citep{Andrews2009} opacity used for more evolved Class II disks and normalized to our wavelength, the masses would decrease by a factor of 0.44; however, the \citet{Andrews2009} opacity may not be suitable for young, embedded disks.  We report the masses calculated from the normalized \citet{Ossenkopf1994} $\kappa_{\nu}$ for easier comparison with previous observations in the literature.
 
   \begin{table*}[t!]
\caption{Resolved Source Data}
\begin{center}
\begin{tabular}{lccccccc}
\hline
Source & Class  & Size & Disk P.A. & Disk Inc.  & $F_{8 mm}$ & $M_{d}$ & $F_{>1700k\lambda}$\\
&   &  (mas$\times$mas) & ($^{\circ}$) & ($^{\circ}$)  & ($\mu$Jy) & (M$_{\odot}$) & ($\mu$Jy)\\
\hline
\hline
 \multicolumn{8}{|l|}{Modeled Disk Candidates} \\
 \hline
SVS13B &  0 & 163$\times$80 & 71.4$\pm$2.5 & 61 & 1352.7$\pm$11.6 & 0.14 - 0.29 & 82.0 \\
Per-emb-50 &  I &  137$\times$53 & 170.0$\pm$0.3  & 67 &1664.9$\pm$12.5 &  0.18 - 0.36 & 133.2 \\ 
Per-emb-14 &  0 &  174$\times$76 & 12.7$\pm$0.9  & 64 & 882.1$\pm$13.0 & 0.09 - 0.19 & 68.8\\ 
Per-emb-30 &  0 &  87$\times$74 & 40.0$\pm$23.0  &31 & 957.0$\pm$9.4 &  0.10 - 0.21 & 130.9 \\ 
HH211-mms &  0 &  93$\times$59 & 34.8$\pm$9.6  & 51& 867.5$\pm$8.1 &  0.09 - 0.19 & 42.9 \\ 
IC348 MMS & 0 &  145$\times$105 & 70.8$\pm$2.2  & 44 & 1126.5$\pm$10.3 & 0.12 - 0.24 & 0.0\\ 
Per-emb-8 &  0 &  111$\times$84 & 116.1$\pm$2.8  & 41 & 1120.7$\pm$10.3 &  0.12 - 0.24 & 126.5\\ 
Per-emb-25 & 0/I& 91$\times$55 & 11.1$\pm$16.0 & 52 & 613.6$\pm$27.6 & 0.06 - 0.13 & 95.0 \\
NGC 1333 IRAS1 A &I & 77$\times$44 & 34$\pm$1.8 & 55 & 586.2$\pm$12.4 & 0.06 - 0.13 & 0.0\\
Per-emb-62 & I & 106$\times$65 & 107.7$\pm$2.8 & 52 & 730.9$\pm$12.6 & 0.08 - 0.16 & 121.1\\
Per-emb-63 & I& 122$\times$36 & 109.7$\pm$4.1 & 73 & 270.6$\pm$10.9 & 0.03 - 0.06 & 0.0\\
SVS13C & 0& 275$\times$70 & 95$\pm$1.3 &75 & 2128.0 $\pm$11.5 & 0.22 - 0.46 & 27.4\\
NGC 1333 IRAS4A& 0 & 250$\times$205 & 95.7$\pm$5.2 & 35 & 14836.1$\pm$43.2 & 1.57 - 3.20 & 81.0 \\
NGC 1333 IRAS2A & 0/I & 65$\times$45 & 110.9$\pm$12.3 & 46 & 1935.1$\pm$8.9 & 0.20 - 0.42 & 494.8\\
\hline
 \multicolumn{8}{|l|}{Asymmetric Disk Candidates (Cannot be Modeled)} \\
 \hline
IRAS 03292+3039 & 0& ...  & ... & ... & 3289.3$\pm$15.4 & 0.35 - 0.71 &... \\
IRAS 03282+3035 &0 &... & ...  & ...& 1481.2$\pm$12.0 & 0.16 - 0.32 &... \\
Per-emb-18 & 0& ... & ... & ...& 636.5$\pm$11.8 & 0.07 - 0.14 & ...\\
L1448 IRS3B& 0& ...&... &... & 95.6$\pm$9.5 & 0.01 - 0.02 & ...\\
\hline
 \multicolumn{8}{|l|}{Resolved Sources Determined Not to be Disk Candidates by Model} \\
 \hline
NGC 1333 IRAS4B &0 & 344$\times$224 & 90.9$\pm$6.6 & 49 & 1537.2$\pm$14.0 & 0.16 - 0.33 & 0.0\\
NGC 1333 IRAS1 B & I& 67$\times$52 & 32.1$\pm$9.6 & 39 & 296.0$\pm$14.3 & 0.03 - 0.06 & 72.3\\
B5-IRS1& I& 120$\times$83 & 118$\pm$6.2 & 46& 255.8$\pm$11.1 & 0.03 - 0.06 & 116.9\\
\hline
\end{tabular}\\[5.0pt]
 \end{center}
 Sizes represent deconvolved sizes and are measured from image-plane 2D Gaussian fits.  Position angles are measured counterclockwise from north, also from 2D Gaussian fits.  Uncertainties on the deconvolved sizes are $\sim$5.0 mas. Uncertainties on inclinations are $\sim$10$^{\circ}$.  IRAS4A and IRAS4B measurements were made with baselines $<350$ k$\lambda$ excluded to better filter out envelope emission.  Fluxes are measured from observations, and masses are estimated from the observed fluxes. $F_{>1700k\lambda}$ is the flux estimated from only the longest baselines, representing the lower-limit on the free-free point-source component of the emission.
\label{diskfulltab2} 
\end{table*}

\section{MODELING THE {\it U,V}-DATA} \label{vandam:modeling}

For each source, we fitted an axisymmetric intensity profile to the continuum emission to model the 8 mm A+B array data, assuming the data is optically thin.  We assumed our 8 mm data is optically thin, because significant optical depth effects are not expected at this long wavelength \citep[e.g.,][]{Testi2003,Isella2009}.  While we did not model the envelope here, we did take into account a fixed free-free component of the emission. 
 For all sources we used CASA task FIXVIS to place the extended source we wish to model at the phase center of the visibility data set to reduce ringing in the final real component vs. {\it u,v}-distance plots (see Appendix \ref{vandam:bestfits}). 
 We assumed the disks are circularly symmetric and geometrically thin, and deprojected the visibility data to  fixed position and inclination angles determined via image-plane 2D Gaussian 
fitting of the disk candidates (Table \ref{diskfulltab3}).  The visibility data was
azimuthally averaged in the {\it u,v}-plane and binned in linearly spaced bins with width 50 k$\lambda$  
for {\it u,v}-distances from 0 to 1500 k$\lambda$.   For {\it u,v}-distances from 1500 to 4000 k$\lambda$, we switched to log-spaced bins in order to boost the signal-to-noise level at large {\it u,v}-distances.  For bright sources with high signal-to-noise, we used 30 log-spaced bins for the long baselines, and we used 20 log-spaced bins for the long baselines in dimmer sources with lower signal-to-noise levels at large {\it u,v}-distances.

 For sources with binary components or other nearby sources in the field of view, we subtracted the other sources in the field from the {\it u,v}-data before modeling to reduce ringing from off-center sources in the field when we deproject, azimuthally average, and bin the data.  Fewer residuals from non-disk components allows for a better fit of the disk model to the {\it u,v}-data of the extended VANDAM sources.  We used the CASA task CLEAN with a region around only the companions we wish to subtract out and with option usescratch=True to save model visibilities of the companions to the model data column of the MS file.  We then used the task UVSUB to subtract the model data column from the corrected data column with the residuals of only the extended source we wish to model written to the corrected data column.  
 
We used a C-based implementation of \texttt{emcee}, an affine-invariant Markov chain Monte Carlo ensemble sampler \citep{Goodman2010,ForemanMackey2013}, to fit the real components of the deprojected, averaged, and binned profile to a simple disk model. We assumed the imaginary components were zero in the model because we assumed symmetry and the sources were at the phase center.  We chose a model which imitates a Shakura-Sunyaev 
disk \citep{Shakura1973} with a power law temperature profile.  We used a model disk surface brightness profile of
\begin{equation}
I(r)_{disk}\propto\left(\frac{r}{R_{c}}\right)^{-\left(\gamma+q\right)}\exp{\bigg\{-\left(\frac{r}{R_{c}}\right)^{\left(2-\gamma\right)}}\bigg\}.
\label{diskeqnfull}
\end{equation}
$I(r)_{disk}$ is the radial 
surface brightness distribution of the disk, $q$ takes into account the temperature structure of the disk, $\gamma$ is the inner-disk surface density power-law, and
$r$ is radius.  $R_{c}$ is a characteristic radius at which there is a significant
 drop off of disk flux, a proxy for outer disk radius. 
This intensity profile was applicable to our data because 
our 8 mm data is in the Rayleigh-Jeans tail of the dust emission. 
Flux is yielded by $F = \int_{0}^{\infty} I(r)~2\pi r~dr$.
Free parameters in the modeling were flux, disk radius, and the power-law of the inner-disk surface density.  In order to avoid over-fitting marginally-resolved data while exploring a physically reasonable parameters space of $q$, we fitted models for fixed values of 
$q=[0.25,0.50,0.75,1.00]$.

To account for a lower limit on a free-free point-source component arising from shocks in protostellar jets  \citep{Anglada1998}, we included a fixed linear component in the model.  We calculated the average of the real components of data having {\it u,v}-distance $>$1700 k$\lambda$ since point sources in the image-plane have constant flux at all {\it u,v}-distances.  The visibility profiles become flat at values $>$1700 k$\lambda$ in all sources (see visibility plots in Appendix \ref{vandam:bestfits})  The calculated average point-source components which we attribute to free-free emission are reported in (Table \ref{diskfulltab2}).  For sources where the average real component of the binned data at values $>$1700 k$\lambda$ is less than or equal to zero, we do not subtract a lower-limit free-free point source component.

We do not model an envelope component because for the majority of sources a disk model can explain most of the continuum flux.  To illustrate this, we plot visibility profiles corresponding to $R^{-1.5}$ and $R^{-2.0}$ envelope volume density profiles \citep[respectively, free-fall collapse and  singular isothermal sphere envelopes][]{Shu1977}  alongside the visibility plots of the observations (see Figure \ref{SVS13Buvdist} and Appendix \ref{vandam:bestfits}).  For the envelope visibility profiles, we adopted a value of $q=0.4$, which is typical for envelope emission \citep{Looney2003}.  Observations consistent with envelopes are expected to fall between the two envelope profile curves.  For all but one modeled source, envelope profiles alone cannot account for the vast majority of the dust emission.  For the bright sources IRAS4A and IRAS4B, we do not fit the shortest ($<$350 k$\lambda$) baselines (corresponding to emission from scales larger than 0.71$^{\prime\prime}$ or 165 AU) to remove a majority of the envelope contamination.  This does not completely eliminate the envelope emission, but it removes enough large-scale emission to model a disk component.  

In Figure  \ref{SVS13Buvdist}, we also plot an example of a Gaussian profile that reproduces the parameters of the image-plane Gaussian fit for SVS13B.  The data points are better described by our adopted disk model than the Gaussian profile, a quality shared by the candidate disks in our sample.  Gaussian profiles are not typically realistic for protostellar disk or compact circumstellar structures \citep[e.g.,][]{Harsono2014,Harvey2003}; hence, we do not conduct a detailed analysis of Gaussian fits in {\it u,v}-space in this study.

To quantify the smallest-recoverable model disk radius, we generated 36 synthetic disks with radii varying from 4.0 to 14.0 AU in steps of 2.0 AU and fluxes varying from 100.0 to 350.0 $\mu$Jy in steps of 50.0 $\mu$Jy.  We fixed the position angles of all synthetic disks to 0$^{\circ}$ and inclination angles to 45$^{\circ}$.  We adopted values of $q=0.25$ and $\gamma=0.3$ for the synthetic disks, which are typical values recovered from modeling the observations (Table \ref{diskfulltab3} and Section \ref{vandam:discussion:diskprop}).  We produced images of the synthetic disks and added a noise component with an rms of 15 $\mu$Jy from the robust=0.25 residual map of Per-emb-14, representative of the noise level in our observations.  We transformed the image-plane data to the \textit{u,v}-plane and deprojected, azimuthally averaged, and radially binned the data.  For modeling the synthetic disks, we followed the standard fitting procedure for the observations described above.  Uncertainties on the modeled radii were calculated from 90\% confidence intervals.  
We considered a model successful if (1) the synthetic radius is within the 90\% confidence interval of the modeled radius, and if (2) the percentage change between the best-fit model radius and both the minimum and maximum 90\% confidence levels ($\frac{uncertainty}{R_{best}}\times100$) is $<$30\%.  For synthetic disks with input radii of 4.0 or 6.0, the conditions were not satisfied for criteria (1), and for synthetic disks with fluxes of 100.0 and 150.0 $\mu$Jy, the conditions were not satisfied for criteria (2).  We conclude that our model is sensitive to disks with radii $\geq$ 8.0 AU and fluxes $\geq$200.0 $\mu$Jy.  A radius of 8 AU is 1.5 beams resolved across the major axis of the disk (our beam is 12 AU); thus we can accurately model marginally-resolved sources. 

We also performed a similar study of 36 synthetic disks to quantify the reliability of fitting large disk radii.  We generated synthetic disks with radii of 30, 40, 50, 60, 70, and 80 AU and fluxes varying from 100.0 to 350.0 $\mu$Jy in steps of 50.0 $\mu$Jy.  We fixed the position and inclination angles, q, and $\gamma$ of the synthetic disks with the same values as the study for the smallest-recoverable model disk radius, and we used the same procedure to add a noise component and fit the synthetic disks.  Using the same criteria for a successful model fit to a synthetic disk as described above, we found that we can accurately model disks with radii $\leq$ 60.0 AU for fluxes $\geq$200.0 $\mu$Jy.  For larger synthetic disks, the surface brightness becomes too low to accurately model the radii with the fluxes we tested, and the fitting procedure begins to significantly under-fit the disk radii.  Since our modeled disk candidates all have fluxes $>$200.0 $\mu$Jy (Table \ref{diskfulltab2}) and have modeled radii $\leq$ 42.2 AU (Table \ref{diskfulltab3}), for our sample the recovered model disk radii are accurately described by our fitting procedure.

\section{FULL VANDAM SURVEY DISK MODELING RESULTS} \label{vandam:results}

\begin{figure}[h]
        \centering
                \includegraphics[width=0.8\textwidth]{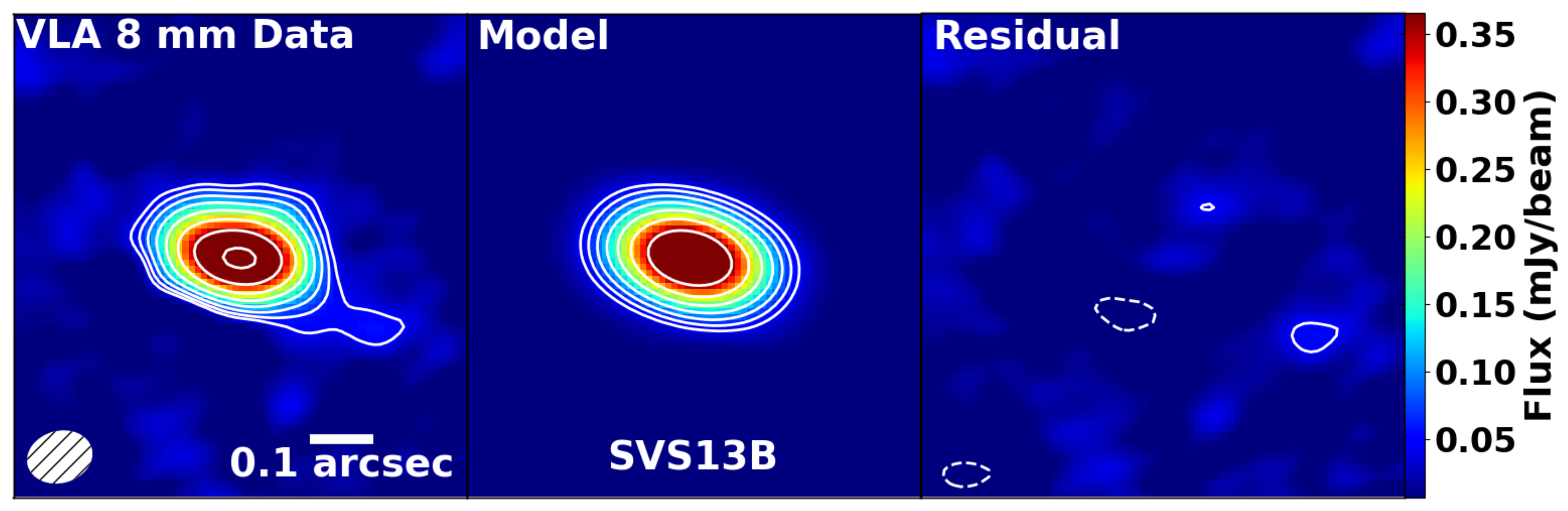}
        \caption{VLA A+B array data (left), $q=1.0$ model from {\it u,v}-plane best-fit (center), and
residual (right) of SVS13B. Images were produced with robust = 0.25 weighting. Contours start at 3$\sigma$ ($\sigma$ $\sim$15$\mu$Jy) with a factor of $\sqrt{2}$ spacing. The synthesized beam is in the lower left.}
        \label{SVS13Bdmr}
\end{figure}

\begin{figure}[h]
        \centering
                \includegraphics[width=0.7\textwidth]{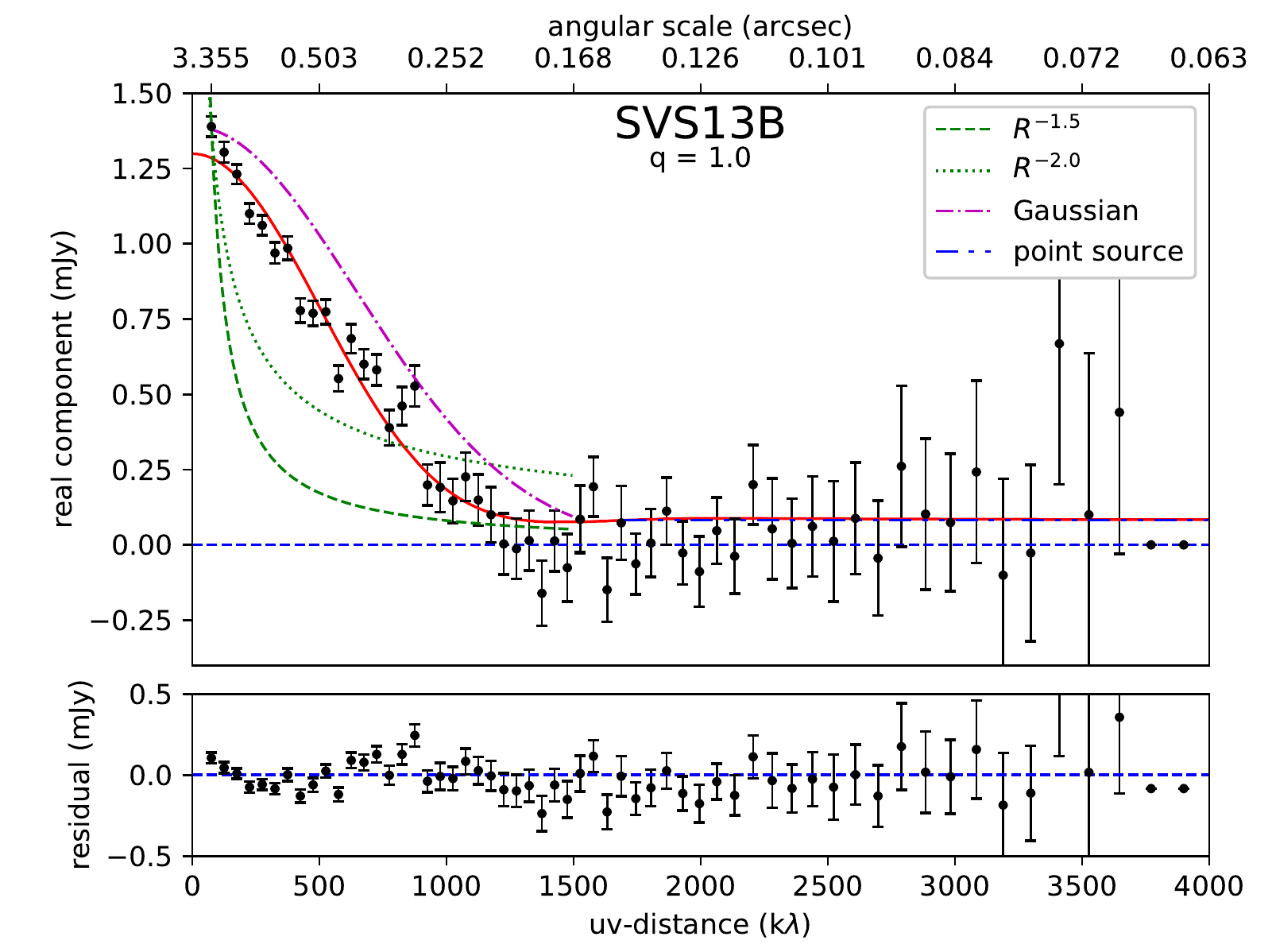}
                \caption{Real vs {\it u,v}-distance plot of 8 mm data for SVS13B. Top:  real component of data. The blue dashed line indicates real component of zero.  The red solid line is the best-fit
        model. The green dashed and dotted lines correspond to R$^{-2.0}$ and R$^{-1.5}$ envelope visibility profiles. The magenta dot-dashed line is a Gaussian profile that corresponds to the image-plane Gaussian fit.    Bottom: residual of real component minus model.}
        \label{SVS13Buvdist}
\end{figure}

After we completed modeling the candidate disks, we generated synthetic disk images with the parameters of the best-fit model for each source. We determined the best-fit disk models from the lowest $\chi^{2}_{reduced}$ value (i.e.~the maximum likelihood) of all models we fitted to the data. We Fourier transformed the best-fit model synthetic disk, with the same sampling at the same {\it u,v}-points as the data, to produce model visibilities.  We generated residual visibilities by subtracting the model visibilities from the data visibilities.  We then imaged the model and residual visibilities using the same weighting as the data to produce synthetic maps of the best-fit disk models and residuals of each modeled source.   We produced images with robust=0.25 weighting, which is a trade off between slightly higher resolution at the expense of slightly worse sensitivity.  With robust=0.25 weighting, the disks are extended and the outer parts of the disks are relatively well-detected over the noise level. While we do not model the data in the image plane, examining the results in both the image and {\it u,v}-planes is useful to study the full extent of the disks. See Figures \ref{SVS13Bdmr},  \ref{SVS13Buvdist}, and  Appendix \ref{vandam:bestfits} for plots of modeling results both in image and {\it u,v}-planes.

 The results of the disk modeling are listed in Table \ref{diskfulltab3}.  For most sources, $\chi^{2}_{reduced}$ values are near 1, indicating that our disk model accounts for the majority of the emission from the candidate disks.  Thus, these sources are likely to be Class 0 and Class I disks rather than simply dominated by inner envelope structure \citep{Chiang2008}.  The majority of the modeled candidate disks have modeled disk radii larger than 10 AU.  $R$ $=$ 10 AU is an upper limit predicted by magnetic braking models \citep{Dapp2010} during the Class 0 stage.  
 
  Theory predicts values of $q$ to be near 0.5 \citep[e.g.,][]{Chiang1997} for more-evolved Class II protostars.  Our protostars are significantly younger, in the Class 0 and I stage; for our sample, values of $q$ $<$0.5 are more likely. Because in the early protostellar phases the mass reservoir of the envelope is large, radiation is reprocessed from the protostar and directed back onto the disk.  This increases the brightness of the outer disk relative to the inner disk, resulting in a flattened brightness distribution
\citep{Dalessio1996}.   Indeed, the majority of our sources have lowest $\chi^{2}_{reduced}$ across all disk-only models when $q=0.25$.

\startlongtable
\begin{deluxetable}{lcccccc}
\tablecaption{Best-fit Disk Modeling Results for Full VANDAM Survey \label{diskfulltab3}}
\tablehead{
\colhead{Source} & \colhead{q} & \colhead{$\gamma$} & \colhead{$R_{c}$}  &\colhead{$F_{fit}$} & \colhead{$M_{fit}$} &\colhead{$\chi^{2}_{\mathrm{reduced}}$} \\
\colhead{} & \colhead{} & \colhead{} & \colhead{(AU)} &\colhead{($\mu$Jy)}& \colhead{(M$_\odot$)} & \colhead{} 
}
\startdata
SVS13B & 0.25 & 0.21$^{+0.23}_{-0.20}$ & 24.3$_{-1.7}^{+2.1}$ & 1309$_{-18}^{+19}$& 0.14 - 0.28& 2.194 \\[0.0ex]
 & 0.50& 0.42$^{+0.25}_{-0.21}$  &  25.5$_{-1.5}^{+1.9}$  & 1305$_{-22}^{+21}$& 0.14 - 0.28& 2.185  \\[0.0ex] 
 & 0.75& 0.63$^{+0.24}_{-0.22}$ & 26.5$_{-1.4}^{+1.6}$  & 1302$_{-26}^{+26}$& 0.14 - 0.28& 2.175 \\[0.0ex]
 & 1.00& 0.85$^{+0.26}_{-0.23}$ & 27.3$_{-1.2}^{+1.4}$  & 1299$_{-31}^{+30}$& 0.14 - 0.28& 2.164  \\[0.0ex]
\cline{1-7}
Per-emb-50 & 0.25& 0.08$^{+0.02}_{-0.16}$ & 21.9$_{-0.9}^{+0.8}$ & 1617$_{-21}^{+21}$&0.17 - 0.35& 1.556  \\[0.0ex]
 & 0.50&  0.26$^{+0.15}_{-0.17}$ & 23.3$_{-1.0}^{+1.1}$ & 1616$_{-24}^{+24}$&0.17 - 0.35& 1.558 \\[0.0ex]
 & 0.75& 0.44$^{+0.16}_{-0.17}$  & 24.6$_{-1.1}^{+1.4}$ & 1614$_{-29}^{+30}$&0.17 - 0.35& 1.560 \\[0.0ex]
 & 1.00&  0.64$^{+0.16}_{-0.18}$ & 25.7$_{-1.3}^{+1.4}$ & 1613$_{-35}^{+34}$&0.17 - 0.35& 1.563 \\[0.0ex]
\cline{1-7}
Per-emb-14 & 0.25& -0.11$^{+0.16}_{-0.00}$ &28.5$_{-2.1}^{+2.3}$ & 877$_{-21}^{+20}$&0.09 - 0.19& 1.110  \\[0.0ex]
  & 0.50& 0.09$^{+0.08}_{-0.21}$ &  30.6$_{-2.3}^{+2.8}$& 877$_{-19}^{+21}$&0.09 - 0.19& 1.114  \\[0.0ex]
  & 0.75& 0.27$^{+0.17}_{-0.24}$ & 32.5$_{-2.8}^{+2.2}$ & 876$_{-22}^{+23}$&0.09 - 0.19& 1.119  \\[0.0ex]
  & 1.00& 0.48$^{+0.19}_{-0.23}$ & 33.9$_{-3.1}^{+3.6}$ & 874$_{-26}^{+25}$&0.09 - 0.19& 1.123 \\[0.0ex]
\cline{1-7}
Per-emb-30 & 0.25&0.02$^{+0.18}_{-0.31}$  & 14.0$_{-0.9}^{+1.0}$& 948$_{-15}^{+16}$&0.10 - 0.20& 1.100 \\[0.0ex]
 & 0.50& 0.20$^{+0.04}_{-0.32}$  & 14.9$_{-1.1}^{+1.9}$ & 948$_{-27}^{+22}$&0.10 - 0.20&1.102 \\[0.0ex]
 & 0.75& 0.39$^{+0.30}_{-0.34}$  & 15.8$_{-1.3}^{+1.9}$  & 948$_{-33}^{+28}$&0.10 - 0.20&1.104  \\[0.0ex]
 & 1.00&0.59$^{+0.14}_{-0.33}$  & 16.5$_{-1.6}^{+31.3}$ & 944$_{-28}^{+33}$&0.10 - 0.20&1.107  \\[0.0ex]
\cline{1-7}
HH211-mms & 0.25& 0.48$^{+0.40}_{-0.78}$ & 10.5$_{-0.8}^{+0.8}$& 737$_{-16}^{+20}$&0.08 - 0.16& 1.009  \\[0.0ex]
 & 0.50& 0.65$^{+0.43}_{-0.82}$  & 11.0$_{-0.9}^{+1.0}$ & 737$_{-24}^{+28}$&0.08 - 0.16& 1.009 \\[0.0ex]
 & 0.75& 0.81$^{+0.42}_{-0.79}$ &11.5$_{-1.2}^{+1.2}$ & 737$_{-30}^{+37}$&0.08 - 0.16& 1.009  \\[0.0ex]
 & 1.00& 1.01$^{+0.44}_{-0.81}$  &11.9$_{-1.3}^{+1.4}$ & 737$_{-36}^{+42}$&0.08 - 0.16& 1.009  \\[0.0ex]
\cline{1-7}
IC348 MMS & 0.25& -0.58$^{+0.11}_{-0.11}$ & 25.7$_{-2.2}^{+2.8}$ & 1044$_{-26}^{+27}$& 0.11 - 0.23& 1.085  \\[0.0ex]
  & 0.50& -0.39$^{+0.19}_{-0.11}$ & 29.0$_{-2.6}^{+3.2}$& 1039$_{-22}^{+22}$&0.11 - 0.22 & 1.096  \\[0.0ex]
  & 0.75&  -0.19$^{+0.11}_{-0.27}$ & 31.6$_{-2.9}^{+4.1}$  & 1035$_{-25}^{+25}$&0.11 - 0.22& 1.107  \\[0.0ex]
  & 1.00& 0.02$^{+0.07}_{-0.11}$ & 33.7$_{-3.1}^{+4.3}$  & 1031$_{-28}^{+27}$&0.11 - 0.22& 1.118  \\[0.0ex]
\cline{1-7}
Per-emb-8 & 0.25& 0.01$^{+0.16}_{-0.19}$  &19.0$_{-1.1}^{+1.2}$ & 1076$_{-19}^{+21}$&0.11 - 0.23& 1.099  \\[0.0ex]
 & 0.50&   0.20$^{+0.17}_{-0.20}$   & 20.2$_{-1.3}^{+1.4}$& 1074$_{-19}^{+19}$&0.11 - 0.23&  1.107  \\[0.0ex]
 & 0.75&   0.40$^{+0.17}_{-0.21}$  &21.2$_{-1.4}^{+1.6}$  & 1071$_{-21}^{+22}$&0.11 - 0.23& 1.114  \\[0.0ex]
 & 1.00& 0.61$^{+0.17}_{-0.20}$ & 22.1$_{-1.6}^{+1.8}$ & 1070$_{-24}^{+24}$&0.11 - 0.23& 1.122  \\[0.0ex]
 \cline{1-7}
Per-emb-25 & 0.25&  0.15$^{+0.50}_{-0.35}$  & 28.4$_{-3.3}^{+4.1}$ & 704$_{-23}^{+23}$&0.07 - 0.15& 1.321  \\[0.0ex]
 & 0.50&   0.36$^{+0.49}_{-0.34}$   &  27.5$_{-3.6}^{+4.6}$& 702$_{-20}^{+20}$&0.07 - 0.15& 1.327  \\[0.0ex]
 & 0.75&   0.56$^{+0.49}_{-0.34}$  & 26.3$_{-4.0}^{+5.0}$  & 702$_{-21}^{+21}$&0.07 - 0.15& 1.334  \\[0.0ex]
 & 1.00&  0.78$^{+0.52}_{-0.35}$ &  25.0$_{-4.5}^{+5.5}$ & 700$_{-24}^{+22}$&0.07 - 0.15& 1.340 \\[0.0ex]
  \cline{1-7}
NGC 1333 IRAS1 A& 0.25&  0.52$^{+1.58}_{-0.66}$  & 11.1$_{-1.5}^{+2.1}$ & 414$_{-15}^{+16}$&0.04 - 0.09&  1.087 \\[0.0ex]
 & 0.50&  0.68 $^{+2.59}_{-0.99}$   &  11.6$_{-2.2}^{+7.4}$& 414$_{-30}^{+31}$&0.04 - 0.09& 1.089  \\[0.0ex]
 & 0.75&   0.88$^{+2.24}_{-0.73}$  & 12.0$_{-2.2}^{+3.3}$  & 414$_{-28}^{+33}$&0.04 - 0.09&  1.093 \\[0.0ex]
 & 1.00&  1.06$^{+2.26}_{-1.10}$ &  12.4$_{-2.9}^{+11.7}$ & 414$_{-59}^{+43}$&0.04 - 0.09& 1.096 \\[0.0ex]
  \cline{1-7}
Per-emb-62 & 0.25&  0.04$^{+0.33}_{-0.27}$  & 21.5$_{-2.0}^{+2.6}$ & 680$_{-24}^{+25}$&0.07 - 0.15& 1.059  \\[0.0ex]
 & 0.50&   0.22$^{+0.34}_{-0.28}$   &  23.0$_{-2.3}^{+3.0}$& 680$_{-26}^{+25}$&0.07 - 0.15& 1.058  \\[0.0ex]
 & 0.75&   0.41$^{+0.35}_{-0.29}$  & 24.2$_{-2.6}^{+3.5}$  & 680$_{-28}^{+27}$&0.07 - 0.15&  1.058 \\[0.0ex]
 & 1.00&  0.61$^{+0.35}_{-0.29}$ &  25.3$_{-2.9}^{+3.9}$ & 679$_{-29}^{+29}$&0.07 - 0.15&1.057  \\[0.0ex]
  \cline{1-7}
Per-emb-63 & 0.25&  -0.56$^{+0.54}_{-0.38}$  & 20.6$_{-3.7}^{+7.8}$ & 354$_{-18}^{+19}$&0.04 - 0.08& 1.080  \\[0.0ex]
 & 0.50&   -0.38$^{+0.58}_{-0.42}$   &  23.2$_{-4.6}^{+3.3}$& 356$_{-28}^{+25}$&0.04 - 0.08& 1.079  \\[0.0ex]
 & 0.75&   -0.18$^{+0.58}_{-0.42}$  & 25.5$_{-5.3}^{+4.7}$  & 355$_{-37}^{+33}$&0.04 - 0.08&  1.078 \\[0.0ex]
 & 1.00& 0.04$^{+0.57}_{-0.42}$ & 27.0$_{-6.2}^{+5.2}$ & 355$_{-41}^{+35}$&0.04 - 0.08& 1.077 \\[0.0ex]
  \cline{1-7}
SVS13C & 0.25&  -0.32$^{+0.11}_{-0.10}$  & 32.9$_{-1.4}^{+1.6}$ & 2317$_{-22}^{+22}$&0.24 - 0.50&  1.638 \\[0.0ex]
 & 0.50&   -0.13$^{+0.12}_{-0.11}$   &  35.8$_{-1.8}^{+2.1}$& 2312$_{-28}^{+28}$&0.24 - 0.50& 1.656  \\[0.0ex]
 & 0.75&   0.08$^{+0.13}_{-0.12}$  & 38.2$_{-2.2}^{+2.6}$  & 2305$_{-38}^{+37}$&0.24 - 0.50&  1.674 \\[0.0ex]
 & 1.00&  0.29$^{+0.15}_{-0.12}$ &  40.2$_{-2.8}^{+3.0}$ & 2300$_{-45}^{+50}$&0.24 - 0.50& 1.691 \\[0.0ex]
  \cline{1-7}
NGC 1333 IRAS4A & 0.25&  0.07$^{+0.03}_{-0.03}$  & 38.1$_{-1.0}^{+1.1}$ & 8743$_{-299}^{+344}$&0.92 - 1.89& 1.642  \\[0.0ex]
 & 0.50&  0.27 $^{+0.03}_{-0.03}$   &  39.8$_{-1.0}^{+1.0}$& 8488$_{-238}^{+260}$&0.90 - 1.83&1.588   \\[0.0ex]
 & 0.75&   0.48$^{+0.03}_{-0.02}$  & 41.2$_{-1.0}^{+1.0}$  & 8263$_{-204}^{+219}$&0.87 - 1.78& 1.553  \\[0.0ex]
 & 1.00&  0.69$^{+0.03}_{-0.02}$ &  42.2$_{-0.9}^{+0.9}$ & 8072$_{-175}^{+185}$&0.85 - 1.74& 1.535 \\[0.0ex]
   \cline{1-7}
NGC 1333 IRAS2A& 0.25&  1.65$^{+1.74}_{-0.72}$  & 9.1$_{-0.4}^{+0.4}$ &1670$_{-23}^{+33}$&0.18 - 0.36&  3.853 \\[0.0ex]
 & 0.50&   1.86$^{+2.40}_{-0.42}$   &  9.3$_{-1.2}^{+1.0}$& 1669$_{-37}^{+28}$&0.18 - 0.36&  3.855 \\[0.0ex]
 & 0.75&   1.99$^{+2.51}_{-0.73}$  & 9.5$_{-1.3}^{+1.4}$  & 1669$_{-49}^{+59}$&0.18 - 0.36&  3.861 \\[0.0ex]
 & 1.00&  2.18$^{+2.35}_{-1.81}$ &  9.7$_{-1.9}^{+1.1}$ & 1670$_{-57}^{+59}$&0.18 - 0.36&  3.864\\[0.0ex]
      \cline{1-7}
      NGC 1333 IRAS4B  & 0.25&  -0.77$^{+0.17}_{-0.17}$  & 16.0$_{-2.8}^{+7.8}$ & 1071$_{-58}^{+147}$&0.11 - 0.23& 1.009   \\[0.0ex]
 & 0.50&   -0.59$^{+0.16}_{-0.14}$   &  18.8$_{-3.5}^{+8.1}$& 1065$_{-35}^{+65}$&0.11 - 0.23& 1.009   \\[0.0ex]
 & 0.75&   -0.40$^{+0.15}_{-0.12}$  & 21.1$_{-3.9}^{+6.7}$  & 1059$_{-34}^{+35}$&0.11 - 0.23& 1.009  \\[0.0ex]
 & 1.00&  -0.19$^{+0.13}_{-0.11}$ &  23.0$_{-3.9}^{+6.4}$ & 1053$_{-34}^{+32}$&0.11 - 0.23&  1.010 \\[0.0ex]
  \cline{1-7}
NGC 1333 IRAS1 B& 0.25&  42.02$^{+10.92}_{-1.55}$  & 6.8$_{-2.4}^{+160.6}$ & 264$_{-29}^{+94}$&0.03 - 0.06& 1.302   \\[0.0ex]
 & 0.50&   42.46$^{+12.25}_{-2.08}$   &  6.8$_{-1.3}^{+8.8}$& 264$_{-41}^{+29}$&0.03 - 0.06& 1.302  \\[0.0ex]
 & 0.75&  53.96$^{+14.61}_{-2.00}$  & 6.8$_{-1.9}^{+106.2}$  & 263$_{-70}^{+36}$&0.03 - 0.06& 1.302  \\[0.0ex]
 & 1.00& 36.71$^{+8.72}_{-1.47}$ &  6.8$_{-2.7}^{+970.4}$ & 265$_{-71}^{+45}$&0.03 - 0.06&1.302  \\[0.0ex]
  \cline{1-7}
B5-IRS1 & 0.25&  -1.41$^{+4.20}_{-0.22}$  & 55.8$_{-79.3}^{+37383.5}$ & 279$_{-82}^{+272}$&0.03 - 0.06& 1.050  \\[0.0ex]
 & 0.50&   -1.21$^{+5.30}_{-0.30}$   &  88.6$_{-25.7}^{+395.2}$& 275$_{-37}^{+141}$&0.03 - 0.06&  1.049 \\[0.0ex]
 & 0.75&   -0.99$^{+5.47}_{-0.35}$  & 116.8$_{-25.9}^{+728.8}$  & 276$_{-46}^{+49}$&0.03 - 0.06& 1.048  \\[0.0ex]
 & 1.00&  -0.76$^{+2.77}_{-0.27}$ &  135.9$_{-74.0}^{+1686.0}$ & 277$_{-48}^{+57}$&0.03 - 0.06&1.047  \\[0.0ex]
\enddata
\tablecomments{Values of $q$ are fixed.  Values of $\gamma$, $R_{c}$, and $F_{fit}$ are determined from best-fit models. $M_{fit}$ is the estimated model mass calculated from $F_{fit}$, with the range given from varying $T_{d}$, as described in Section \ref{vandam:masses}.  Uncertainties reflect 90\% confidence intervals.}
\end{deluxetable}
 
 When we varied the fixed values of $q$ with each model, the best fit $\gamma$  changed enough that even when including uncertainties, the values of $\gamma$  for models with $q$ between $0.25$ and $1.00$ will not be fully in agreement.  Despite the uncertainty of  $\gamma$ in these sources as $q$ varies, $R_{c}$---our proxy for outer disk radius---typically remains in agreement across all best fit models for each candidate disk.  Negative values of $\gamma$ are unphysical for smooth disks, indicating that the surface brightness would increase with increasing radius as distance from the central protostar increased. While a negative value of $\gamma$ could describe a disk with a cavity around the protostar \citep[e.g.,][]{Tazzari2017}, because we have marginally-resolved emission for many sources, we consider a cavity unlikely to be detected with our data.  Thus, we require at least one disk model to have a positive value of  $\gamma$ for the source to be considered a candidate disk.   For most sources, we find positive values of $\gamma$ in at least one of the best fit models where $q=[0.25,0.50,0.75,1.00]$.  These trends, along with the $\chi^{2}_{reduced}$ values mostly near 1 and relatively empty residual maps in the image plane (Appendix \ref{vandam:bestfits}), indicate that Class 0 and I disks are likely present in the Perseus molecular cloud.  

Some sources have positive central residuals which remain after subtracting the model from the data.  In the bright sources NGC 1333 IRAS4A and NGC 1333 IRAS4B---for which we attempted to control for the envelope flux by removing the short-spacing baselines for the fit (see Section \ref{vandam:modeling})---the positive central residuals are most likely due to remaining envelope flux not accounted for by the short-spacing baseline cut.  Less bright sources with positive residuals could have a minimal envelope contribution or may be better described by a more physical disk model with more parameters (e.g.~Figures \ref{Per8dmr} and \ref{Per25dmr}).  A few sources have negative residuals that are near the central emission peak, but the negative dips are slightly off center (e.g.~Figures \ref{Per35Admr} and \ref{IRAS2Admr}), which probably arises from minor asymmetries in the sources.  In nearly all sources, the majority of the emission remains well-described by our adopted disk profile, and we note the few cases of larger deviations from our model in Section \ref{vandam:results:marginal}. 

\subsection{Sources Not Well Described by the Disk Model} \label{vandam:results:marginal}

We performed the disk modeling on all extended sources from the VANDAM survey which were nearly axisymmetric.  Not all of these extended sources are disk candidates:  disk modeling has revealed that NGC 1333 IRAS1 B and B5-IRS1 are not well-described by a disk profile.  The results of the disk models have marginally resolved disks or are unphysical for disks, and their values of $R_{c}$ have large uncertainties (Table \ref{diskfulltab3}).     NGC 1333 IRAS1 B was marginally resolved and has extremely high, unphysical values of $\gamma$, and a disk radius less than a half beam.  For NGC 1333 IRAS1 B, the model, with its steep $\gamma$ and small radius, is approximating more closely an envelope profile than a disk profile, therefore we do not consider it a disk candidate.  B5-IRS1 has no positive values for $\gamma$ for any value of model $q$ and is therefore inconsistent with a smooth disk profile and not a candidate disk. 

NGC 1333 IRAS4B has a dense envelope of which we remove a majority of the emission by applying a \textit{u,v}-cut to the inner 350 k$\lambda$ during the imaging process.  We do not fit our disk model to these inner baselines (Section \ref{vandam:modeling}).  We note that in the 8 mm data, the peak of the emission is slightly offset from candidate disk center.  Both the offset peak and the offset from phase center likely contribute to the ringing seen in Figure \ref{Per13uvdist}.    Although the value of $\chi^{2}_{reduced}$ is rather low ($\sim$1.009), all values of $q$ give negative best-fit values for $\gamma$, inconsistent with a disk profile.  The deprojected and averaged visibility profile falls between the envelope profiles for free-fall collapse and a singular isothermal sphere, and is consistent with an envelope profile (Figure \ref{Per13uvdist}).   Thus, we do not consider NGC 1333 IRAS4B to be a candidate disk, and it is more likely to be dominated by envelope emission at the resolution of our observations.

\subsection{Description of Candidate Disk Modeling Results} \label{vandam:results:candidates}

The best-fit disk models for the candidate disks modeled in \citet{SeguraCox2016} are reported and described in that work.  We include the best-fit disk models of those candidate disks in Table \ref{diskfulltab3} with the rest of the VANDAM extended source model results for completeness.  A description of information previously known about each source is available in Appendix  \ref{vandam:gallery}.  Plots of the best fit results for each source in the image- and {\it u,v}-planes are included in Appendix \ref{vandam:bestfits}, for the full sample of VANDAM extended sources.   SVS13B and Per-emb-8 have minor extensions to the southwest, contributing to minor ringing in their respective visibility profiles.  Per-emb-30 also has ringing in the visibility profile, likely due in-part to the small peak of emission to the southeast of the source.  The candidate disk of IC348 MMS is irregularly shaped, also producing ringing in the visibility profile.

Per-emb-25 was estimated to have a disk along the north-south direction from image-plane Gaussian fitting. Gaussian fitting for the inclination and position angles is most uncertain in this source due to the asymmetric extensions protruding from the central protostar; however, the estimated disk position angle is roughly perpendicular to a known jet \citep{Dunham2014b}.   The best-fit models of Per-emb-25 have $\chi^{2}_{reduced}\sim1.33$, an intermediate value among the candidate disks.  A central residual component is seen in the Figure \ref{Per25dmr}, likely due to a small unmodeled envelope component.  The modeled flux of the disk also falls below the zero-baseline flux (Figure \ref{Per25uvdist}), indicating that the inner few baselines may have a small envelope component.  Figure \ref{Per25uvdist} also shows ringing, especially at long baselines, likely due to the asymmetric dust extensions seen in the 8 mm data.

Per-emb-62 and Per-emb-63 and NGC 1333 IRAS1 A all have relatively low $\chi^{2}_{reduced}$ values near 1.10 despite mild ringing in the visibility profile. Per-emb-62 and Per-emb-63 have $R_{c}$ slightly larger than 20 AU with irregular emission in the immediate vicinity surrounding the protostars, contributing to the ringing in Figures \ref{Per62uvdist} and \ref{Per63uvdist}.   NGC 1333 IRAS1 A has a small peak of emission to the northwest (seen in Figure \ref{Per35Admr}) as well as a companion separated by 1.908$^{\prime\prime}$ \citep{Tobin2016a} that was subtracted from the visibility profile further from the candidate disk, both of which probably contribute to the ringing seen.

The 8 mm emission from SVS13C is extended in both the east-west and north-south directions (Figure \ref{SVS13Cdmr}).  A free-free jet is present in the system along the north-south direction \citep[][see also Appendix \ref{vandam:gallery:binaries:svs13c}]{Tychoniec2018}.  The north-south emission is non-Gaussian and irregular, making subtraction of the north-south emission difficult.  Because the east-west emission likely arises from a disk, we chose not to apply a \textit{u,v}-cut by removing the shortest ($<$350 k$\lambda$) baselines to remove the larger scale north-south emission.  A \textit{u,v}-cut would also remove the east-west emission to which we apply a disk model.  We chose to proceed with the standard modeling procedure described in Section \ref{vandam:modeling}, without accounting for the jet-like north-south emission.  The resulting best-fit models (Table \ref{diskfulltab3}) have a $\chi^{2}_{reduced}\sim1.65$, an intermediate $\chi^{2}_{reduced}$ value for our modeled sources.  The two model with the lowest $\chi^{2}_{reduced}$ have $q=[0.25,0.50]$ with negative values of $\gamma$.  While negative $\gamma$ could indicate a hole in the innermost unresolved regions of a disk \citep{Tazzari2017}, the close to edge-on orientation of SVS13C ($\sim$75$^{\circ}$ inclination) would cause a hole to be hidden by disk emission.  The models with $q=[0.75,1.00]$ however have positive values of $\gamma$, consistent with a disk profile.  All four best-fit models have $R_{c}\sim35$ AU, revealing SVS13C to be the second largest modeled candidate disk in the VANDAM survey.  

NGC 1333 IRAS4A is by far the brightest candidate disk in our sample (Table \ref{diskfulltab2}) with a known dense envelope \citep[e.g.,][]{Looney2000}. For imaging, we apply a \textit{u,v}-cut to the shortest ($<$350 k$\lambda$) baselines, corresponding to large-scale emission, to remove a majority of the envelope contamination from the source and better reveal the disk component.  As described in Section \ref{vandam:modeling}, we also do not fit  the inner 350 k$\lambda$ baselines to our disk model.  A central component is left in the residuals (Figure \ref{IRAS4Admr}), also seen in the  inner baselines where the zero-baseline model flux does not match the zero-baseline data (Figure \ref{IRAS4Auvdist}).  We attribute this discrepancy to the unmodeled envelope component.  The best fit model has a relatively steep $q=1.00$ value for an embedded source, possibly influenced by the dense envelope.

NGC 1333 IRAS2A is the smallest candidate disk we model (Table \ref{diskfulltab3}).  The small disk radius is reflected in Figure \ref{IRAS2Auvdist}, with the disk-like gentle slope feature extending further in {\it u,v}-distance than any candidate disk source.  NGC 1333 IRAS2A has a small asymmetric extension in the northwest direction, likely leading to the low-amplitude ringing in the visibility profile.  A small residual is seen toward the center of the disk in the image plane, which we attribute to a small amount of unmodeled envelope emission.  NGC 1333 IRAS2A has the highest $\chi^{2}_{reduced}$ value of all modeled sources ($\sim$3.853), and with a modeled disk radius of just $\sim$9 AU, the disk diameter is barely larger than the beam.  We do consider NGC 1333 IRAS2A to be a candidate disk, but we note that this source may be a barely resolved disk with some envelope contamination to consider.

\section{DISCUSSION}     \label{vandam:discussion}

\subsection{Candidate Disk Properties}    \label{vandam:discussion:diskprop}

The 18 VANDAM candidate disks have estimated masses of 0.01--3.2 \Ms~ ($M_{d}$; Table \ref{diskfulltab2}).  As discussed in Section \ref{vandam:masses}, the estimated mass of L1448 IRS3B is likely under-estimated, and NGC 1333 IRAS4A is an unusual outlier with $M_{d}$ an order of magnitude larger than all other sources.  The remaining 17 candidate disks have $M_{d}$ values of 0.03--0.71 \Ms.  Our values for the estimated masses of the candidates are all larger than the Minimum Mass Solar Nebula, the 0.01 \Ms~ amount of material expected to be required to form the planets in our own Solar System \citep{Weidenschilling1977}, indicating that these disks have the potential to eventually form planets.  The modeled fit fluxes and hence masses calculated from the fit fluxes ($F_{fit}$ and $M_{fit}$ respectively; see Table \ref{diskfulltab3}) for all modeled candidate disks except NGC 1333 IRAS4A are within 0.7 to 1.3 times the measured fluxes and estimated masses from observations ($F_{8mm}$ and $M_{d}$ respectively; see Table \ref{diskfulltab2}), with the largest deviations occurring in sources with smaller modeled radii. The value of $M_{fit}$ of NGC 1333 IRAS4A is a factor of $\sim$0.56 lower compared to the observed $M_{d}$ value, likely because of the remaining envelope emission seen in this source which was not accounted for in the modeling procedure.

When scaled to our opacity, the Class 0 protostar L1527's disk mass is 0.013 \Ms~\citep{Tobin2013b} with T$_{d}$ = 30 K. \citet{Harsono2014} revealed four Class I disks to have masses of 0.004-0.033 \Ms, using T$_{d}$ = 30 K and \citet{Ossenkopf1994} opacities.  Compared to these embedded disks, our disk masses appear to be significantly higher.  One possibility is that our assumption of dust opacity spectral index $\beta$ = 1 is too large.  L1527 was found to have a shallower $\beta$ $\sim$ 0  \citep{Tobin2013b} from $\sim$mm wavelengths, which could be attributed to a population of large ($\sim$cm) dust grains centered in the disk midplane with a smaller scale height than observed at shorter wavelengths.  Our 8 mm data indeed do trace large grains settled in the midplane, indicating that values of $\beta$ near 0 may be more common than previously thought for deeply embedded young disks, as suggested by \citet{Kwon2015}.  If we assume $\beta$ = 0 instead of $\beta$ = 1, our estimated disk masses would change from 0.02--0.71 \Ms to 0.003--0.10 \Ms.  Because we model the disk flux, not mass, any uncertainties in $\beta$ do not impact our modeling results.  

Our best-fit models for 14 candidate disks give -0.58 $<$ $\gamma$ $<$ 1.65 (Table \ref{diskfulltab3}), with the average value of inner-disk surface density power law $\gamma=0.32$ for our Class 0 and I sources. Negative values of $\gamma$ imply increasing disk surface density with radius, incongruent with typical disk profiles.  For all our candidate disks at least one value of $q$, the disk temperature structure parameter, produces a positive best-fit value of $\gamma$.  The average value of $\gamma=0.32$ is a shallower profile than more evolved disks.  Disks around Class II protostars in Ophiuchus yield an a typical value of $\gamma=0.9$ \citep{Andrews2009}.  The steeper values of $\gamma$ in Class II sources indicates that evolved disks are generally  more centrally concentrated than our Class 0 and I disks.

The few Class 0 disks with Keplerian rotation have relatively large radii, though it is unclear if these are typical radii of young disks or if this is simply detection bias towards large and bright sources.  At 1.3 mm, VLA 1623 has $R$ $\sim$189 AU \citep{Murillo2013}, Lupus 3 MMS has $R$ $\sim$100 AU \citep{Yen2017}, and L1527 has $R$ $\sim$54 AU \citep{Ohashi2014}. HH212 has $R$ $\sim$60 AU in 850 $\mu$m ALMA data \citep{Codella2014,Lee2017}.  Our Class 0 and I VANDAM candidate disks have 9.1 AU $<$ $R_{c}$ $<$ 42.2 AU.  The modeled radii (Table \ref{diskfulltab3}) give disk diameters that are a factor of 1 to 1.5 times larger than the deconvolved sizes from the image-plane 2D Gaussian fits (Table \ref{diskfulltab2}). For most sources, the candidate Class 0 and I disks are larger than the expected upper limit of 10 AU from strong magnetic braking models \citep{Dapp2010}. HH211-mms, NGC 1333 IRAS1 A, and NGC 1333 IRAS2A are the three smallest disks with $R_{c}$ $\sim$10 AU.  The remaining candidate disks have $R_{c}$ consistent with the radii of Keplerian Class 0 disks L1527 and HH212.

\subsection{8 mm Emission as a Lower Limit on Dust Disk Radius}    \label{vandam:discussion:lowerlim}

As noted in Appendix \ref{vandam:gallery:singles:per14}, Per-emb-14 was resolved with CARMA in continuum dust emission at 1.3 mm \citep{Tobin2015b} with a dust disk a factor of $\sim$3 larger than our modeled 8 mm continuum radius \citep{SeguraCox2016}.  ALMA 1.3 mm data (Tobin et al.~2018, submitted) also show evidence for more-extended emission at 1.3 mm compared to 8 mm, with image-plane Gaussian fit major axes of the 1.3 mm data 1.7$\times$ to 4.3$\times$ larger than the 8 mm modeled radii presented here for disk candidates NGC 1333 IRAS1 A, SVS13B, NGC 1333 IRAS4A, and NGC 1333 IRAS2A.   A dependance on disk size with wavelength was also found for the more evolved classical T Tauri stars AS 209, CY Tau, and DoAr25 \citep{Perez2012,Perez2015}, with disk size decreasing with longer wavelength observations.  Since the wavelength of thermal emission from dust grains roughly traces the sizes of the dust grains, 8 mm emission traces a population of larger sized grains than in 1.3 mm emission.  The larger 8 mm grains experience radial drift to a larger extent \citep{Perez2012}, forming a more compact disk closer in to the central protostar than smaller grains which remain further from the protostar for longer periods of time \citep[e.g.,][]{Birnstiel2010}.  At shorter wavelengths, near 1 mm, dust emissivity is higher causing the dust emission to be stronger in the outer parts of the disk and more likely to be detected.  We consider our VANDAM modeled disk radii at 8 mm to be extreme lower limits on disk size.  Shorter wavelength observations may be better tracers of the full extent of circumstellar dust disks due to these large-grain radial-drift effects and surface brightness sensitivity limits at 8 mm.

\subsection{Outflow Orientations and Other Indirect Evidence of Disks}    \label{vandam:discussion:outflows}

Nearly all VANDAM candidate disks, except for Per-emb-63, have clearly associated outflows roughly perpendicular (60-90$^{\circ}$) to the major axis of the candidate disks (see Appendix \ref{vandam:gallery} for details).  No outflows are associated with Per-emb-63.  The orientations of outflows have been used as proxies for the disk rotation axis \citep[e.g.,][]{Hull2013}, hence outflows nearly perpendicular to extended continuum emission is a strong indicator of a protostellar disk.  Along with our continuum emission disk modeling, we use the perpendicular outflows as indirect evidence of rotationally supported disks in embedded sources.  

HH211-mms, Per-emb-14, Per-emb-15, Per-emb-25, Per-emb-8, SVS13B have bipolar outflows perpendicular to their candidate disk elongation.   Per-emb-30 and Per-emb-62 both have single monopolar outflows, possibly because of dense gas interacting with unseen components of their bipolar outflows.  Their detected monopolar outflows are both perpendicular to their candidate disk directions.

IC348 mms does not appear to be the central driving source of the bipolar outflow in its binary system though the candidate disk remains perpendicular to the outflow.  NGC 1333 IRAS4A has a close binary companion, each with bipolar outflows perpendicular to the estimated disk position angle of our candidate disk.  The binary system NGC 1333 IRAS2A drives two bipolar outflows, one coming from each close-separation component.  The NGC 1333 IRAS2A candidate disk is perpendicular to the outflow associated with its protostar.  NGC 1333 IRAS1 A drives an outflow almost perpendicular to its candidate disk and has a binary companion, which may be causing the outflow to have an S-shape via gravitational interactions between the binary protostars.  

SVS13C drives a bipolar outflow perpendicular to the candidate disk.  We detect evidence of the outflow in our 8 mm data.  As seen in Figure \ref{SVS13Cdmr}, right panel, the east-west emission component of SVS13C is well modeled with minimal residuals, while the north-south outflow seen in free-free \citep{Tychoniec2018}, remains.  Because we already accounted for a point-source component in our model, we did subtract out any small-scale free-free emission coming from the jet-launching regions of the disk \citep{Anglada1998}, leaving minimal residual at target center.

The non-axisymmetric candidate disks (Appendix \ref{vandam:gallery:weird}) cannot be fit with our modeling procedure, and for most the orientation of the candidate disk is unclear from 8 mm continuum data alone.  IRAS 03292+3039 has a bipolar outflow perpendicular to a velocity gradient across the protostar on 1000 AU scales \citep{Yen2015}.  IRAS 03282+3035 is a very close separation binary, with a velocity gradient along the outflow as well as a gradient perpendicular to the outflow on the southeast side of the envelope \citep{Tobin2011}.  The 8 mm data of Per-emb-18 is highly elongated along the direction perpendicular to the outflow \citep{Davis2008}.  Finally, the triple system L1448 IRS3B has a velocity gradient perpendicular to the outflows that are centered on two of the three triple components \citep{Tobin2016b}.

\subsection{The Frequency of Class 0 and I Candidate Disks}    \label{vandam:discussion:freq}

With the VANDAM survey, we have detected 18 new candidate disks (14 Class 0 and 4 Class I) in the deeply embedded, young protostellar phases.  Our survey has more than doubled the number of known possible disks around Class 0 and I protostars, bringing the total count from $\sim$15 to $\sim$33.  With so many young disks and candidate disks now known, we can characterize typical young embedded disk frequency and dust properties, determine the relative rarity of large embedded disks, look for evolutionary trends between the protostellar phases, and begin to study the role magnetic fields play during the early stages of disk growth.

Of the Class 0 protostars (including Class 0/I sources) in Perseus, 14/43 (33\%)  have candidate disks on scales of 12 AU or larger.  Only 4/37 (11\%) of Class I protostars in our sample have large, resolved candidate disks.  Here we have included both the modeled and unmodeled complicated candidate disks in our counts. We also find that 62/80 (78\%) of Class 0 and I protostars do not have signs of disks within our 8 AU radius modeling limit.  Since disk formation in protostars is expected to be a natural consequence of conservation of angular momentum during core collapse, this implies that at 8 mm most disks in the Class 0 and I phases are small  ($<$10 AU).  The population of unresolved disks may undergo stronger magnetic braking, be subject to other processes limiting disk growth, or the observations may be limited by surface brightness sensitivity.  Small disk size at 8 mm does not necessarily imply that the entire disk is small, because disks may be more extended at shorter wavelengths.

The lower proportion of Class I candidate disks compared to Class 0 candidate disks  is surprising because naively, the disk is expected to grow from the Class 0 to the Class I stage as the envelope is dissipated in part by accreting onto the disk, though no correlations between disk masses or radii have yet been found between the Class 0 and I phases \citep{Williams2011}.  The Class I candidate disks in Perseus may suffer from small-number statistics since only 4 Class I candidate disks were detected at 8 mm and may not reflect typical Class I disk frequency in other molecular clouds. This result only applies at 8 mm and is not a universal result since disk sizes vary with observational wavelength.  

An alternative explanation for the low proportion of VANDAM Class I candidate disks lies in the size of the dust grains seen in our observations. Our data trace large dust grains ($\sim$8 mm), since thermal emission from dust roughly traces the size of the emitting grains.   It is possible that the candidate disks around more evolved Class I protostars have been stable long enough for radial drift \citep{Perez2012} to cause the 8 mm dust grain population to be more centrally concentrated around the protostar relative to Class 0 sources (Figure \ref{tbolvsgamma} may also support this scenario; see Section \ref{vandam:discussion:trends}). Observations made at shorter wavelengths which trace smaller dust grains reflect more extended disk sizes since smaller grains have less pronounced radial drift effects. The detectable 8 mm radius of an embedded disk may shrink as the protostar evolves due to  8 mm grain population radially drifts inwards causing the signal-to-noise ratio in the outer disk to decrease below instrument sensitivity thresholds.  If disks detected at 8 mm do become more centrally concentrated as they evolve with a smaller detectable radius, Class I candidate disks may have had larger 8 mm disks in the past which may have been detected with 12 AU resolution, but at present the 8 mm disks may have shrunk to below resolution or sensitivity limits.

\begin{figure}[!ht]
    \centering
        \begin{minipage}{.47\textwidth}
        \centering
                \includegraphics[width=0.95\textwidth]{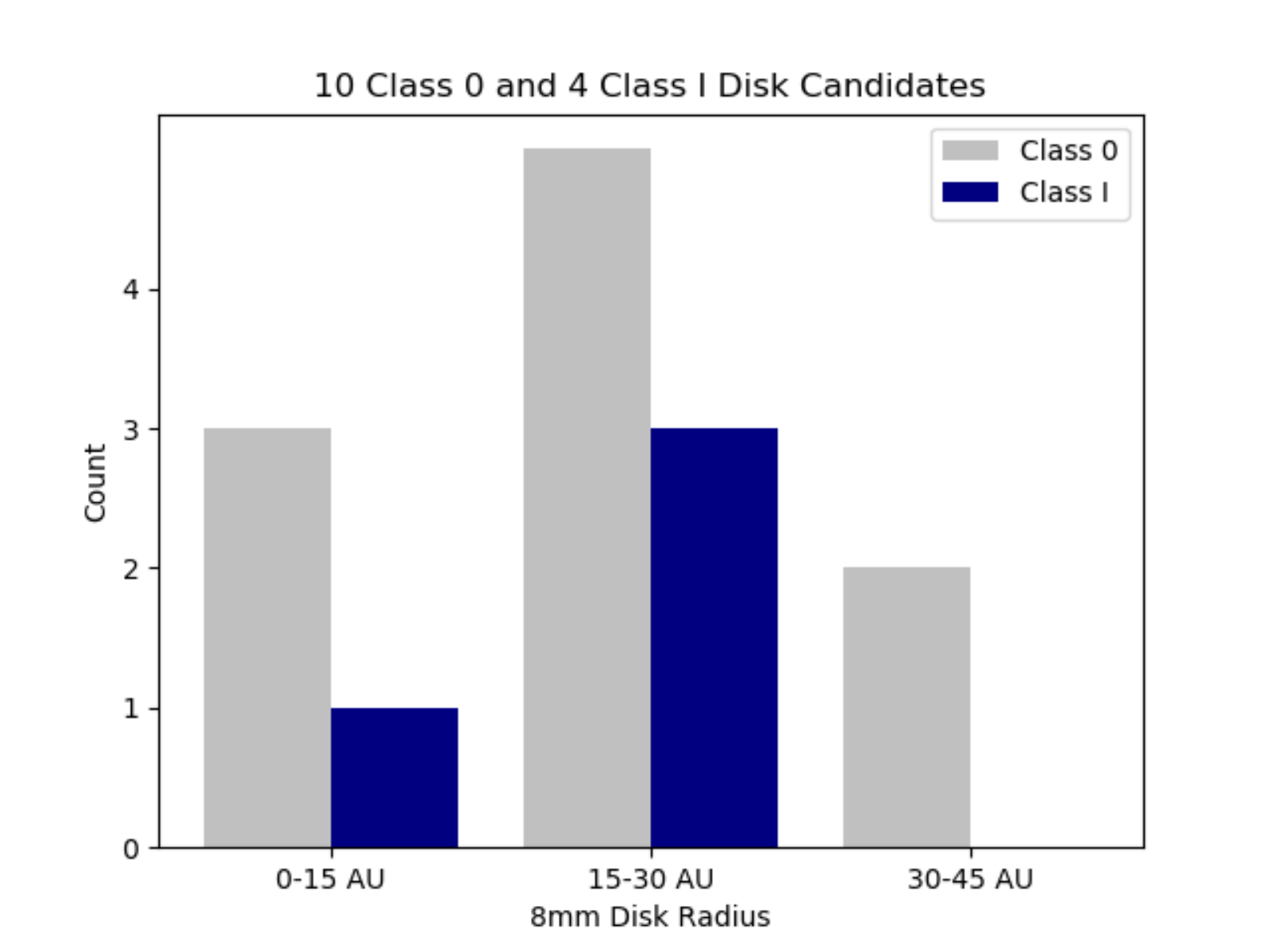}
        \caption{A histogram of the radii of all 14 modeled candidate disks, broken down by protostellar Class. }
\label{diskhisto}
    \end{minipage}%
        \hspace{0.2cm}
    \begin{minipage}{0.47\textwidth}
        \centering
                \includegraphics[width=0.95\textwidth]{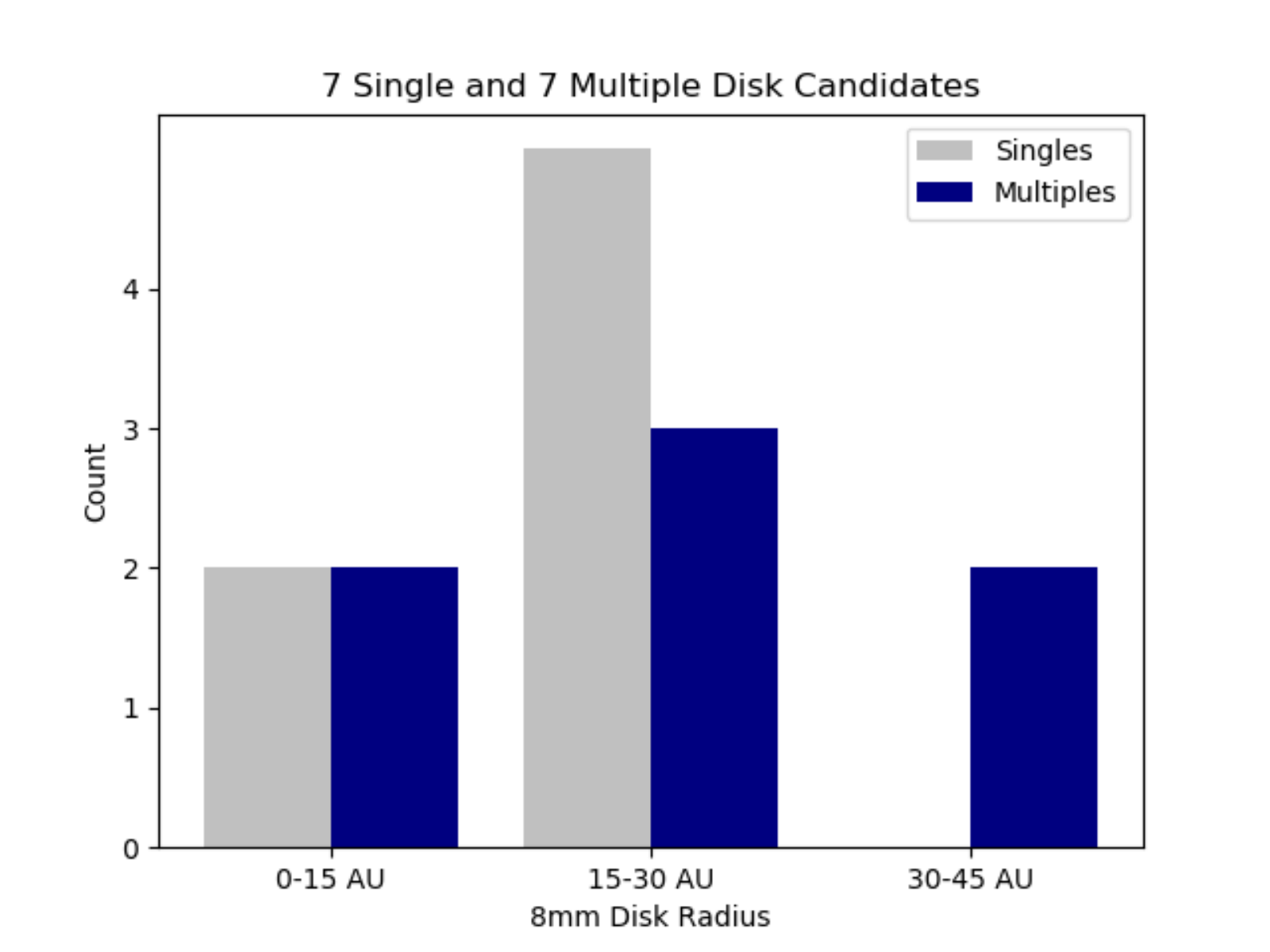}
        \caption{A histogram of the radii of all 14 modeled candidate disks, broken down by multiplicity. }
\label{diskhistomults}
    \end{minipage}
\end{figure}

Figure \ref{diskhisto} shows a histogram of all 14 modeled candidate disk radii, including 10 Class 0 and 4 Class I sources.  The four non-axisymmetric sources could not be modeled to derive a disk radius and are not included in this histogram.  Both the Class 0 and Class I candidate disk distributions peak at 15-30 AU radii, which are well resolved with the 12 AU resolution observations.  The two largest disks, with 8 mm radii 30-45 AU both belong to Class 0 sources.  With so few Class I protostars sampled, any further differences between the Class 0 and I candidate disk radii are difficult to distinguish.  Figure \ref{diskhistomults} shows a histogram of the 14 modeled candidate disk radii, separated by whether the sources are in a single or multiple protostellar system. Seven of the modeled candidate disks are single systems, and 7 belong to multiple systems.  It is unclear if there are variations in the distribution of radius that depend on the multiplicity of the systems.

\subsection{Trends of Candidate Disk Characteristics}    \label{vandam:discussion:trends}

\begin{figure}[h]
        \centering
                \includegraphics[width=0.75\textwidth]{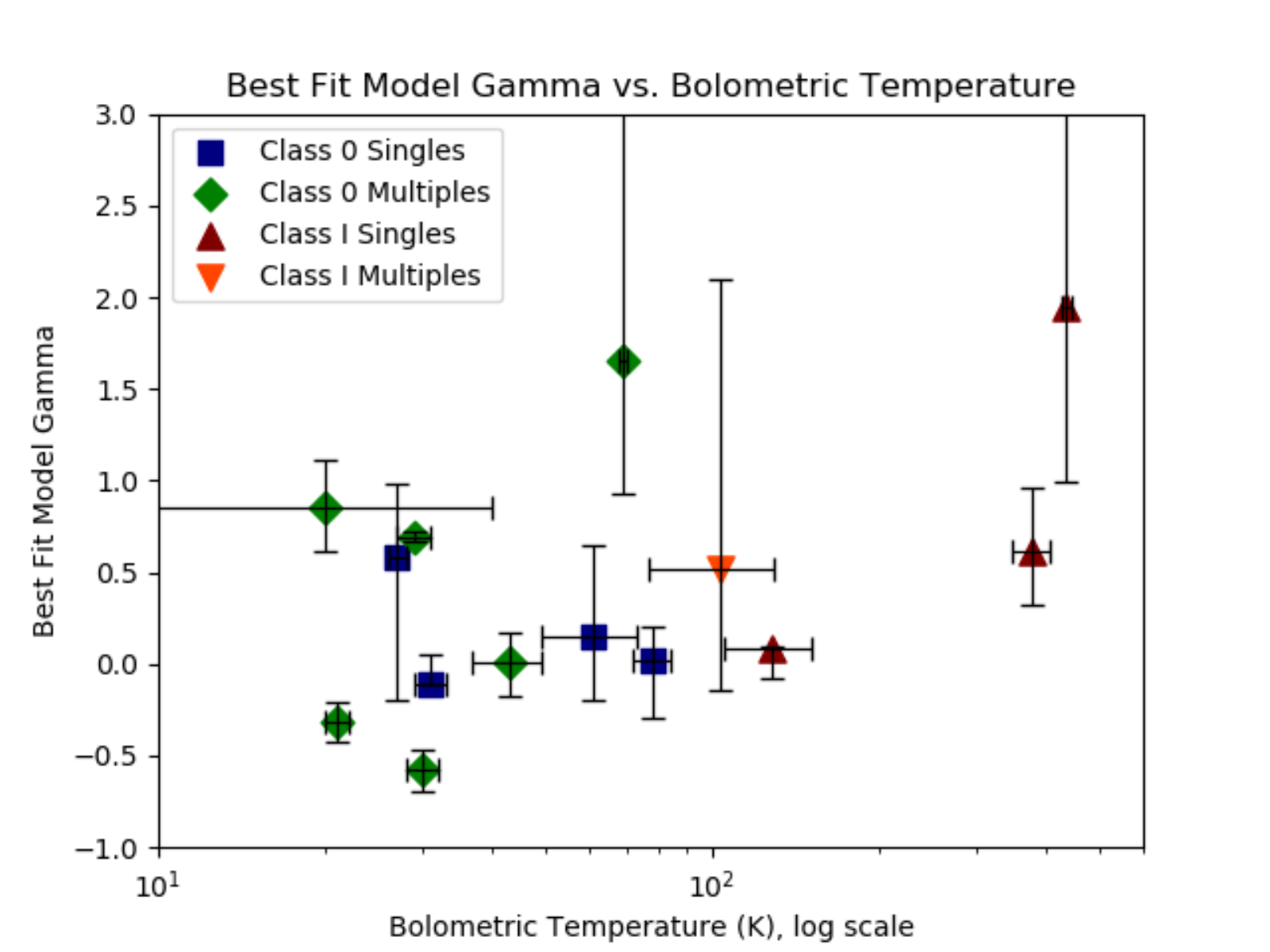}
                 \caption{A mild correlation between T$_{bol}$ and modeled inner-disk surface density power law $\gamma$ is seen by eye, with a weak trend of more evolved sources with higher T$_{bol}$ having higher values of $\gamma$. Only modeled candidate disks are included in this figure.}
\label{tbolvsgamma}
\end{figure}

\begin{figure}[!ht]
    \centering
        \begin{minipage}{.47\textwidth}
        \centering
                \includegraphics[width=0.95\textwidth]{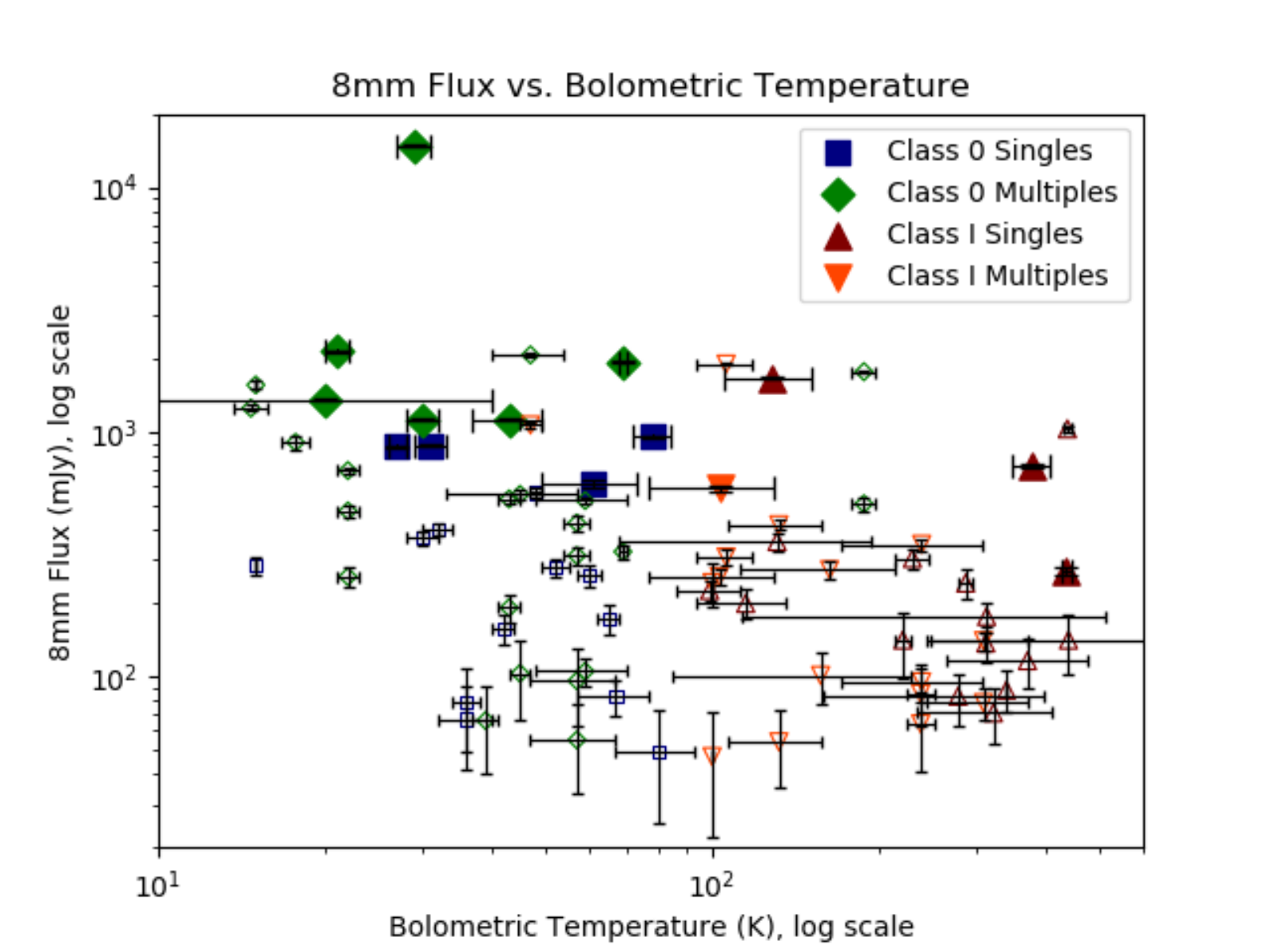}
                \caption{No clear correlations are seen between T$_{bol}$ and 8 mm flux.  Filled, large symbols are candidate disks.  Open, small symbols are unresolved sources.}
	\label{tbolvsflux}
    \end{minipage}%
    \hspace{0.2cm}
    \begin{minipage}{0.47\textwidth}
        \centering
                \includegraphics[width=0.95\textwidth]{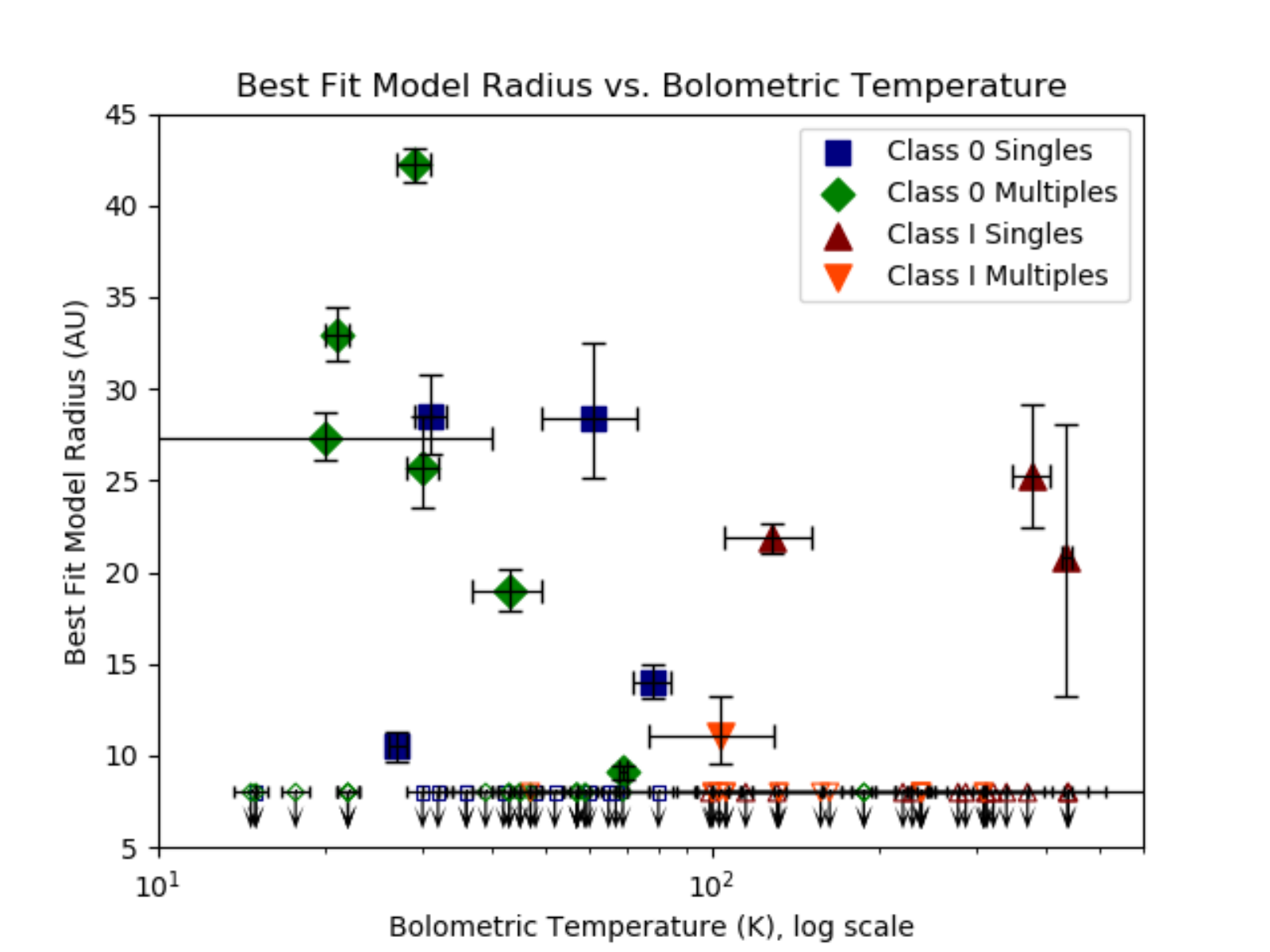}
                 \caption{No clear correlations are seen between T$_{bol}$ and modeled $R_{c}$.  Filled , large symbols are candidate disks.  Open, small symbols are unresolved sources, and the upper limit radii reflect the 8 AU modeling limit.}
	\label{tbolvsradius}
    \end{minipage}
\end{figure}

As seen in Figures \ref{tbolvsgamma}-\ref{tbolvsradius}, no tight correlations between protostellar age, 8 mm measured flux, modeled candidate disk radius, and modeled inner-disk surface density power law $\gamma$ are found by eye. We use bolometric temperature, T$_{bol}$, as an indicator of protostellar evolution to probe protostellar age \citep[]{Chen1995}. In these plots, Class 0/I sources are counted as Class 0 and we include all 14 modeled VANDAM candidate disks.  Bolometric temperatures were taken from \citet{Tobin2016a}.  When possible, we include unresolved sources which are not candidate disks if measurements or upper limits on the plotted parameters exist.  We see no significant differences between single and multiple candidate characteristics with age, nor any significant differences between Class 0 and Class I candidate disk characteristics with age.  

A mild correlation between $\gamma$, the power-law of the inner-disk surface density, and T$_{bol}$ (Figure \ref{tbolvsgamma}) may be present with a tentative trend of higher values of $\gamma$ being found in older sources with higher T$_{bol}$.  Higher values of $\gamma$ indicates that flux drops off faster with increasing radius, meaning 8 mm flux is more centrally distributed for younger sources than more evolved sources with higher T$_{bol}$.  This may be explained as dust grains in more evolved sources having more time to experience radial drift \citep{Perez2012} and concentrate closer to the central protostellar source.  As further evidence, in Section \ref{vandam:discussion:freq} we demonstrated that the typical value of $\gamma$ grows larger from the young Class 0/I stage to the more evolved Class II/III phases.

If there are truly no correlations between 8 mm disk flux or radius and age (Figures \ref{tbolvsflux} and \ref{tbolvsradius}), a process independent of evolution could be setting the disk radii.  Possible processes independent of age that could influence the disk radii include the initial angular momentum imparted onto the disk from the natal protostellar core \citep[e.g.,][]{Terebey1984} or magnetic braking dominating over any evolutionary effects \citep[e.g.,][]{Dapp2010}.  Alternatively using T$_{bol}$ as an indicator of protostellar evolution may be imprecise.  Certainly it is clear that overall there are no major differences in 8 mm fluxes or radii between the Class 0 and Class I phases. 

\begin{figure}[!h]
    \centering
        \begin{minipage}{.47\textwidth}
        \centering
                \includegraphics[width=0.95\linewidth]{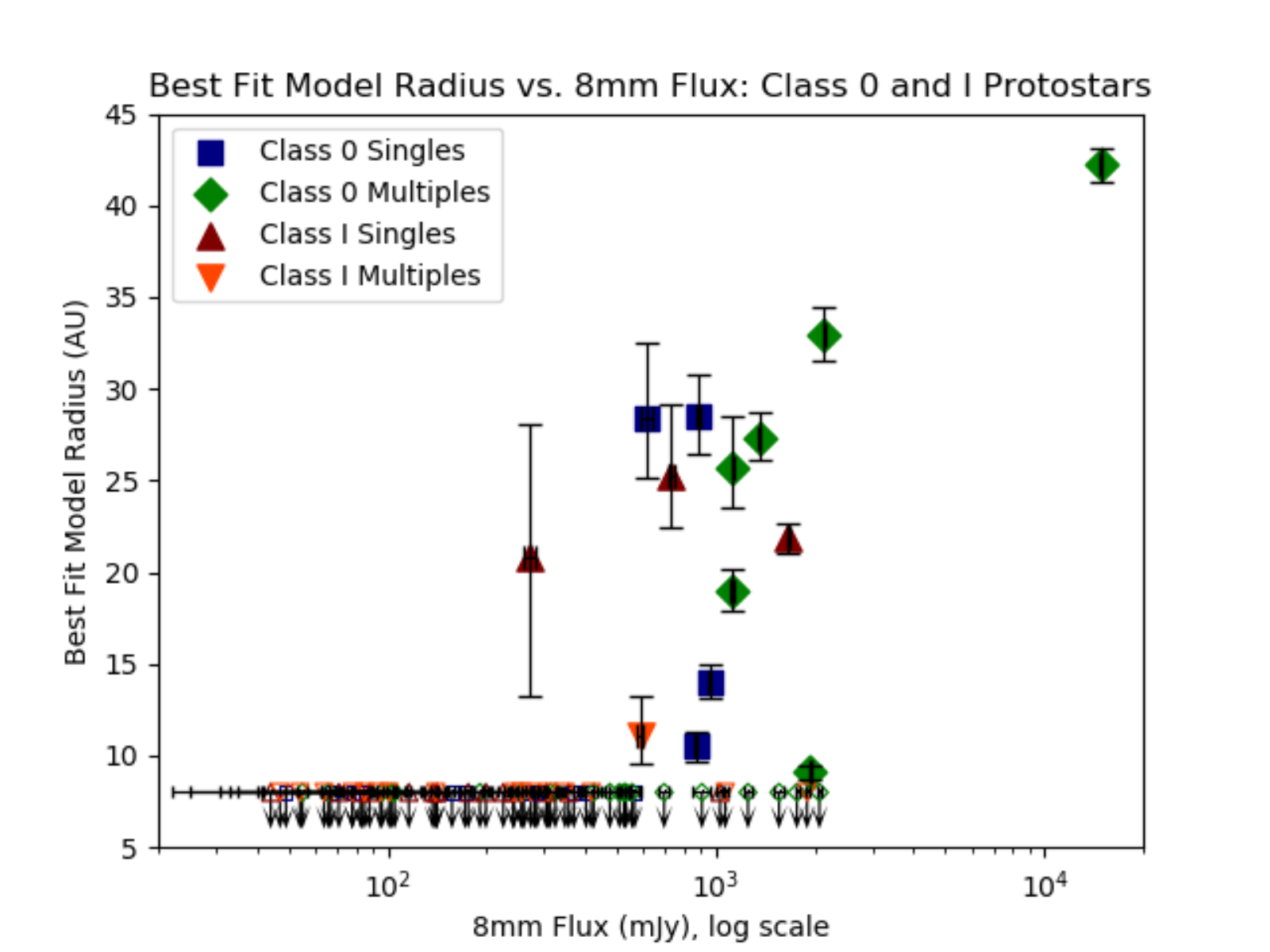}
                 \caption{No clear correlations are seen between 8 mm flux and modeled $R_{c}$.  Filled, large symbols are candidate disks.  Open, small symbols are unresolved sources, and the upper limit radii reflect the 8 AU modeling limit.}
	\label{fluxvsradius}
    \end{minipage}%
        \hspace{0.2cm}
    \begin{minipage}{0.47\textwidth}
        \centering
                \includegraphics[width=0.95\linewidth]{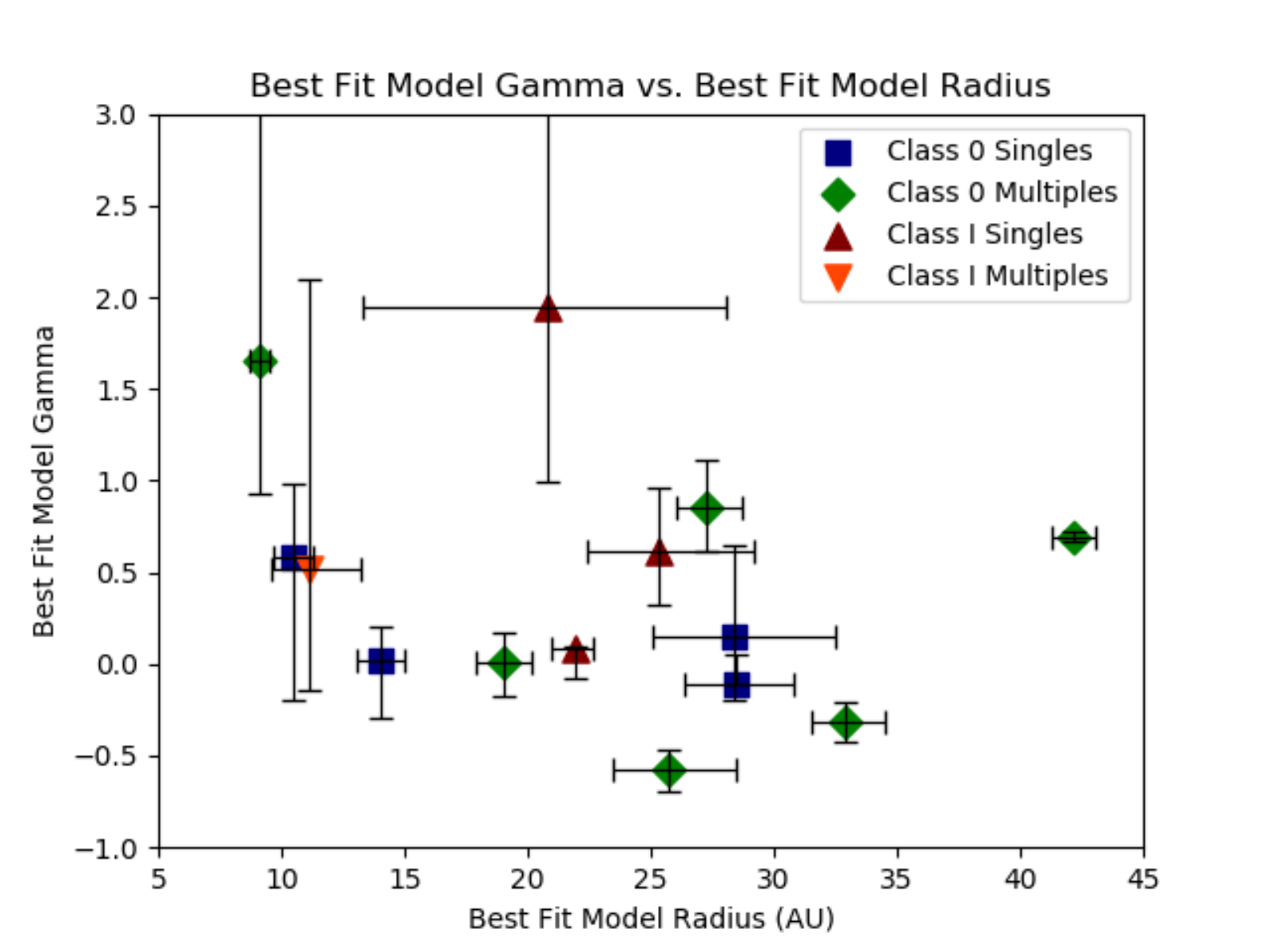}
                 \caption{No clear correlations are seen between modeled $R_{c}$ and  modeled $\gamma$, the power-law of the inner-disk surface density.  Only modeled candidate disks are included in this figure. }
	\label{radiusvsgamma}
    \end{minipage}
\end{figure}

Additionally, in Figures \ref{fluxvsradius}-\ref{radiusvsgamma}, no correlations are seen between measured flux and modeled radius, or modeled radius and modeled $\gamma$.  Again no clear differences are seen in the characteristics between the single and multiple populations or the Class 0 or Class I phases.  For Figure \ref{fluxvsradius}, we include the unresolved sources which we do not consider to be candidate disks, with the radii represented as an upper limit set by the 8 AU modeling limit (see Section \ref{vandam:modeling}).

\begin{figure}[!h]
    \centering
    \begin{minipage}{0.47\textwidth}
        \centering
                \includegraphics[width=0.95\linewidth]{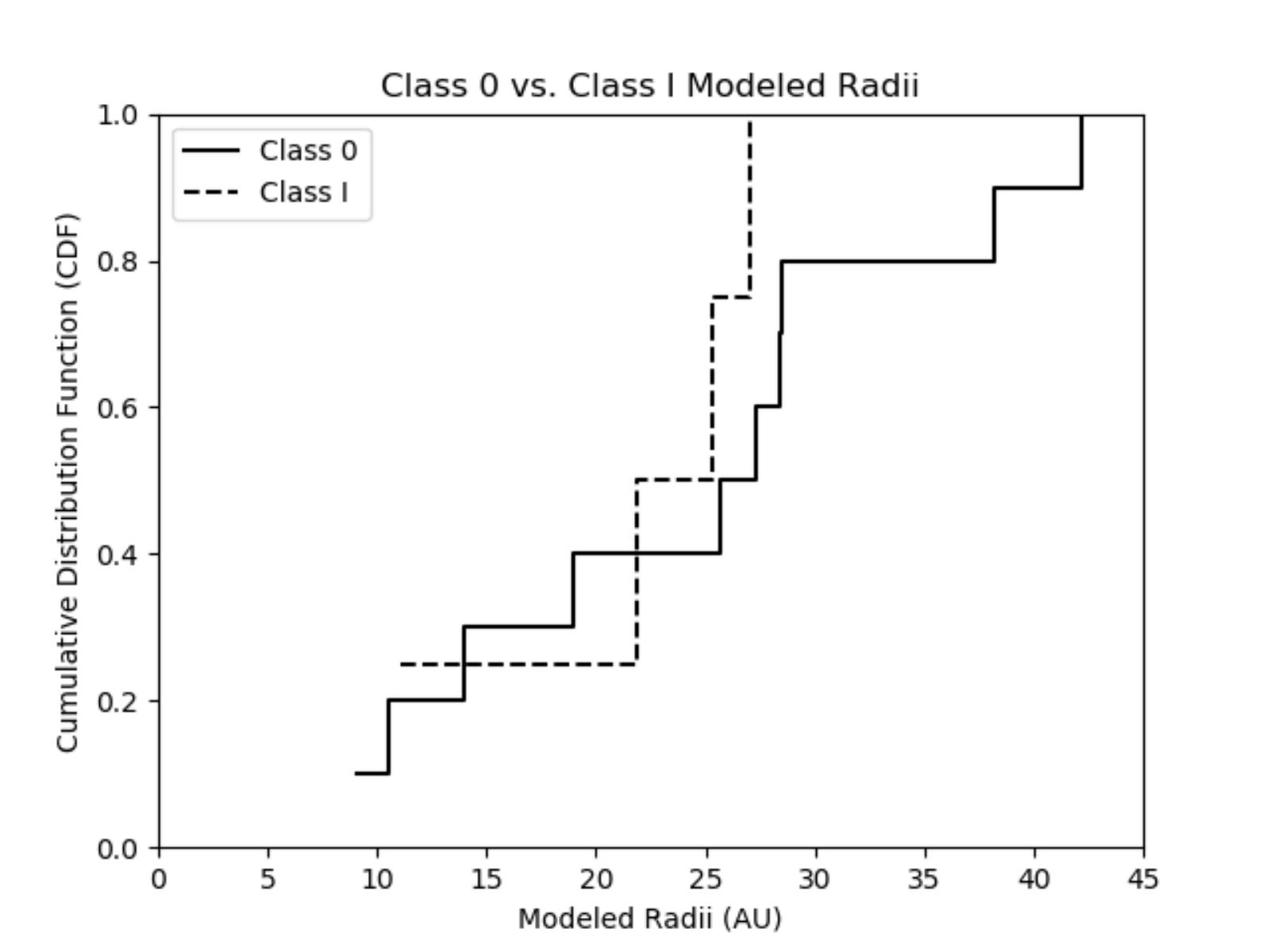}
                 \caption{Empirical cumulative distribution function versus modeled radius for the Class 0 and Class I protostars. p=0.46}
        \label{CDF_classes_radius}
    \end{minipage}%
       \hspace{0.2cm}
    \begin{minipage}{0.47\textwidth}
        \centering
                \includegraphics[width=0.95\linewidth]{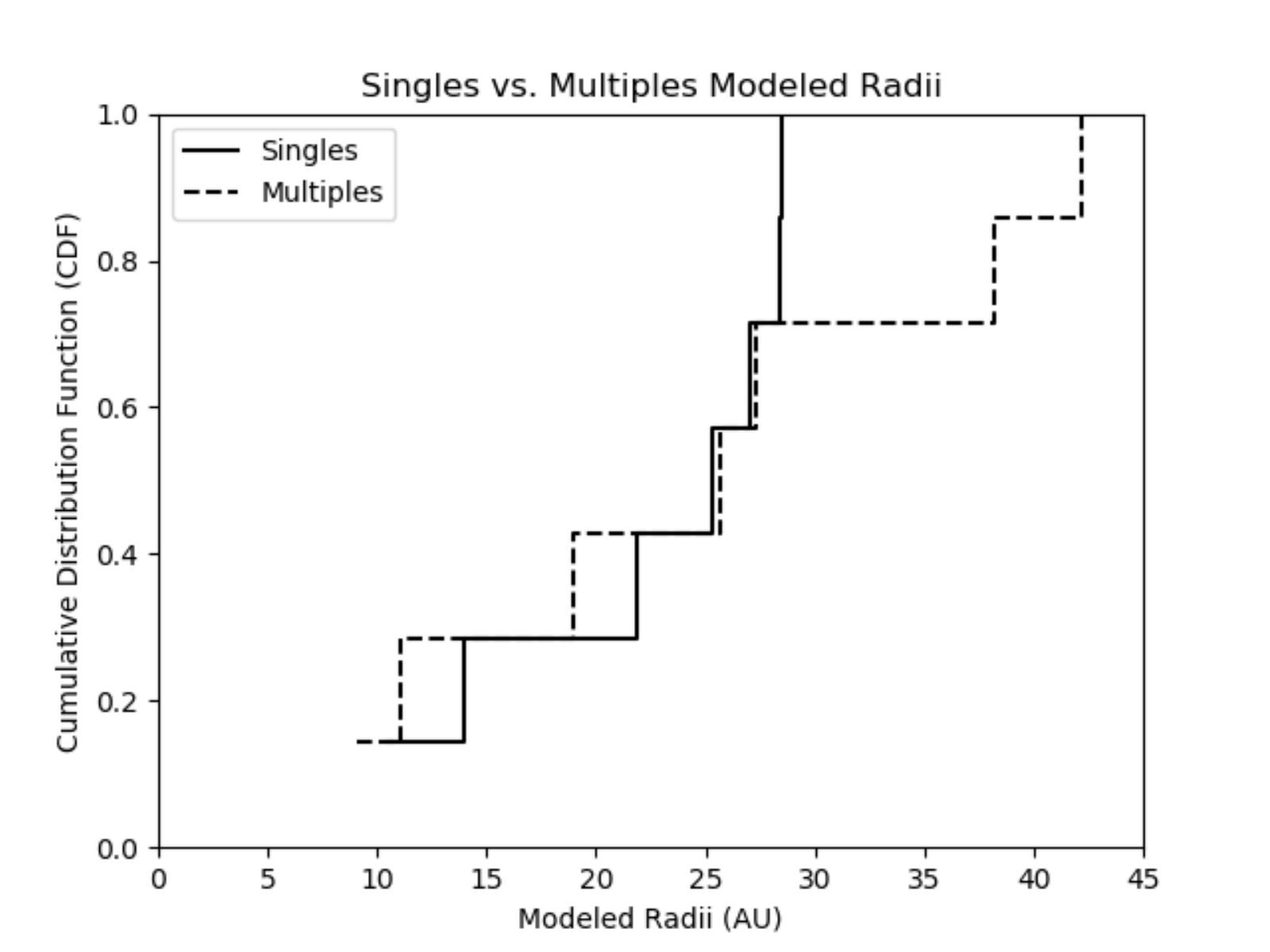}
                 \caption{Empirical cumulative distribution function versus modeled radius for the single and multiple protostars p=0.85}
        \label{CDF_mults_radius}
    \end{minipage}
\end{figure}

\begin{figure}[!h]
    \centering
    \begin{minipage}{.47\textwidth}
        \centering
                \includegraphics[width=0.95\linewidth]{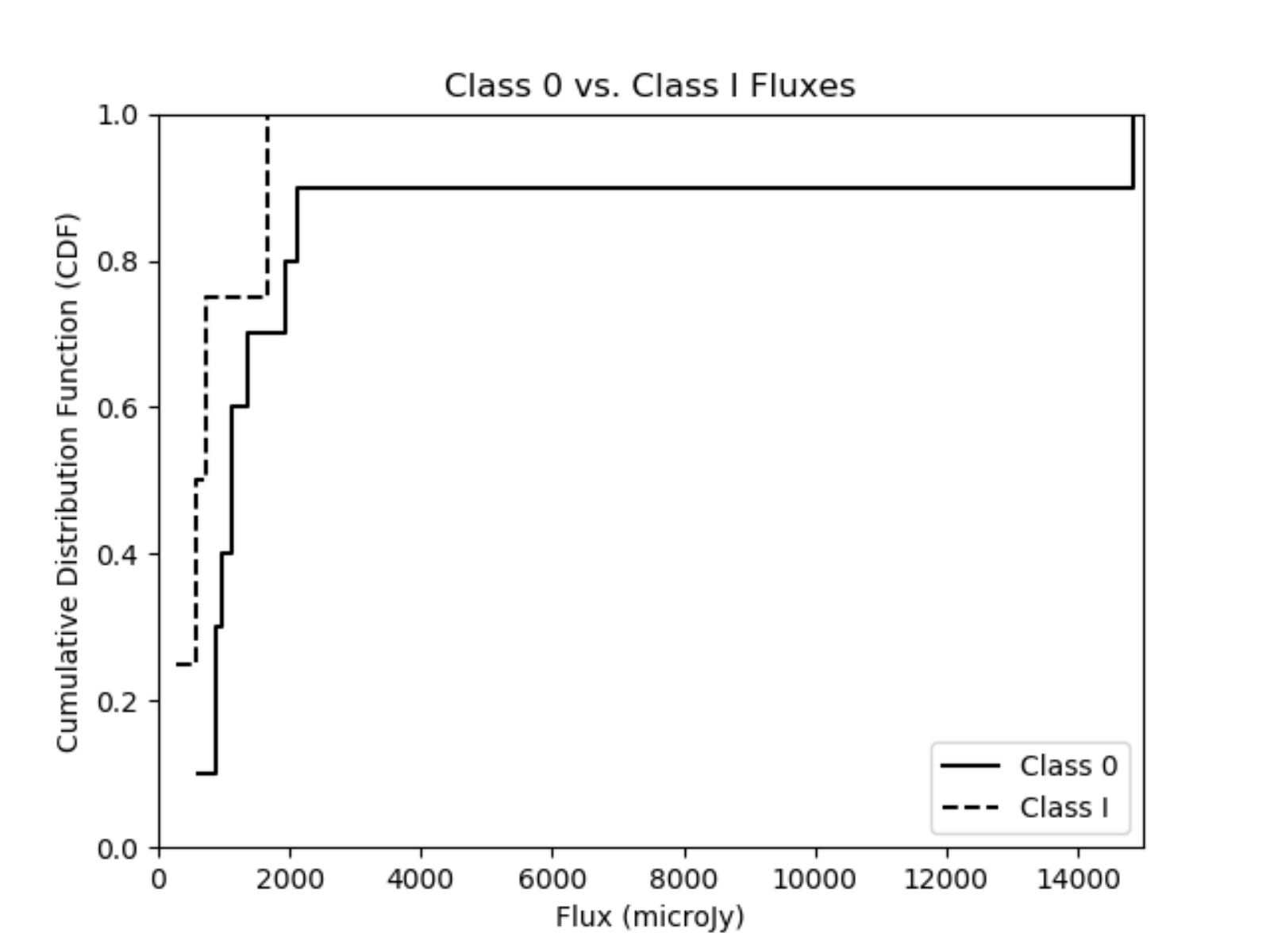}
                 \caption{Empirical cumulative distribution function versus observed flux for the Class 0 and Class I protostars. p=0.35}
        \label{CDF_classes_fluxes}
    \end{minipage}%
        \hspace{0.2cm}
    \begin{minipage}{0.47\textwidth}
        \centering
                \includegraphics[width=0.95\linewidth]{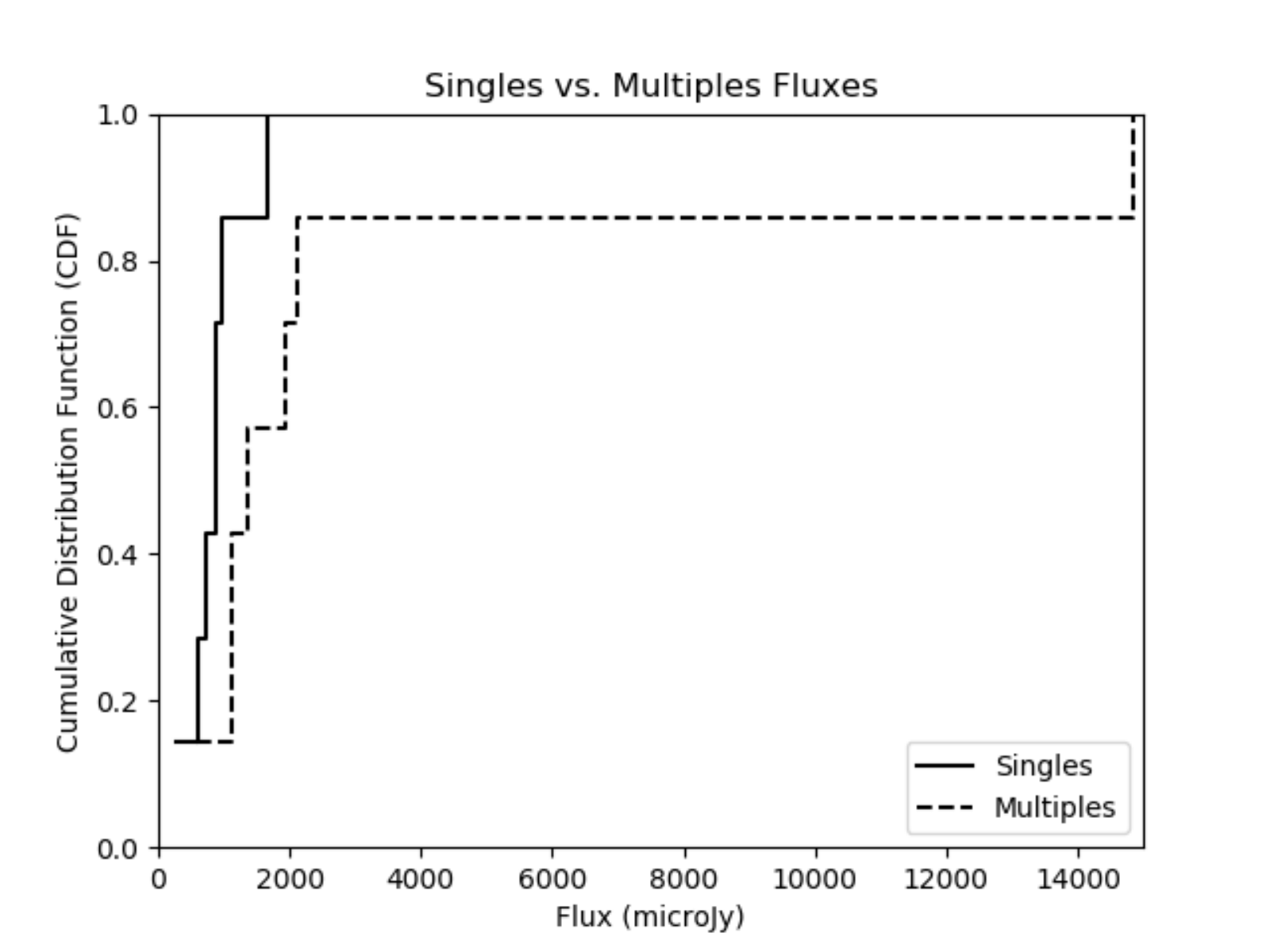}
                 \caption{Empirical cumulative distribution function versus observed flux for the single and multiple protostars. p=0.22}
        \label{CDF_mults_fluxes}
    \end{minipage}
\end{figure}

\begin{figure}[!h]
    \centering
    \begin{minipage}{.47\textwidth}
        \centering
                \includegraphics[width=0.95\linewidth]{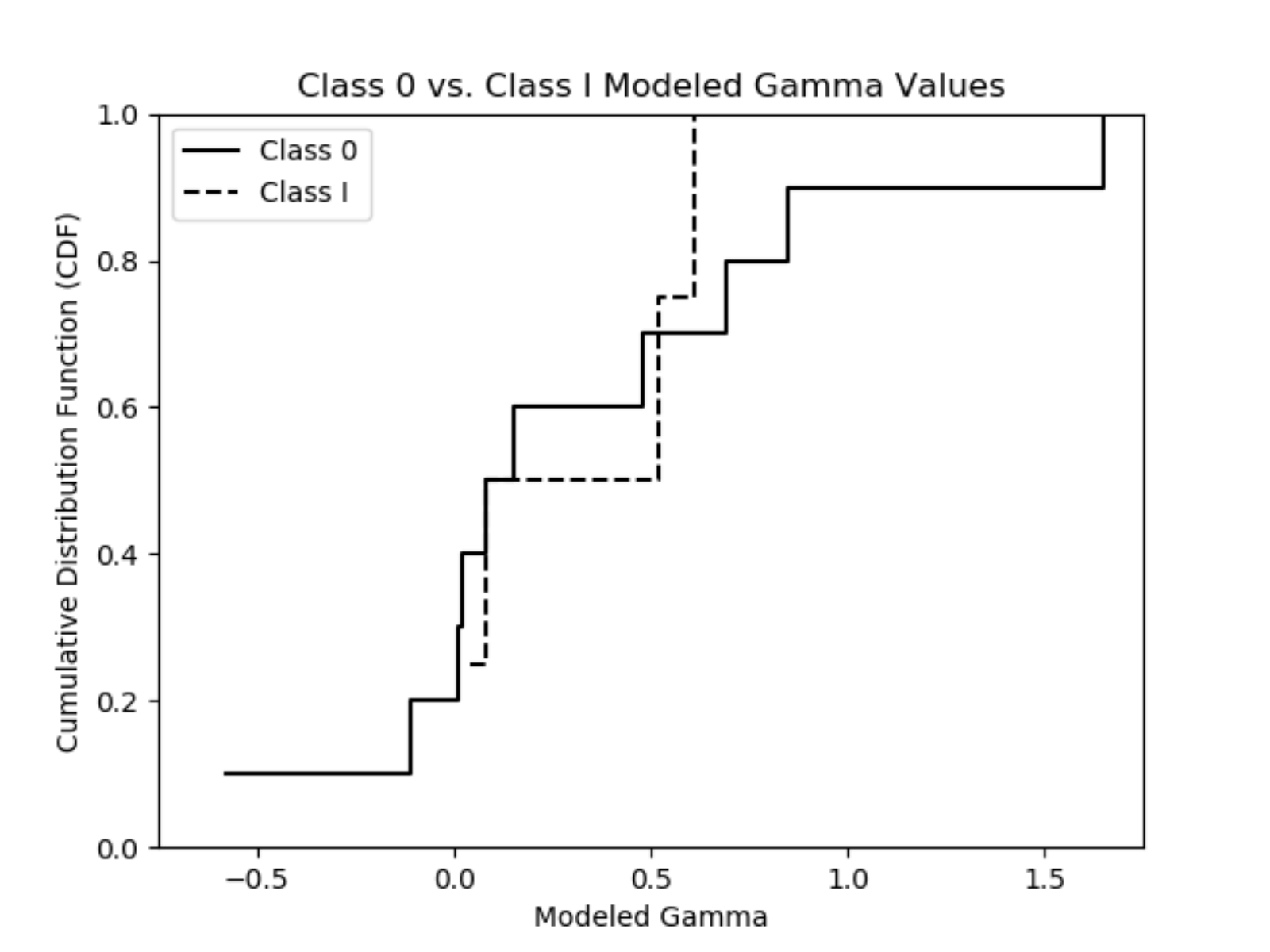}
                 \caption{Empirical cumulative distribution function versus modeled $\gamma$, the power-law of the inner-disk surface density, for the Class 0 and Class I protostars. p=0.62}
        \label{CDF_classes_gammas}
    \end{minipage}%
       \hspace{0.2cm}
    \begin{minipage}{0.47\textwidth}
        \centering
                \includegraphics[width=0.95\linewidth]{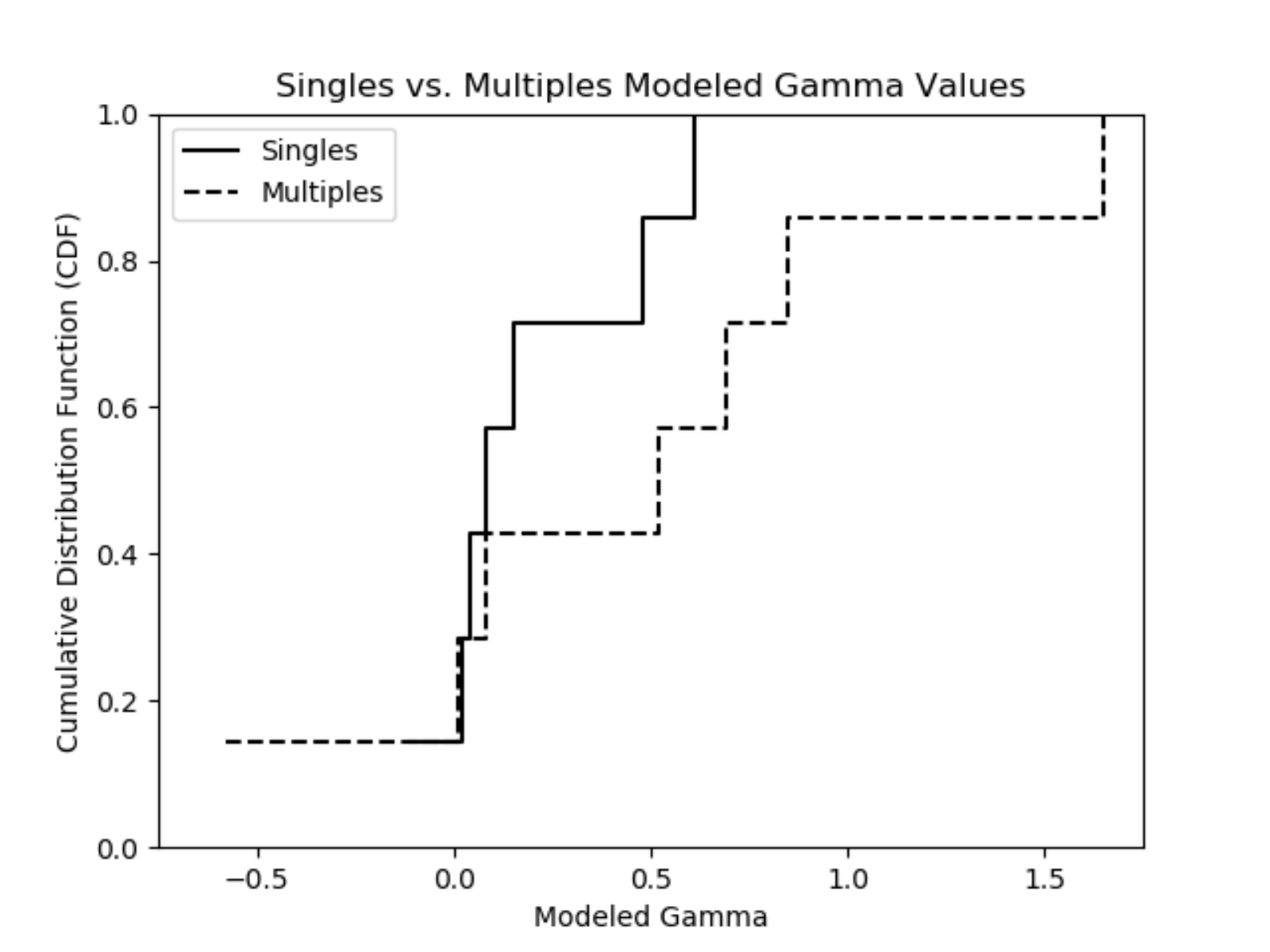}
                 \caption{Empirical cumulative distribution function versus modeled $\gamma$, the power-law of the inner-disk surface density, for the single and multiple protostars. p=0.30}
        \label{CDF_mults_gammas}
    \end{minipage}
\end{figure}

We generated empirical cumulative distribution functions (CDF) for the modeled Class 0 and Class I candidate disks, as well as for the modeled candidate disks which lie in single and multiple systems (Figures \ref{CDF_classes_radius} to \ref{CDF_mults_gammas}) to examine differences between the sub-samples with disk radius, flux, and $\gamma$.  We performed Anderson-Darling (AD) tests \citep{ScholzStephens1987} on each set of empirical CDFs to determine whether the populations are consistent with being drawn from the same distribution.  The AD-test is comparable to  the Kolmogorov-Smirnoff (KS) test but is more statistically robust---especially if differences between the samples are at the ends of the distribution or if there are small yet significant deviations through the whole distribution---because the KS-test uses maximum deviation to calculate the probability, which the AD-test does not rely on.  The results of our AD-tests for the CDFs of disk radius indicate that there is high probability that Class 0 and Class I disk candidates  as well as candidate disks from single and multiple systems are drawn from the same distributions (p-values of 0.46 and 0.85, respectively).  The AD-tests for the CDFs of flux for the Class 0 and Class I candidate disks have a probability of being drawn from the same distribution of 0.35, and for the candidate disks in single and multiple systems have a p-value of 0.22.  Finally, the AD-tests for the CDFs of $\gamma$ for Class 0 and Class I systems have a p-value of 0.62, and there is a p-value of 0.30 for the candidate disks in single and multiple systems.  In all cases, the null-hypothesis that the two sub-samples are drawn from the same distribution cannot be ruled out.  This may indicate that the disk properties of protostars are usually defined early in the Class 0 phase and do not vary greatly through the Class I phase.

\subsection{Candidate Disk Sources with Detected Polarized Emission}      \label{vandam:discussion:polarized}

There are three proposed mechanisms that cause polarized emission in protostellar sources.  Polarization may be caused by dust grains interacting with the magnetic field. Non-spherical spinning dust grains are expected to align their short axis parallel to the magnetic field, resulting in polarization orientation perpendicular to the magnetic field \citep[e.g.,][]{Lazarian2007}.  In the bulk of the disk where the magnetic energy is expected to be smaller than the rotational energy, the magnetic field is expected to be wrapped into a toroidal configuration by disk rotation.   Polarization may also be due to Rayleigh scattering of radiation from the disk and central protostar if grain sizes are not much smaller than the observational wavelength \citep{Kataoka2015}.  A new polarization mechanism of dust continuum emission has been posited: radiative alignment of grains with radiation anisotropy \citep{Tazaki2017}.  This new radiative alignment scenario causes larger grains ($>$mm size), which are not expected to align with the magnetic field, to be aligned by radiative torques from an anisotropic radiation source with their short axes along the direction of the radiative flux, causing polarized emission.  These polarization mechanisms will produce different patterns of polarization morphology and have different efficiencies at varying wavelengths, though determining the definitive contributing mechanisms in any given source is not straightforward and requires detailed modeling \citep{Yang2016}.

Even when a magnetic field is present, dust grains may not align with the magnetic field if high gas densities cause sufficient random collisions to prevent grain alignment.  If submillimeter and millimeter sized grains are settled in the midplane of the disk, gas density is so high that grain alignment with the magnetic field is difficult to achieve due to gaseous dampening even when unusually strong magnetic fields and grains with many superparamagnetic inclusions are taken into account \citep{Tazaki2017}.  While $\sim$mm sized grains do tend to be well settled in the midplane for Class II disks \citep{Pinte2016,Guilloteau2016}, the same is not necessarily true for less evolved Class 0 disks, where grains have grown to millimeter sizes but not yet settled to the midplane \citep{Yang2017}. The Class 0 disk HH212 has resolved, geometrically-thick vertical disk structure in ALMA 850 $\mu$m data \citep{Lee2017}.  The vertical structure of L1527 is also resolved by ALMA and geometrically-thick at $\sim$0.8 mm \citep{Sakai2017}.  The $\sim$mm sized grains further from the midplane in these Class 0 disks will be in a less dense gaseous environment and alignment of grains perpendicular to a toroidal disk magnetic field may still be possible.  Polarization in the envelope, on larger scales than the disk, is likely to be caused by magnetic fields due to the inefficiency of radiative alignment and scattering due to smaller grain sizes and a less dense environment.
Polarized emission has been detected towards Class 0 protostars L1527 and IRAS 16293-2422 B at 1.3 mm and 878 $\mu$m wavelengths respectively \citep{SeguraCox2015,Rao2014}, tracing grain sizes similar to the grains not yet settled in the midplanes of Class 0 disks L1527 and HH212.  Indeed, if the polarization mechanism in L1527 and  IRAS 16293-2422 B is assumed to be purely due to magnetic field alignment, the inferred magnetic field morphologies from polarization in these sources are well-described by disk-wrapped toroidal magnetic fields.  In short, while scattering and radiative alignment likely dominate polarized emission in evolved Class II disks (with $\sim$mm sized grains settled close to the midplane), the geometrically-thick $\sim$mm grain populations in the disks of Class 0 sources mean that for the youngest protostars polarization from magnetic fields cannot be ruled out.

Five of our 18 candidate disks also have observations of polarized emission towards them: HH211-mms,  NGC 1333 IRAS 4A, NGC 1333 IRAS2A,  SVS13B, and L1448 IRS3B (see Appendix \ref{vandam:gallery}).  These sources are all Class 0 or Class 0/I protostars, young sources which may have thick disks and the potential for emission from $\sim$mm sized grains to become polarized via the magnetic field.  We do assume for discussion here that the polarization is purely a signature of the magnetic field in these systems, yet we note that scattering and radiative alignment may also contribute to the polarized emission.  

In \citet{SeguraCox2015}, we suggested a tentative trend of misaligned inferred magnetic field from polarization and rotation axes in Class 0 systems with disks, with the misaligned orientation helping to reduce the effect of magnetic braking and grow disks with $R$ $>$10 AU at early times \citep[e.g.,][]{Joos2012}.  The scenario with aligned inferred magnetic fields and rotation axes would not inhibit magnetic braking effects and a smaller disk would be expected to form.  All five of the candidate disks with polarized data have inferred magnetic fields either perpendicular or misaligned with the rotation axes of the systems.  

 The inferred magnetic field morophology of HH211-mms from the SMA data near disk size scales \citep{Lee2014} resembles with a disk-wrapped toroidal field in the northwest with poloidal field lines expected in envelopes and outflows in the southwest and southeast, though scattering may also play a role in polarization on these scales.  The true magnetic field morphology of the SMA data on inner envelope and disk size scales is poorly determined due to lack of polarization detected across the entire protostar.  The SCUPOL \citep{Matthews2009} and CARMA \citep{Hull2014} inferred magnetic fields are misaligned with both the HH211-mms candidate disk orientation and outflow orientation. The complicated polarization morphology in this source in particular may be partially explained by contributions from either scattering or radiative alignment, though magnetic field contributions cannot be ruled out.

NGC 1333 IRAS4A's polarization morphology was long attributed to magnetic fields, which are misaligned with the outflow of the protostar. The large-scale polarization \citep{Girart2006}, with an hourglass-shaped inferred magnetic field is unlikely to be dominated by scattering due to the column densities and small grain sizes in the envelope \citep{Yang2016}.  The small-scale polarized emission from NGC 1333 IRAS4A \citep{Cox2015} was found to be consistent with a mix of scattering and a toroidal magnetic field wrapped by the disk  \citep{Yang2016}.  Nevertheless, a magnetic field causing the polarized emission cannot be ruled out even on small scales.  If we assume the extreme limit of the polarization being caused by purely magnetic fields, the fields are consistent with large scale vertical-poloidal hourglass fields misaligned with the rotation axis of the disk, transitioning to small scale frozen-in disk-wrapped toroidal fields in the magnetized disk \citep{Hennebelle2009,Kataoka2012}.  As material infalls from the envelope to an embedded disk, frozen-in magnetic field lines will be drawn inwards as well, changing the magnetic field morphology between the envelope and disk  \citep{Li2014}.  In the case of NGC 1333 IRAS 4A, the inferred magnetic field from polarization on small scales is congruent with the idea that misaligned magnetic fields and rotation axes do not inhibit disk growth by magnetic braking in the same way as the aligned scenario.

The inferred magnetic field morphologies from polarization for the five polarized VANDAM candidate disks are consistent with the scenario of perpendicular or misaligned inferred magnetic field orientations compared to the rotation axes of the systems, inhibiting magnetic braking and allowing the disks to grow larger than 10 AU at early times \citep{Hennebelle2009,Joos2012,Li2013,Krumholz2013}. We note that the envelope-scale polarization, which is less likely to be affected by scattering and more likely due to magnetic fields, for all five sources follow this trend.   This tentative trend of misalignment between inferred magnetic fields and outflows in Class 0 sources with disks can be strengthened with previous observations of sources outside the Perseus molecular cloud. In Table 2 of \citet{SeguraCox2015}, the Class 0 sources L1527, IRAS 16293-2422 B, and VLA 1623 also have perpendicular inferred magnetic fields from polarization and outflows with strong evidence of embedded disks \citep{SeguraCox2015}.  Combined this brings the total of known and candidate Class 0 disks with misaligned outflow and inferred magnetic field orientations to eight sources, double the number of to-date Keplerian-confirmed Class 0 disks.  Misalignment between inferred magnetic fields on envelope scales and outflows may indeed be a signpost of disks on smaller scales in Class 0 sources with geometrically-thick disks.  
  
\section{SUMMARY OF VANDAM DISK CANDIDATE RESULTS}    \label{vandam:summary}

With the 12 AU resolution VANDAM survey (with an 8 AU radius modeling limit), we find a total of 18 Class 0 and I candidate disks the Perseus molecular cloud, more than doubling the number of known Class 0 and I disks to-date, bringing the total count from $\sim$15 to $\sim$33.   We were able to model 14 of the 18 candidate disks; four are highly asymmetric and not fit well by our disk model.  Because we do not have small-scale kinematic data to confirm that these are rotationally supported disks, we refer to the VANDAM sources as disk candidates.  

We fit the deprojected, azimuthally averaged, and radially binned VLA 8 mm continuum data in  {\it u,v}-space to a  disk-shaped profile to determine disk candidacy of the extended sources and to begin to model disk properties.  We fix the inclination and position angles of the disks using estimates from the image-plane.  We take into account a point-source component to account for free-free emission from the jet-launching regions of the disks.  NGC 1333 IRAS 4A and NGC 1333 IRAS 4B have obvious envelope contamination in both the image and {\it u,v}-planes.  We apply a \textit{u,v}-cut to the data to account for this in the image plane and do not fit the corresponding removed inner baselines to the disk profile in  {\it u,v}-space.  Other sources have minimal envelope contamination.  Except for sources with envelope contamination or small asymmetric features, the residuals from subtracting the model from the data are $<3\sigma$.  For all but one candidate disk, the major axis of the disk is roughly perpendicular to the outflow axis (a proxy for the rotation axis), as expected for rotating protostellar disks.

33\% of Class 0 (and Class 0/I) protostars and just 11\% of Class I protostars have candidate disks with radii larger than the 8 mm VLA 12 AU resolution data in Perseus.  There is a mild trend that more evolved sources, as gauged by T$_{bol}$, have higher values of $\gamma$ for the 8 mm data; older sources appear to have more centrally concentrated 8 mm dust grain populations.  This may be due to the 8 mm grains in older sources having more time to undergo radial drift towards the central protostar than younger sources, resulting in a steeper inner-disk surface density power law ($\gamma$) for more evolved sources.  Additionally, the two largest candidate disks belong to Class 0 protostars.  78\% of Class 0 and I protostars do not have signs of disks within our 12 AU resolution limit; at 8 mm most disks in the Class 0 and I phases are small  ($<$10 AU).

Our estimated masses of the candidate disks are large compared to masses of known Class 0 and Class I disks, indicating that our assumed value of dust opacity spectral index $\beta$ = 1 is too large.  Most disks have best-fit models with $q$ $<$0.5, typical for embedded disks.  Values of $\gamma$ are lower and more shallow for Class 0 and I candidate disks than in Class II disks, indicating that evolved disks are generally  more centrally concentrated than our Class 0 and I disks.  Modeled radii of the candidate disks are $>$10 AU at 8 mm and comparable to known Class 0 disk radii determined from kinematics at $\sim$mm wavelengths. Per-emb-14 has a $\sim$3$\times$ larger disk at 1.3 mm with a smaller grain population, evidence that our 8 mm data is a lower limit on true disk radius.  Since our 8 mm data trace a population of larger dust grains which radially drift towards the protostar and are lower limits on true disk size, large disks at early times do not seem to be particularly rare.   

To examine how multiplicity and evolution affect disk radius, $\gamma$, and 8 mm flux, we performed AD tests on CDFs for the modeled Class 0 and Class I candidate disks, as well as CDFs for our modeled disks which are found in single and binary systems. In all cases, we cannot rule out the hypothesis that Class 0 and Class I protostars or single and multiple systems are drawn from the same distribution. Disk properties may be defined early in the Class 0 phase, without much variation through the Class I phase.

Five of the 18 candidate disk sources also have polarization detections.  In the extreme case of ignoring scattering and radiative alignment, which may contribute to polarized emission but do not preclude magnetic fields contributing to the polarization signal in geometrically-thick Class 0 disks, the five candidate disks with polarized emission have inferred magnetic field morphologies misaligned with the outflows/rotation axes of the systems.   
 Three additional Class 0 disks from other molecular clouds have misaligned inferred magnetic field and outflow directions, bringing the total number to eight.  Misalignment between inferred magnetic fields on envelope scales and outflows may be used in the future as indirect evidence of possible geometrically-thick disks on smaller scales in Class 0 protostars.


\acknowledgments

DMSC was supported by NRAO Student Observing Support grant SBC NRAO 2015-06997.  ZYL is supported in part by NASA 80NSSC18K1095 and NNX14AB38G and NSF AST-1815784 and AST-1716259.  CM acknowledges support from NSF grant AST-1313428. 
 The National Radio Astronomy
Observatory is a facility of the National Science Foundation
operated under cooperative agreement by Associated Universities, Inc.
This research made use of APLpy, an open-source plotting package for Python hosted at http://aplpy.github.com.


\bibliography{dom_bib_file}
\bibliographystyle{apj}


\clearpage

\appendix

\section{IMAGES OF VANDAM EXTENDED SOURCES}    \label{vandam:extended_images}

 \begin{figure}[h]
        \centering
                \includegraphics[width=0.8\textwidth]{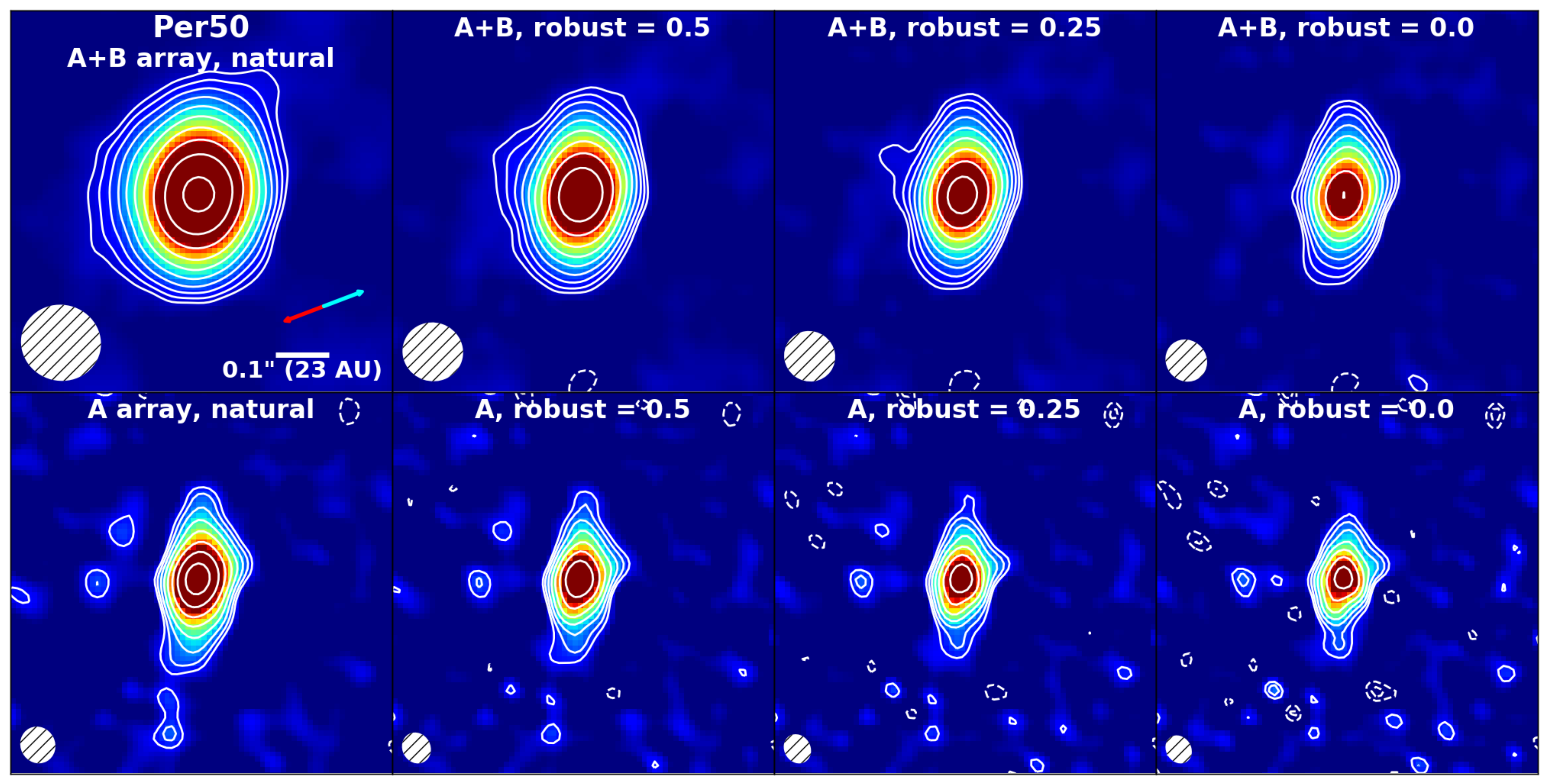}
        \caption{VLA A+B array data (top row), and A-array only data (bottom row) of Per-emb-50, all plotted with the same physical size scale. Images were produced with varying robust weighting values, labeled at the top of each panel. Contours start at 3$\sigma$ ($\sigma$ $\sim$15$\mu$Jy) with a factor of $\sqrt{2}$ spacing. The synthesized beam is in the lower left.  Outflow orientations are indicated by the red and blue arrows in the lower right corner of the upper left-most panel.}
        \label{Per50_2x4}
\end{figure} 

 \begin{figure}[h]
        \centering
                \includegraphics[width=0.8\textwidth]{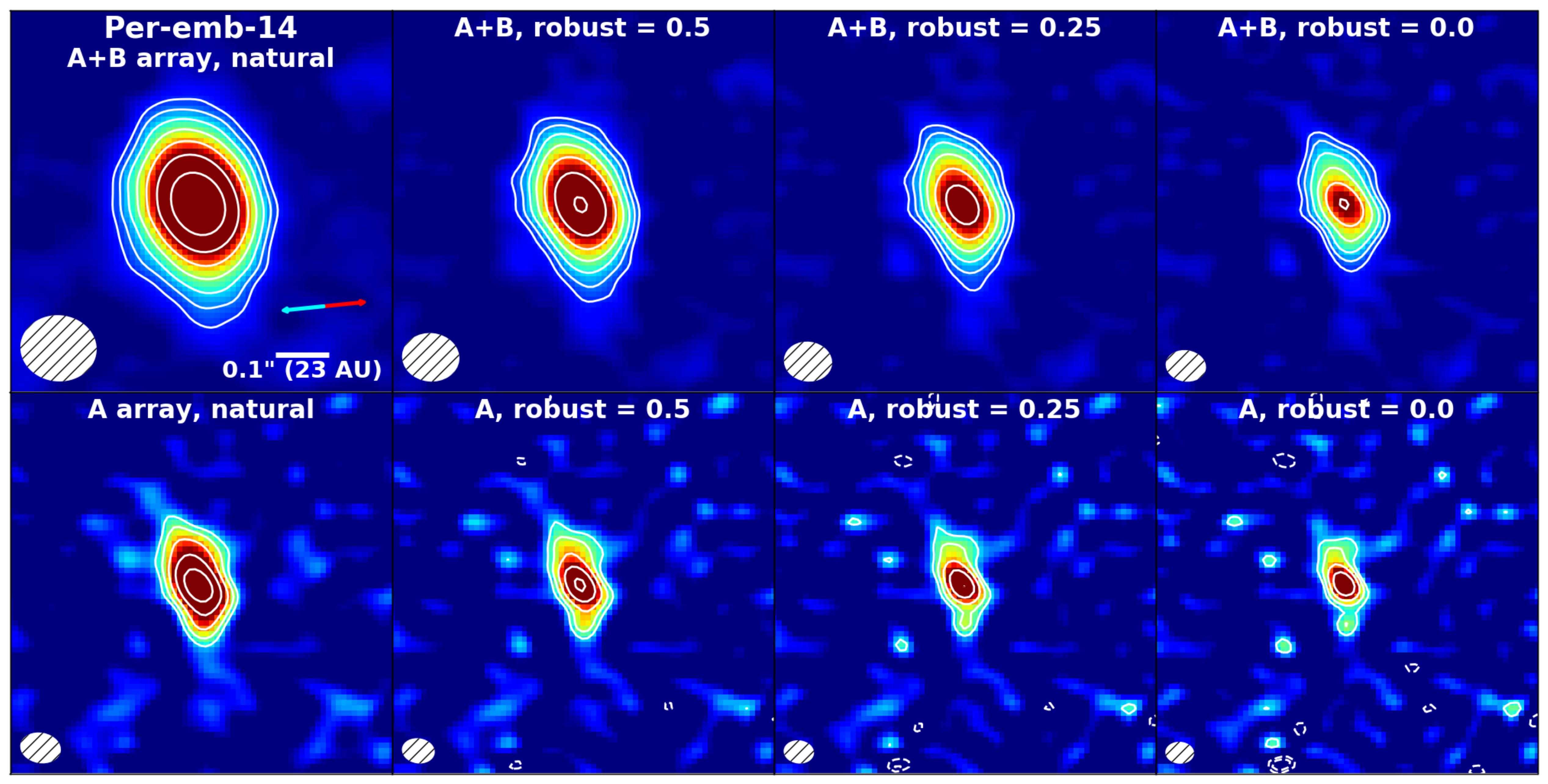}
        \caption{Same as Figure \ref{Per50_2x4}, for Per-emb-14.}
        \label{Per14_2x4}
\end{figure} 

\clearpage

 \begin{figure}[h]
        \centering
                \includegraphics[width=0.8\textwidth]{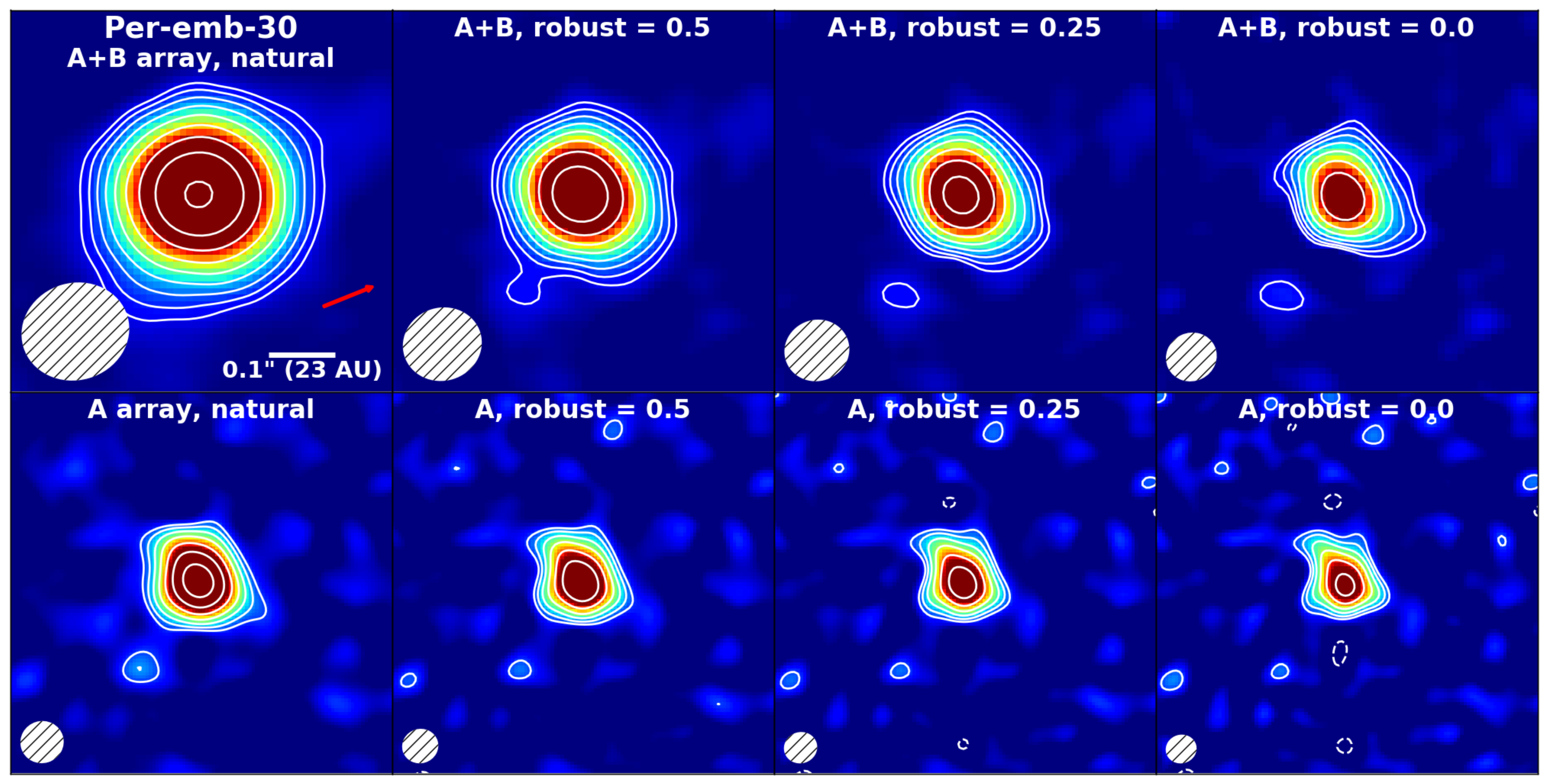}
        \caption{Same as Figure \ref{Per50_2x4}, for Per-emb-30.}
        \label{Per30_2x4}
\end{figure}

 \begin{figure}[h]
        \centering
                \includegraphics[width=0.8\textwidth]{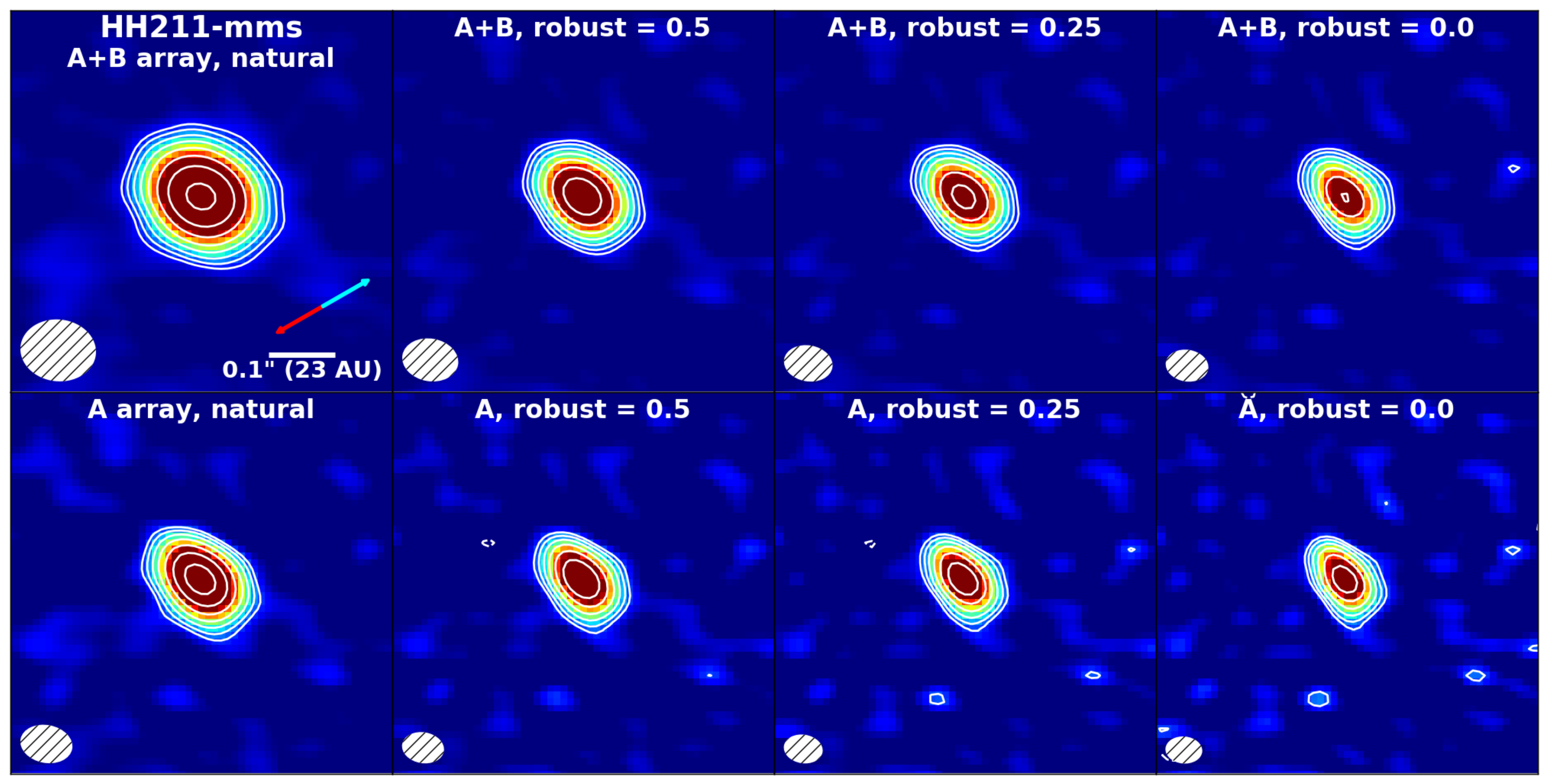}
        \caption{Same as Figure \ref{Per50_2x4}, for HH211-mms.}
        \label{Per1_2x4}
\end{figure} 

\clearpage

 \begin{figure}[h]
        \centering
                \includegraphics[width=0.8\textwidth]{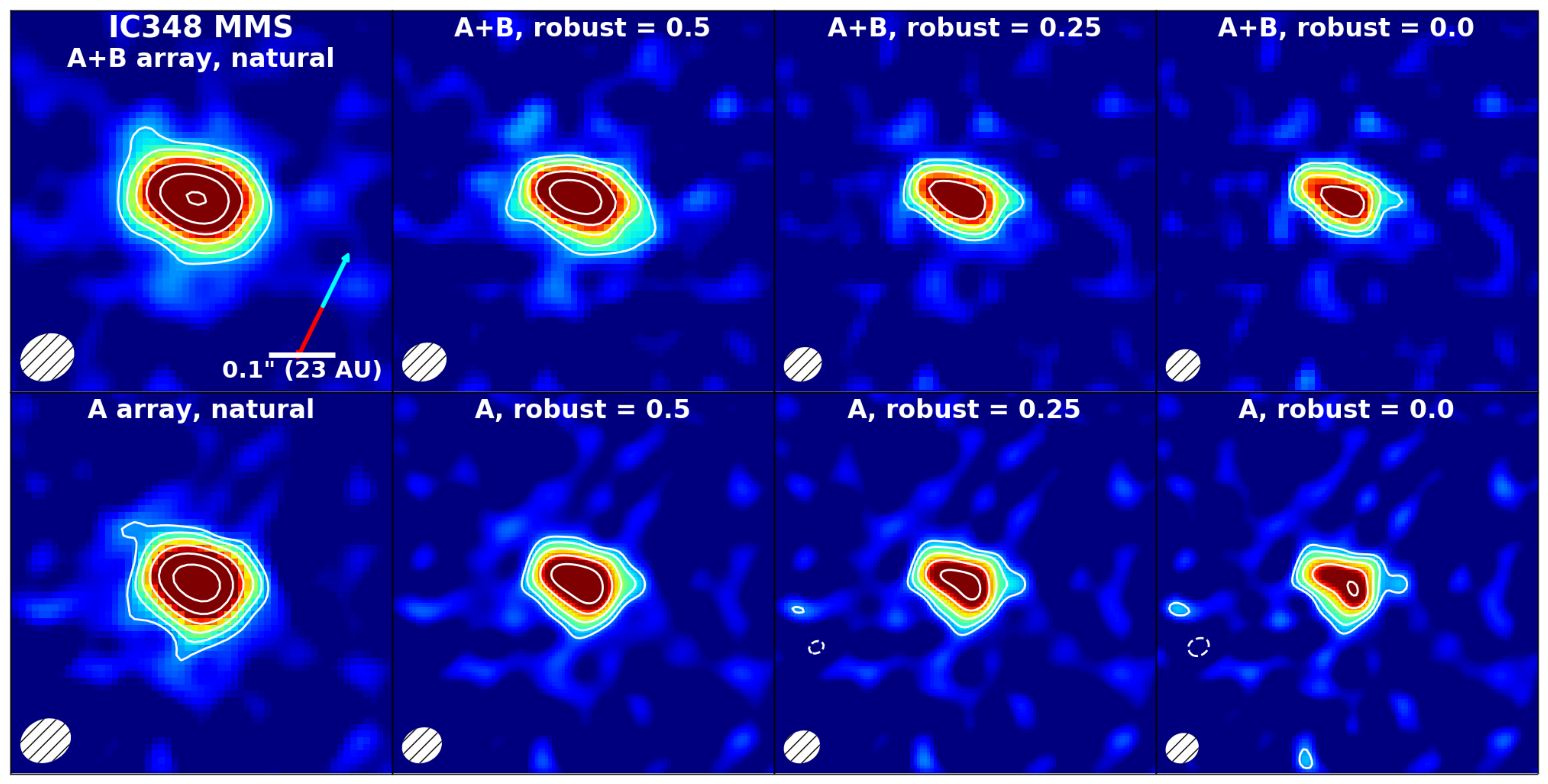}
        \caption{Same as Figure \ref{Per50_2x4}, for IC348 MMS.}
        \label{Per11_2x4}
\end{figure}

 \begin{figure}[h]
        \centering
                \includegraphics[width=0.8\textwidth]{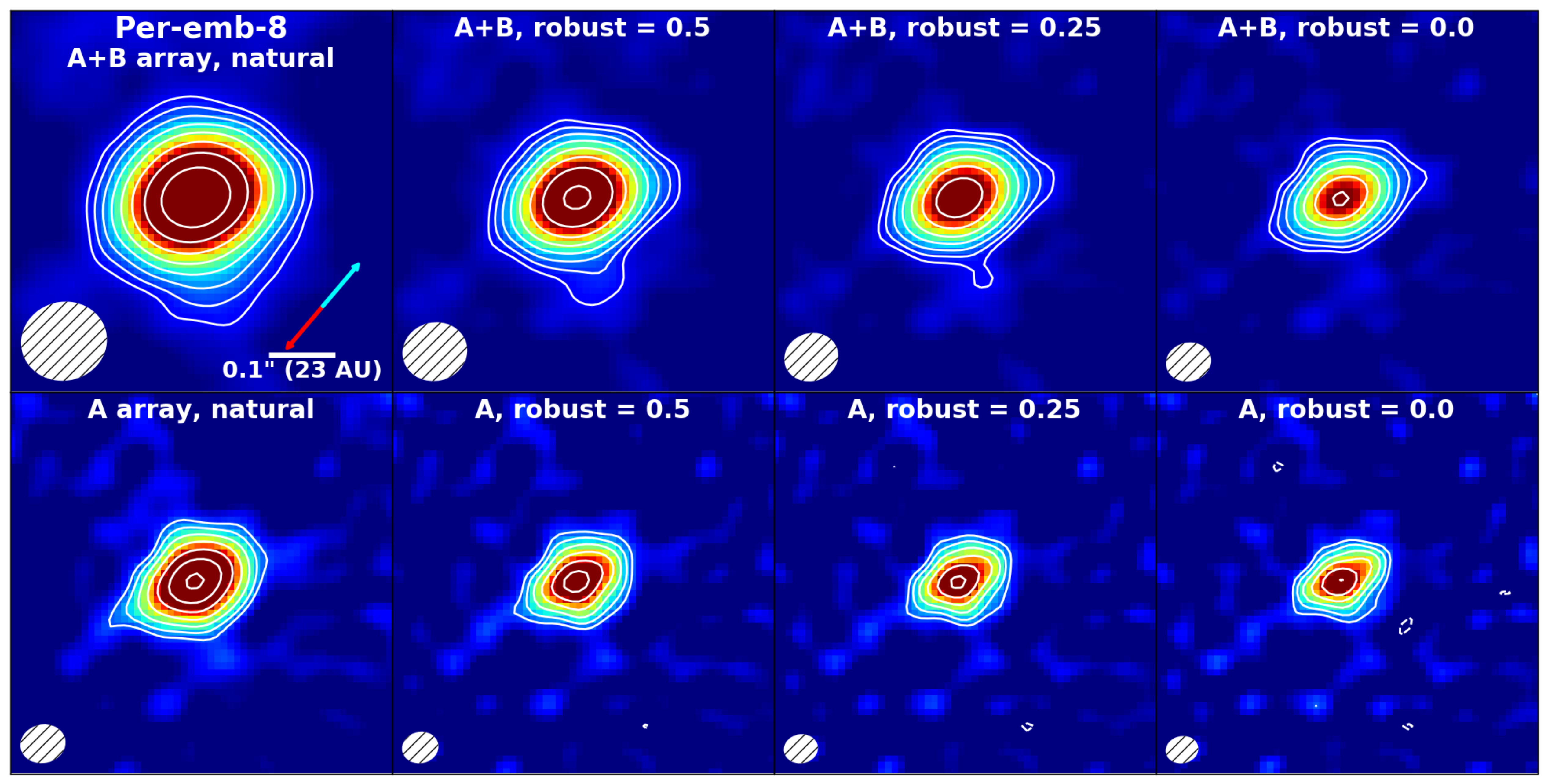}
        \caption{Same as Figure \ref{Per50_2x4}, for Per-emb-8.}
        \label{Per8_2x4}
\end{figure} 

\clearpage

 \begin{figure}[h]
        \centering
                \includegraphics[width=0.8\textwidth]{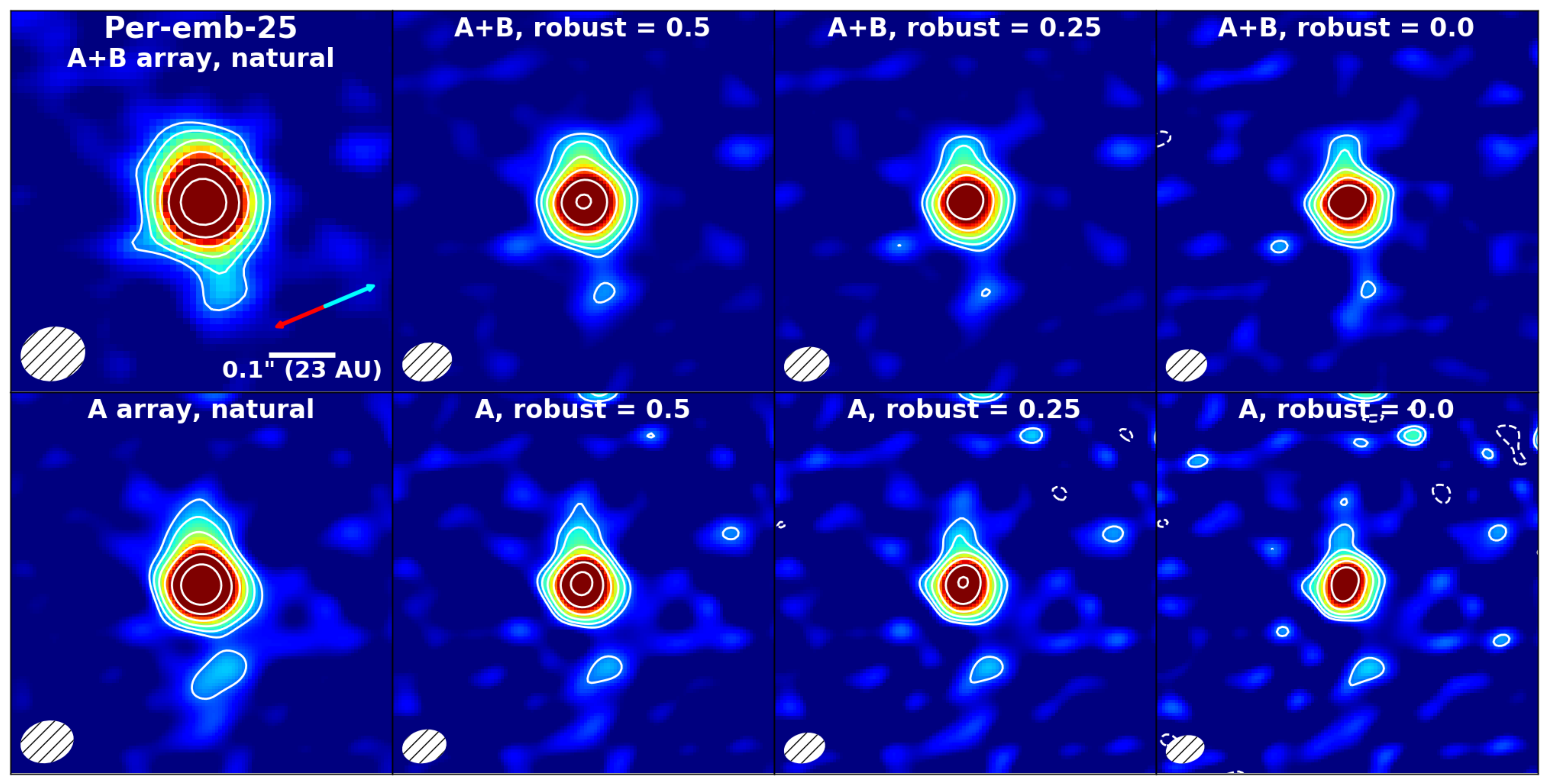}
        \caption{Same as Figure \ref{Per50_2x4}, for Per-emb-25.}
        \label{Per25_2x4}
\end{figure}

 \begin{figure}[h]
        \centering
                \includegraphics[width=0.8\textwidth]{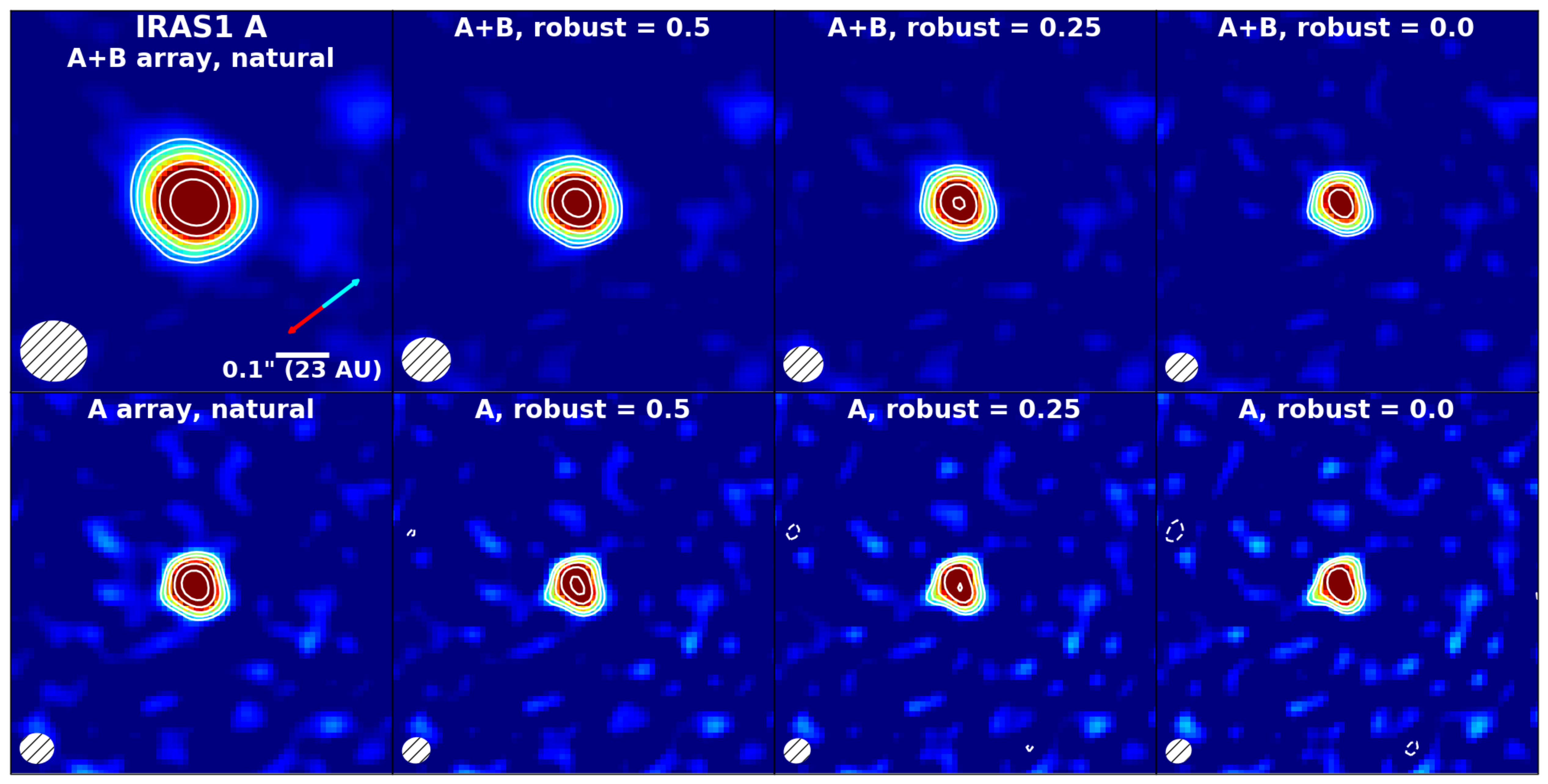}
        \caption{Same as Figure \ref{Per50_2x4}, for NGC 1333 IRAS1 A.}
        \label{Per35A_2x4}
\end{figure} 

\clearpage

 \begin{figure}[h]
        \centering
                \includegraphics[width=0.8\textwidth]{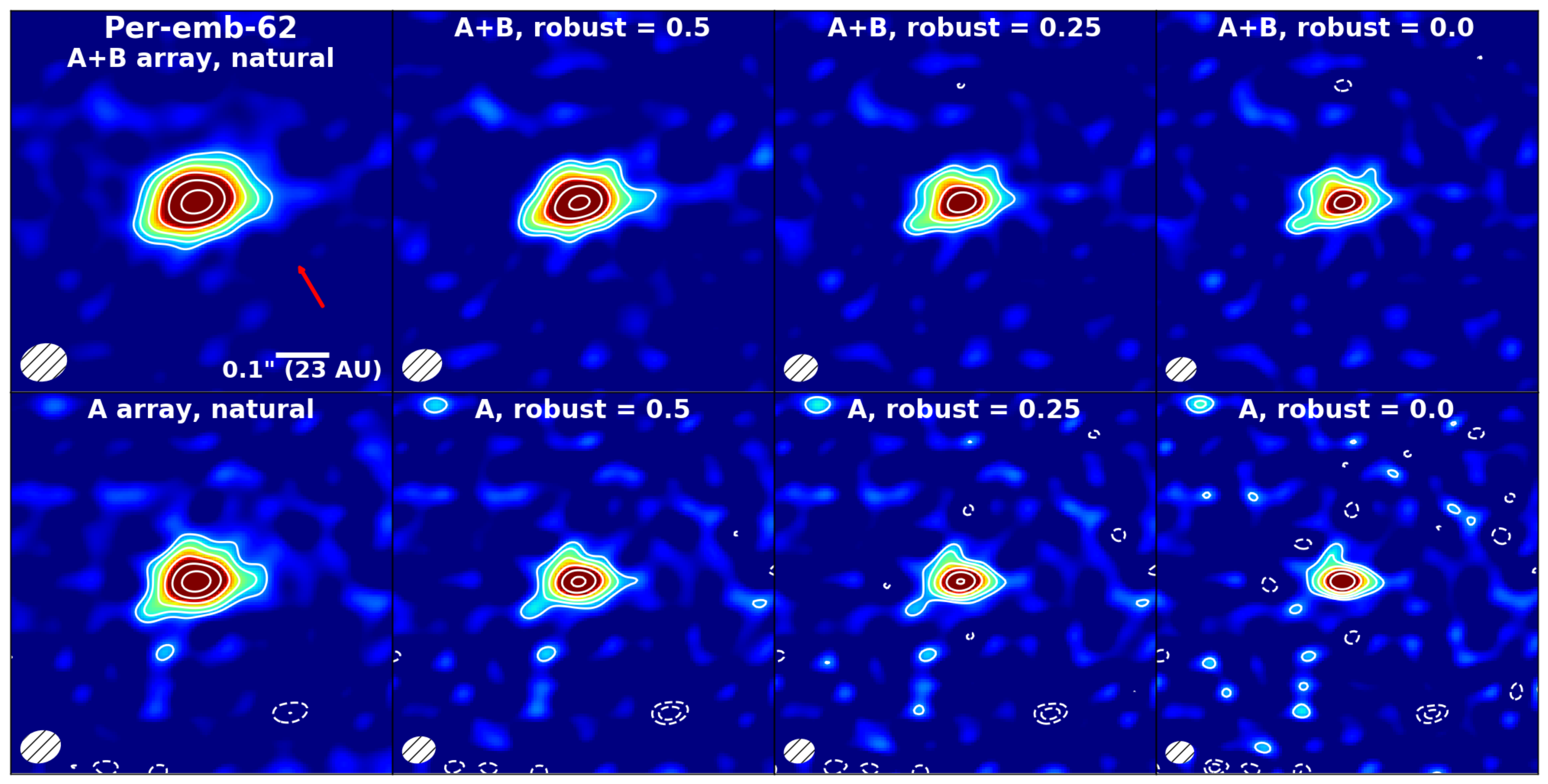}
        \caption{Same as Figure \ref{Per50_2x4}, for Per-emb-62.}
        \label{Per62_2x4}
\end{figure}

 \begin{figure}[h]
        \centering
                \includegraphics[width=0.8\textwidth]{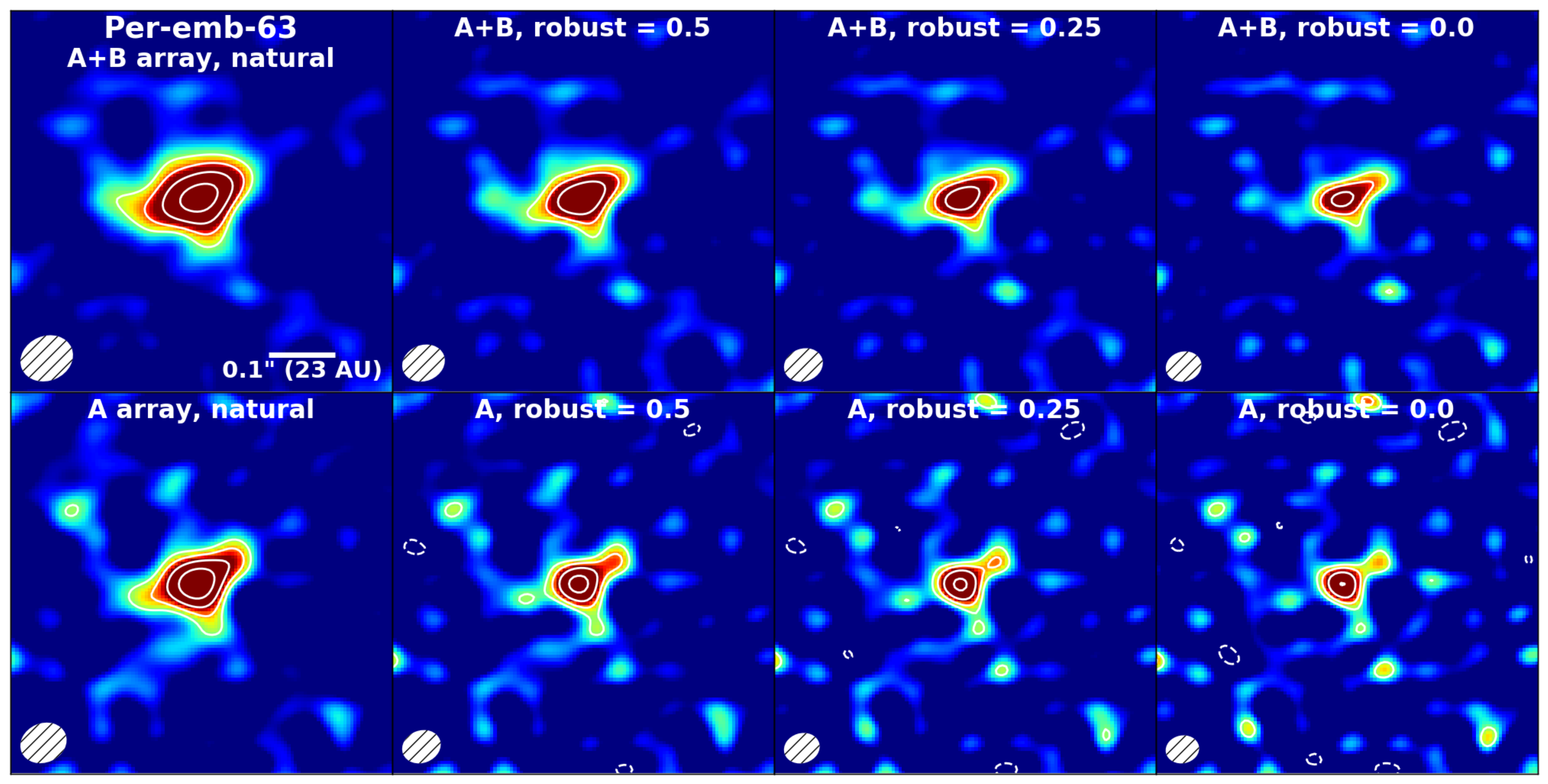}
        \caption{Same as Figure \ref{Per50_2x4}, for Per-emb-63.}
        \label{Per63_2x4}
\end{figure} 

\clearpage

 \begin{figure}[h]
        \centering
                \includegraphics[width=0.8\textwidth]{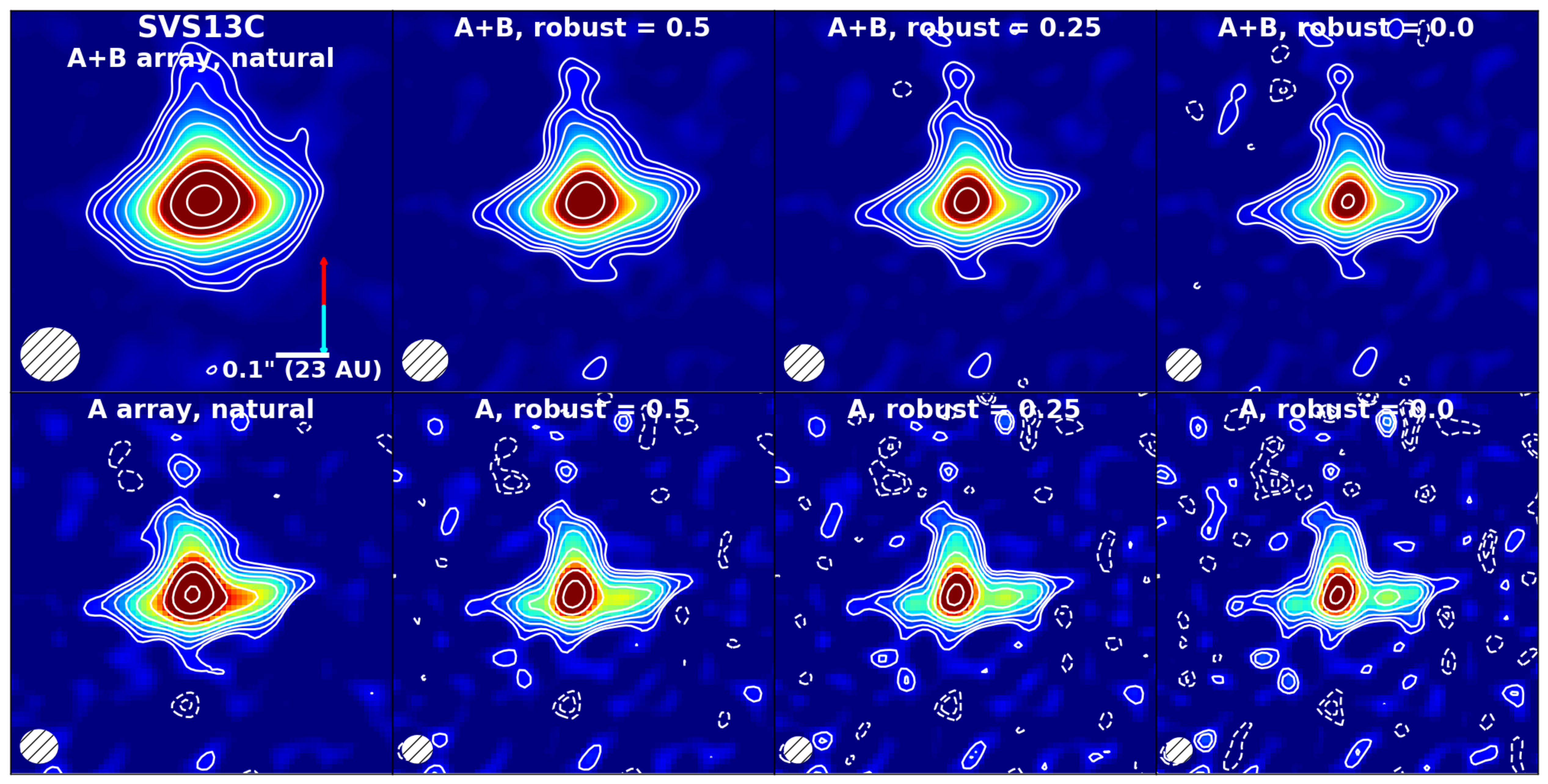}
        \caption{Same as Figure \ref{Per50_2x4}, for SVS13C.}
        \label{Per109_2x4}
\end{figure}

 \begin{figure}[h]
        \centering
                \includegraphics[width=0.8\textwidth]{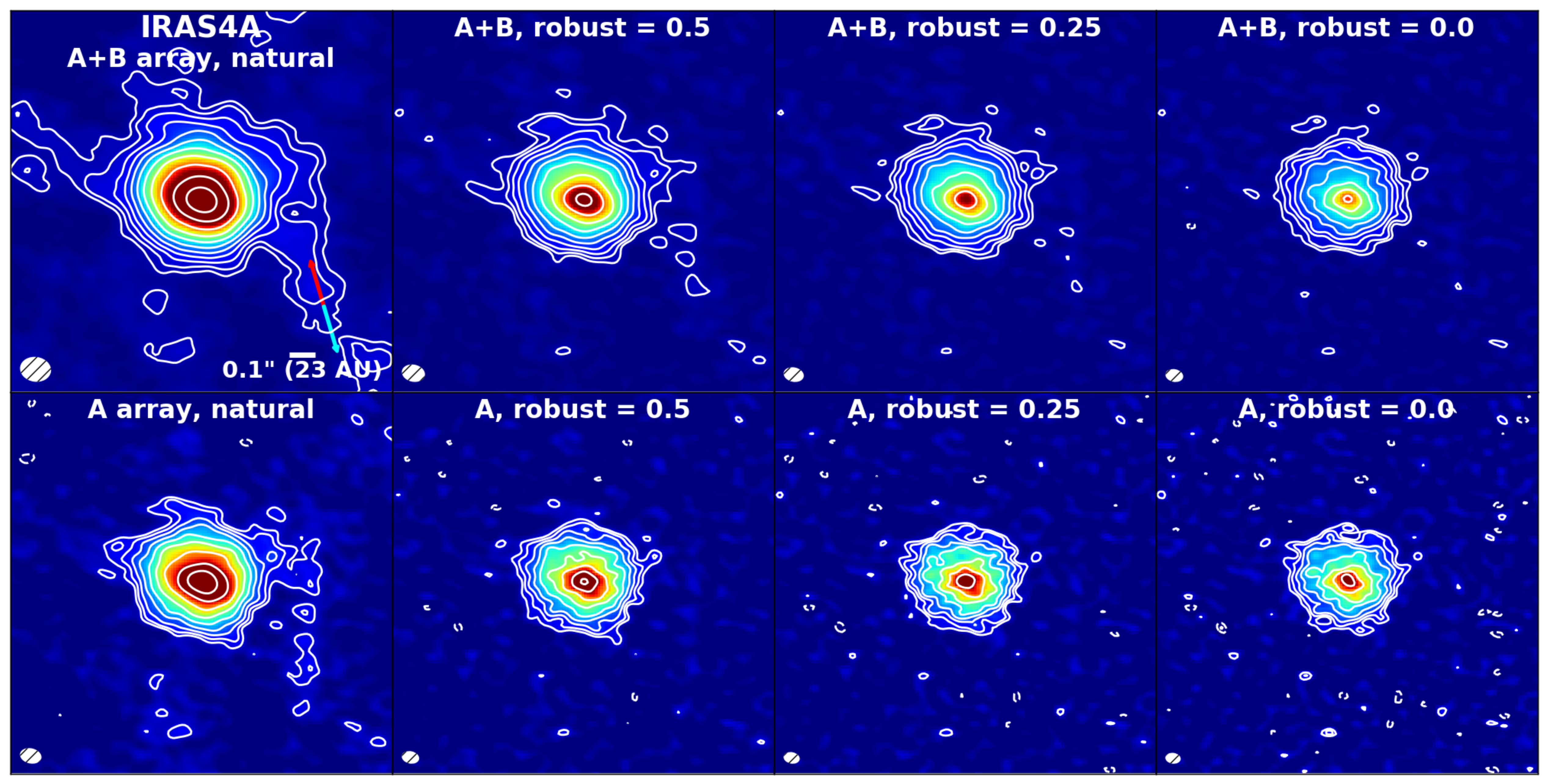}
        \caption{Same as Figure \ref{Per50_2x4}, for NGC 1333 IRAS4A.}
        \label{Per12_2x4}
\end{figure} 

\clearpage

 \begin{figure}[t]
        \centering
                \includegraphics[width=0.8\textwidth]{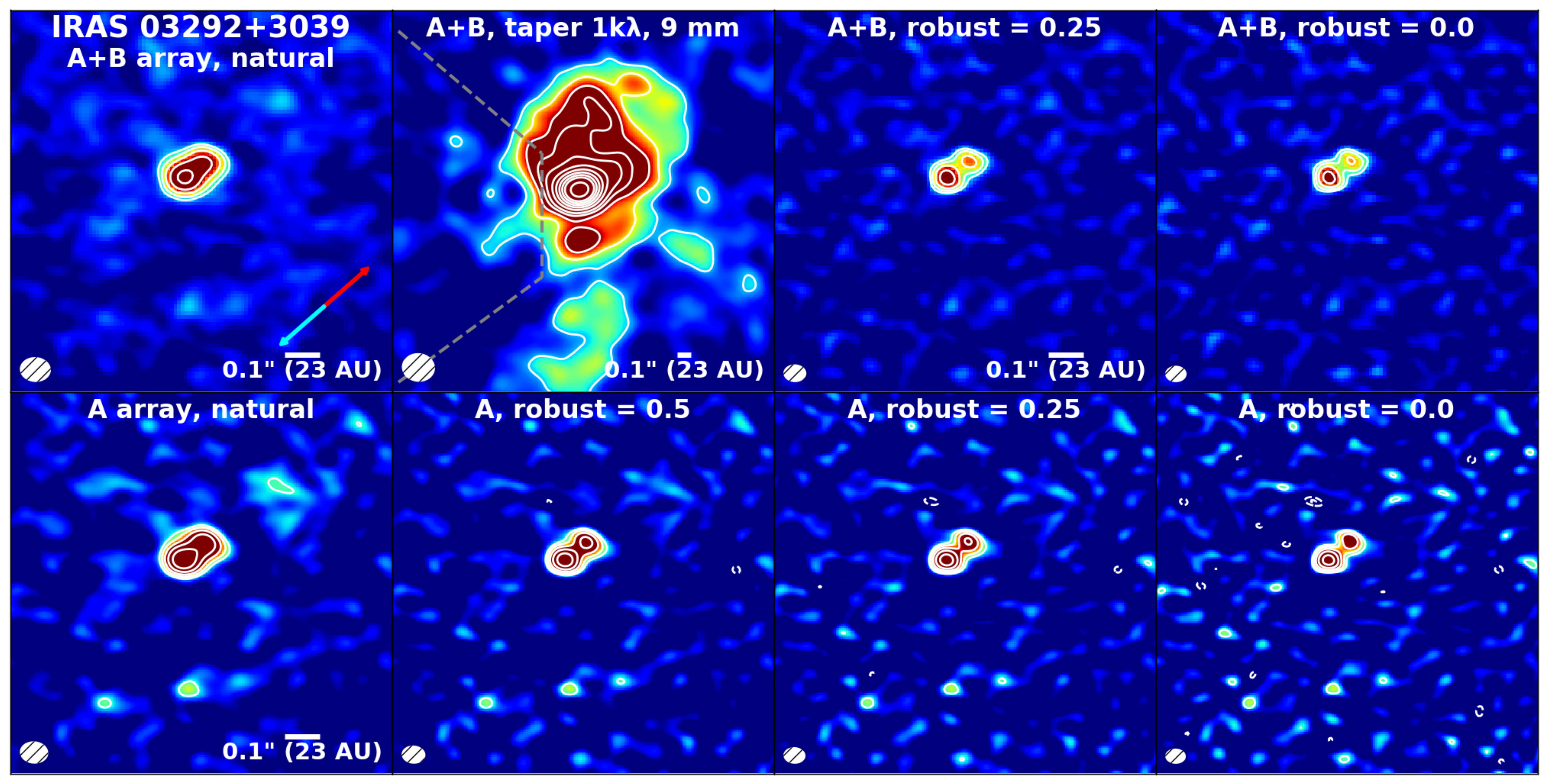}
        \caption{Same as Figure \ref{Per50_2x4}, for IRAS 03292+3039.  Here, for the upper middle-left panel, we replace the usual robust = 0.5 image with a zoomed-out image with a 1 k$\lambda$ taper applied to highlight the low-surface brightness extended emission around the central protostars.  Note the changing scale bars.  For images without scale bars, refer to the next nearest scale bar to the left.}
        \label{Per2_2x4}
\end{figure}

 \begin{figure}[h]
        \centering
                \includegraphics[width=0.8\textwidth]{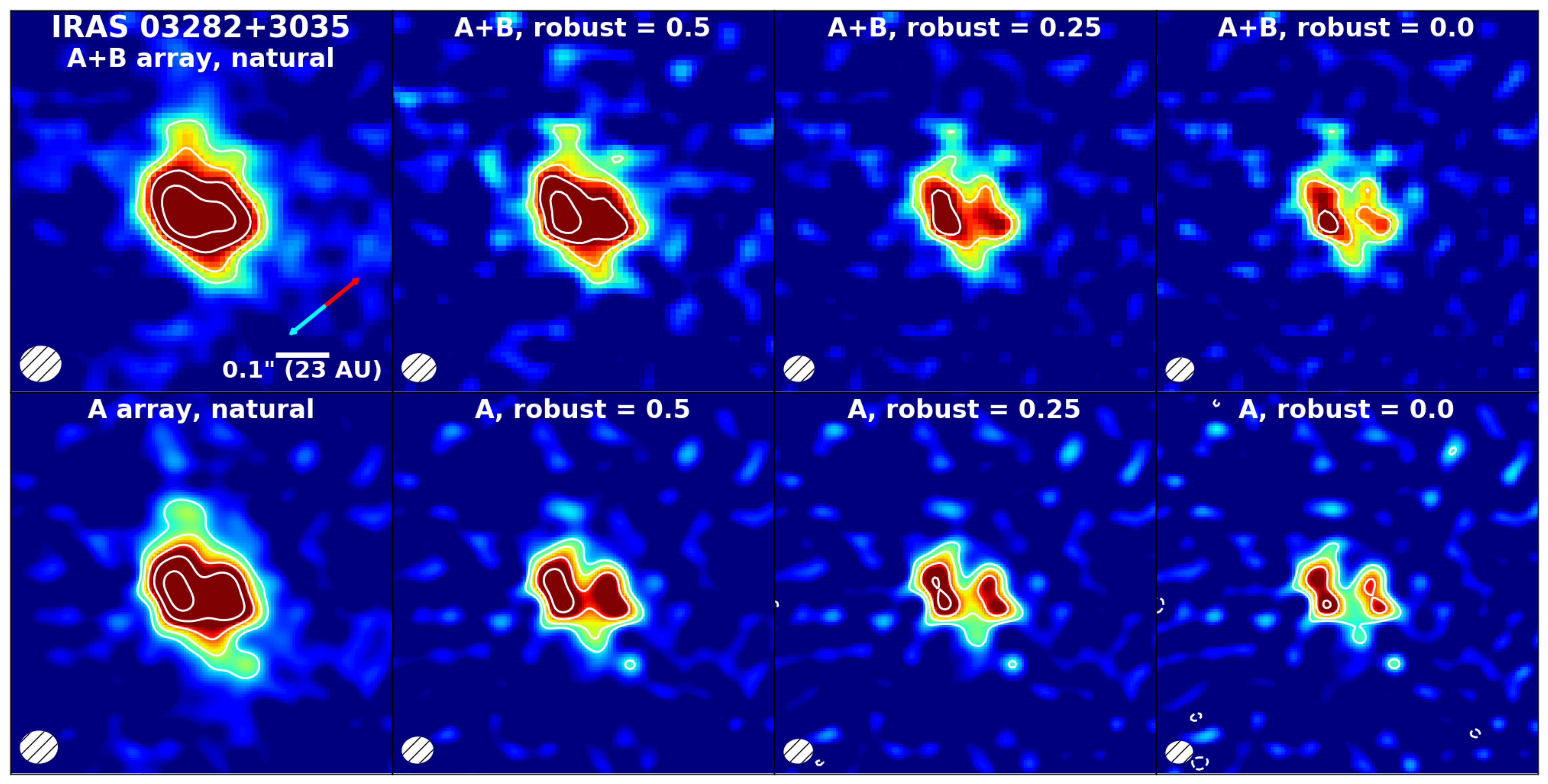}
        \caption{Same as Figure \ref{Per50_2x4}, for IRAS 03282+3035.}
        \label{Per5_2x4}
\end{figure} 

\clearpage

 \begin{figure}[h]
        \centering
                \includegraphics[width=0.8\textwidth]{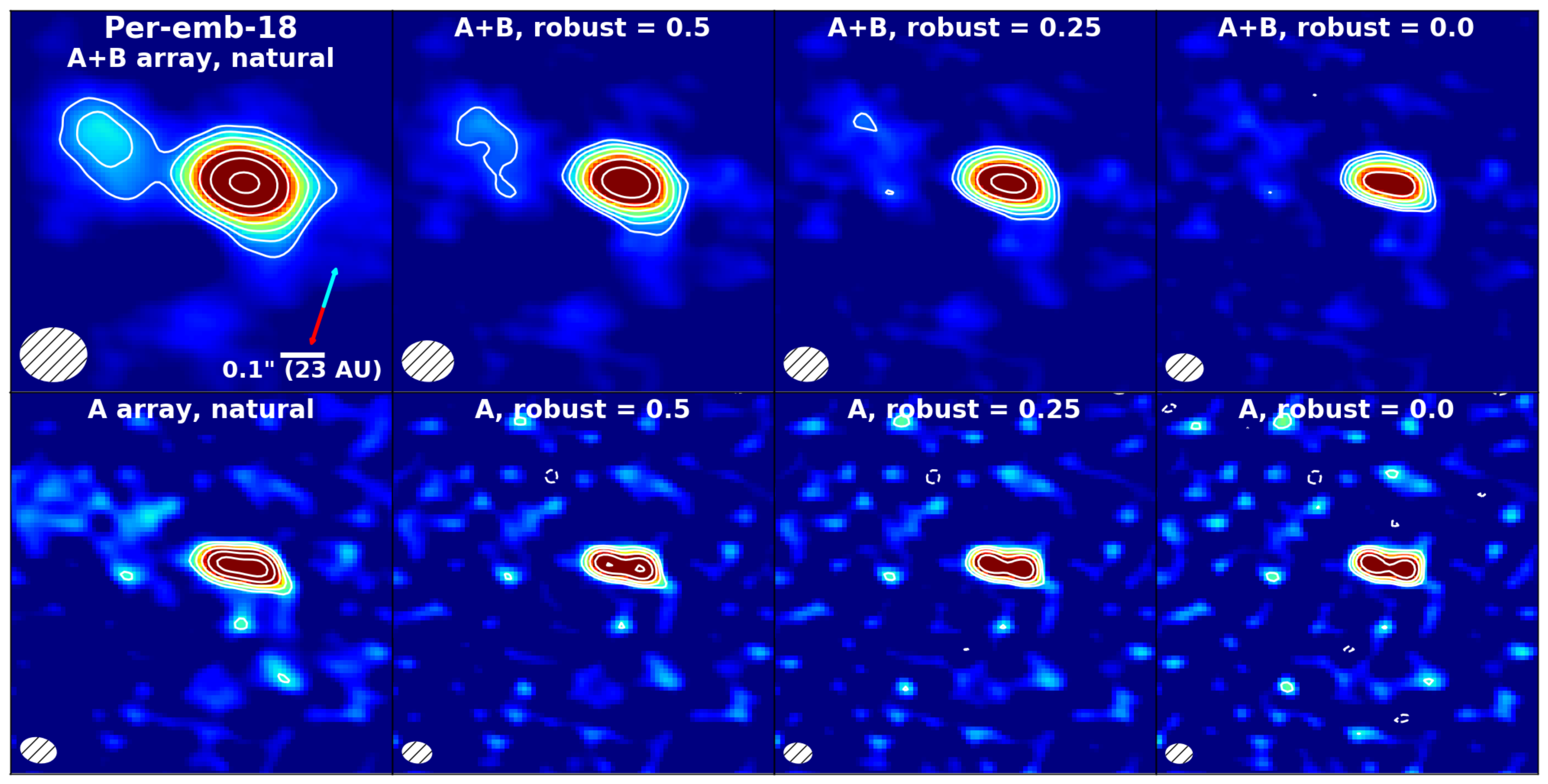}
        \caption{Same as Figure \ref{Per50_2x4}, for Per-emb-18.}
        \label{Per18_2x4}
\end{figure}

 \begin{figure}[h]
        \centering
                \includegraphics[width=0.8\textwidth]{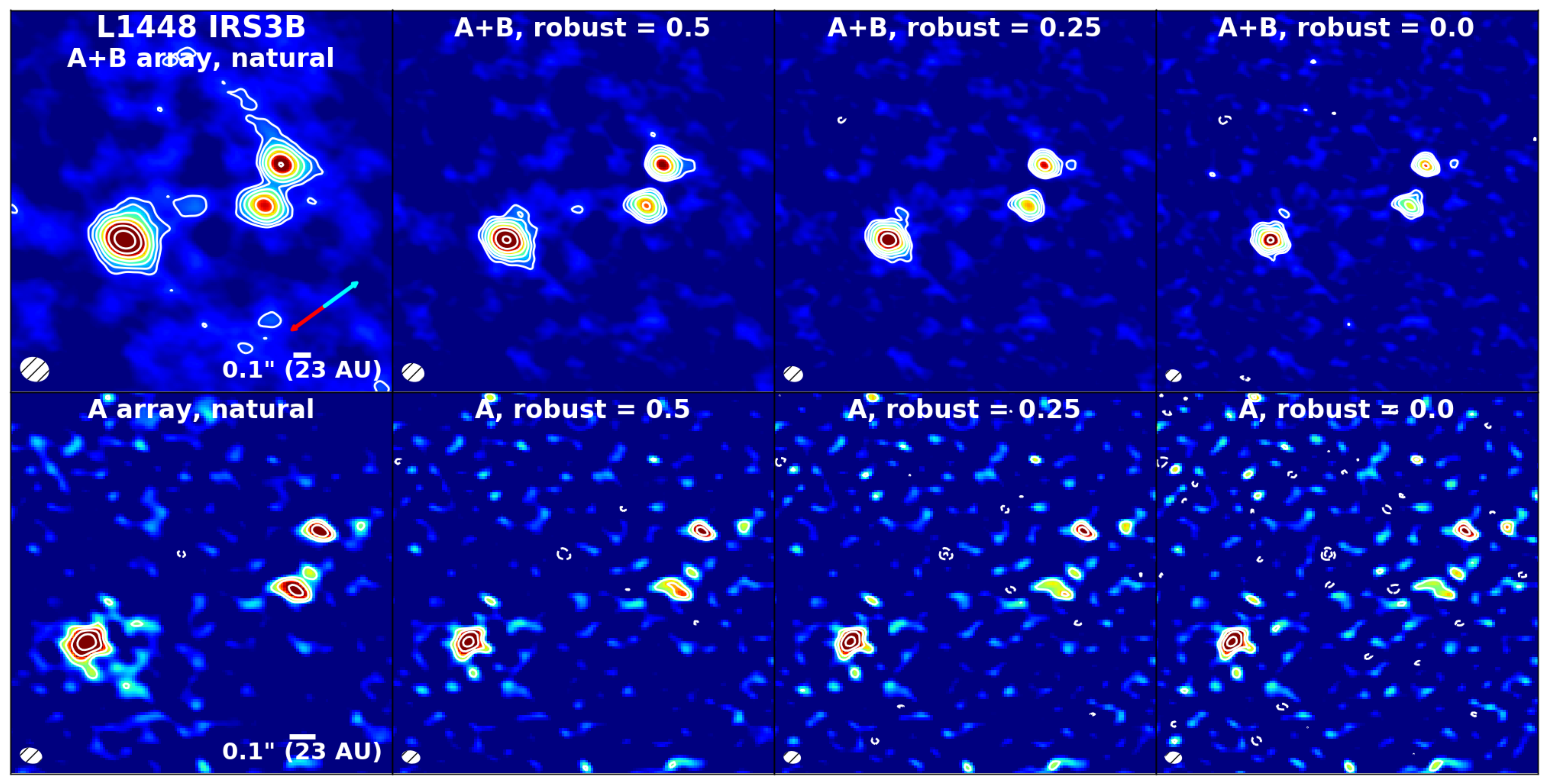}
        \caption{Same as Figure \ref{Per50_2x4}, for L1448 IRS3B. Note the changing scale bars between the upper and lower rows.}
        \label{Per33_2x4}
\end{figure} 

\clearpage

 \begin{figure}[h]
        \centering
                \includegraphics[width=0.8\textwidth]{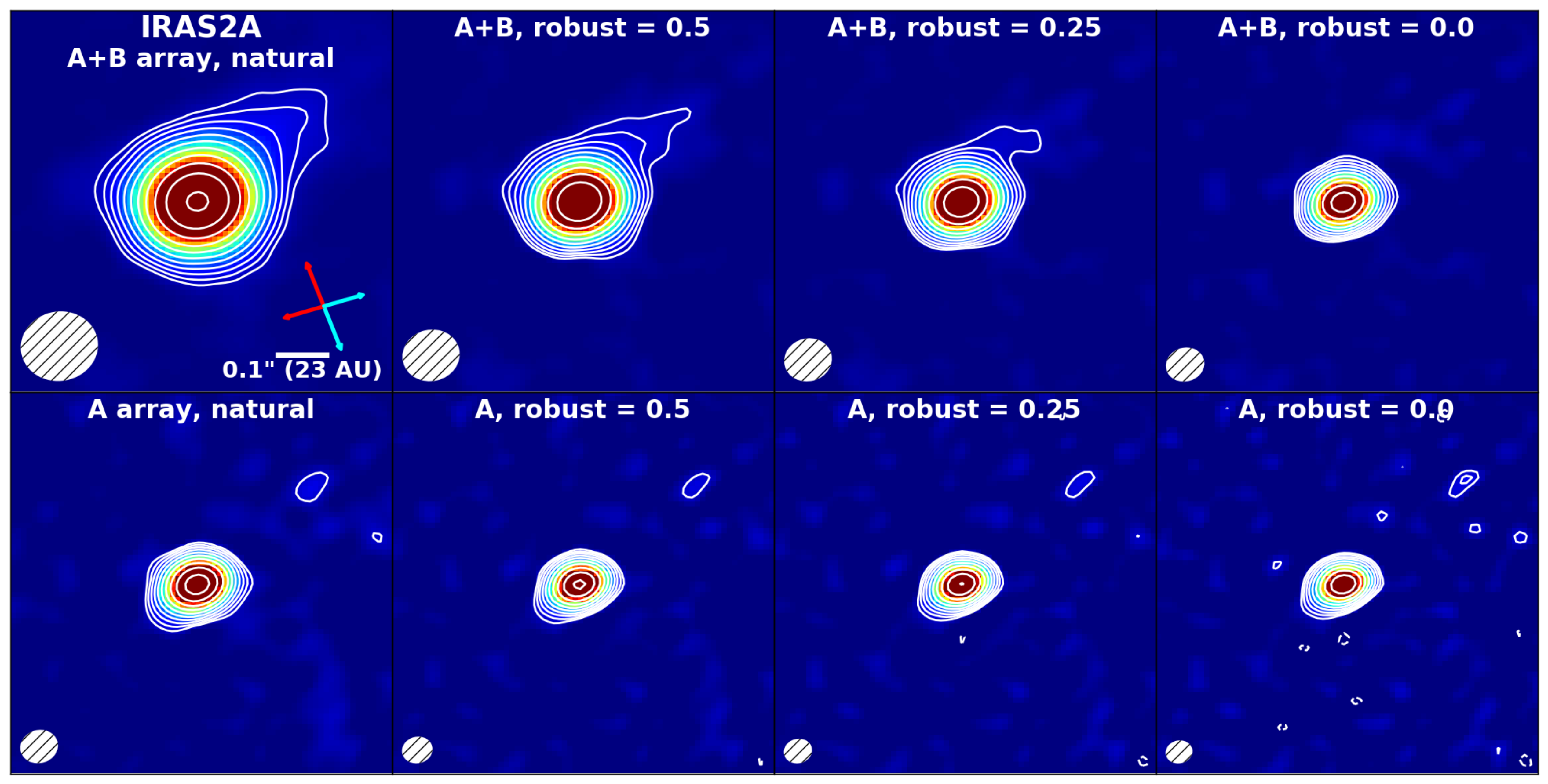}
        \caption{Same as Figure \ref{Per50_2x4}, for NGC 1333 IRAS2A.}
        \label{Per27_2x4}
\end{figure}

 \begin{figure}[h]
        \centering
                \includegraphics[width=0.8\textwidth]{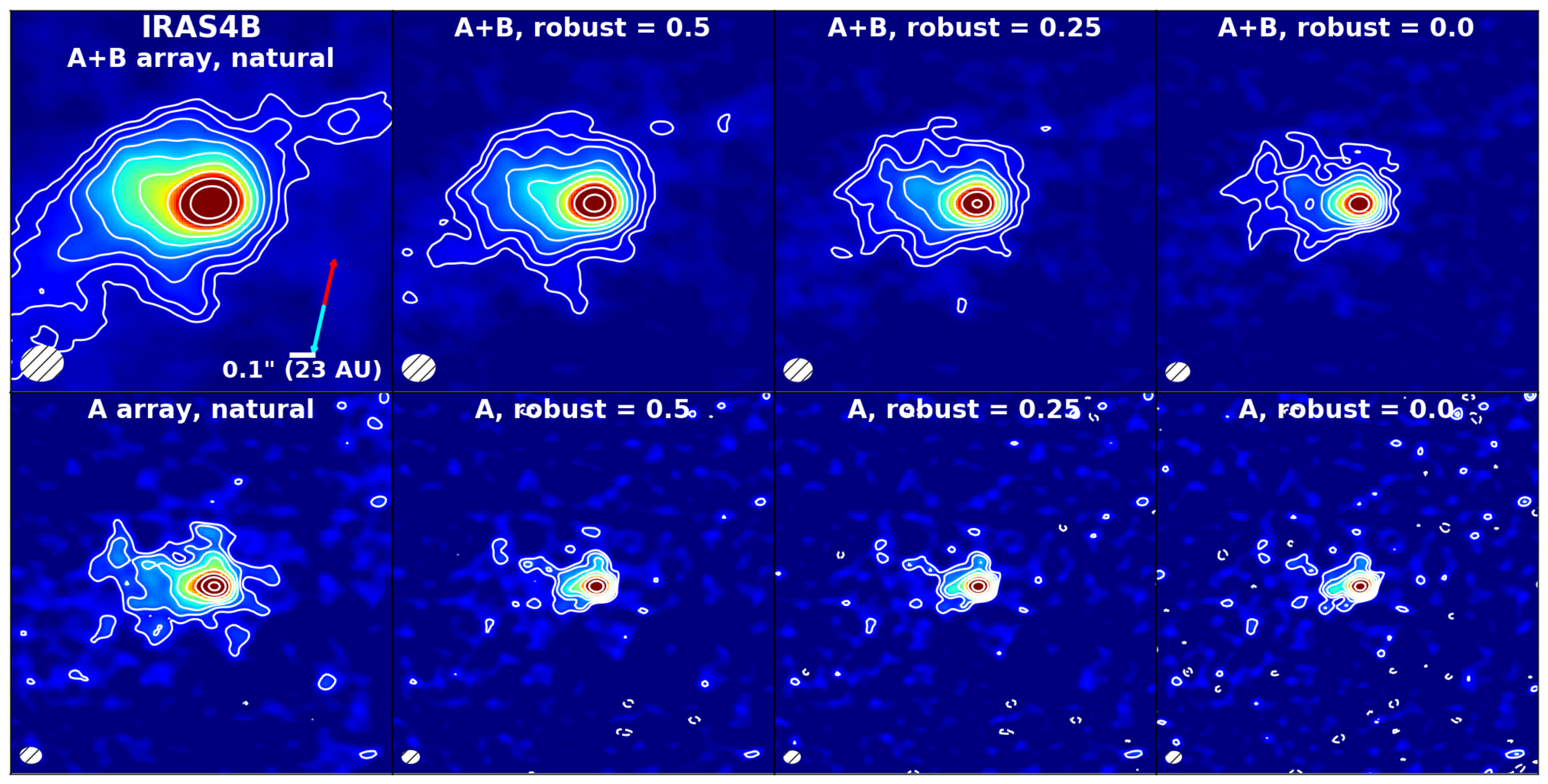}
        \caption{Same as Figure \ref{Per50_2x4}, for NGC 1333 IRAS4B.}
        \label{Per13_2x4}
\end{figure} 

\clearpage

 \begin{figure}[h]
        \centering
                \includegraphics[width=0.8\textwidth]{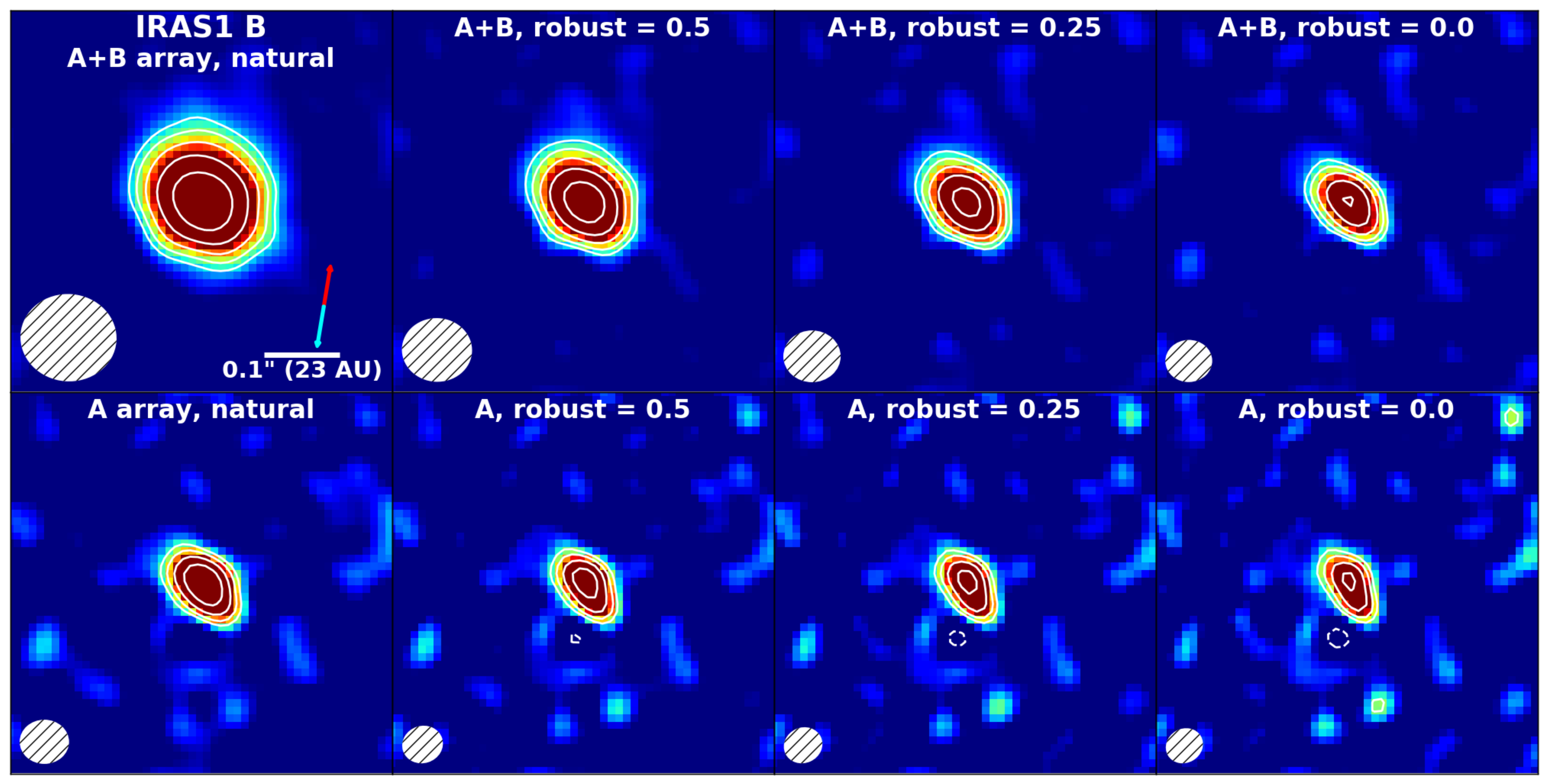}
        \caption{Same as Figure \ref{Per50_2x4}, for NGC 1333 IRAS1 B.}
        \label{Per35B_2x4}
\end{figure}

 \begin{figure}[h]
        \centering
                \includegraphics[width=0.8\textwidth]{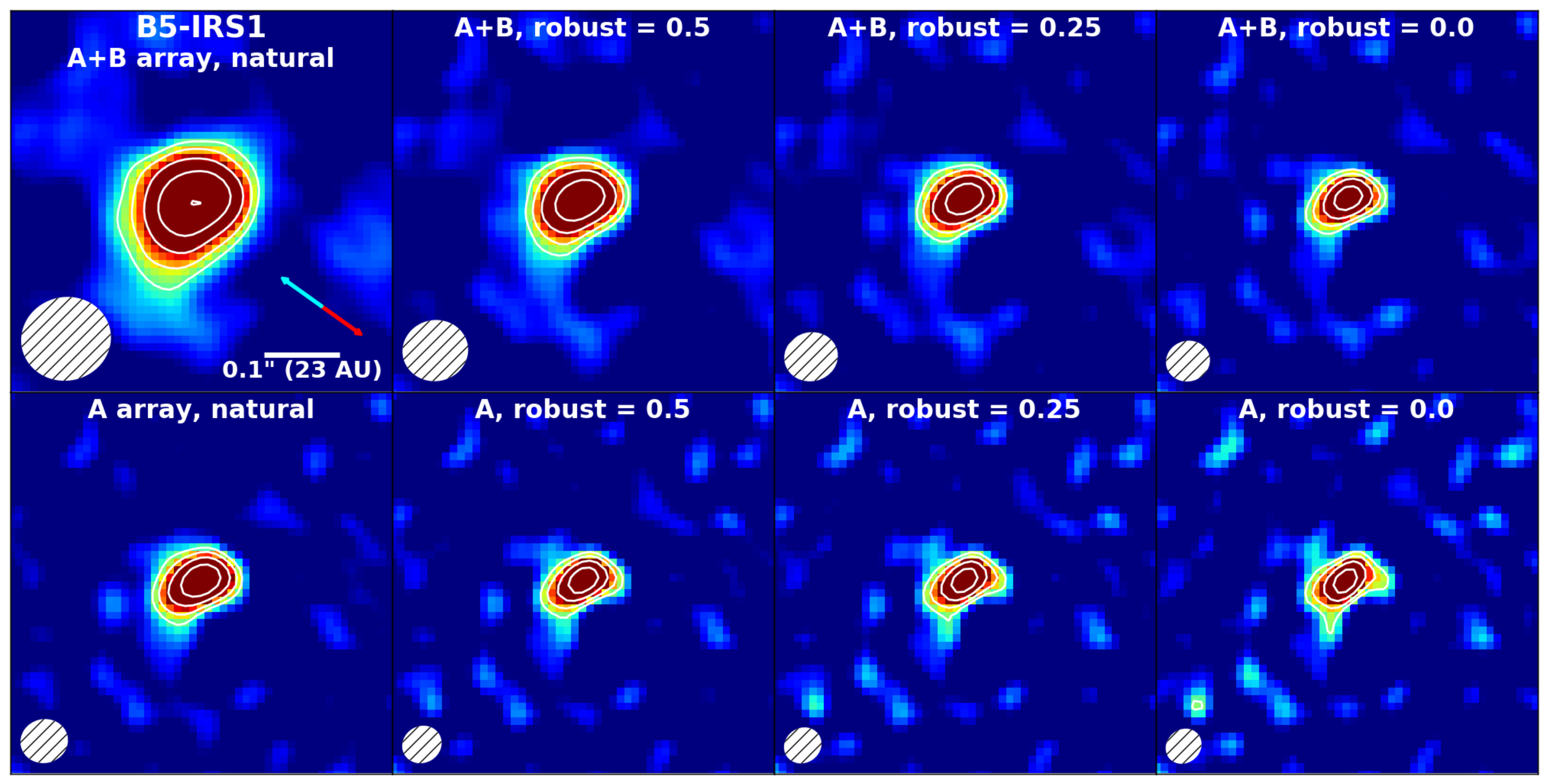}
        \caption{Same as Figure \ref{Per50_2x4}, for B5-IRS1.}
        \label{Per53_2x4}
\end{figure} 

\clearpage

\section{GALLERY OF CANDIDATE DISK SOURCES} \label{vandam:gallery}

In this Section we report the available information from literature for each VANDAM candidate disk source.  We include available information  on first observations, young stellar object classification, outflows, indirect evidence of disks, past polarization observations, and notable features.  The young stellar object classifications are based on T$_{bol}$ from \citet{Enoch2009}, \citet{Sadavoy2014}, and \citet{Young2015} unless otherwise noted.    Outflow position angles are measured counterclockwise from north.  Previous polarization observations are typically interpreted as tracing the magnetic fields of the environment surrounding the protostars, with the inferred magnetic field morphology rotated 90$^{\circ}$ from the polarized morphology \citep{SeguraCox2015}, though we discuss other possible polarization mechanisms in Section \ref{vandam:discussion:polarized}.  Analysis and interpretation of polarization results from the VANDAM survey will be discussed in Cox et al.~2018 (in preparation).

\subsection{Modeled Single Sources} \label{vandam:gallery:singles}
Sources with no detected companions within 10,000 AU.

\subsubsection{Per-emb-50} \label{vandam:gallery:singles:per50}
The Class I protostar Per-emb-50 is located in the NGC 1333 region and was first detected in the near-infrared \citep{Lada1996}, with the detection confirmed later at submillimeter wavelengths with the JCMT \citep{Hatchell2005}.  Per-emb-50 has no detected companions.  Outflows were detected in CO with a position angle of $\sim$72$^{\circ}$ \citep{Curtis2010}, roughly perpendicular to the position angle of the 8 mm VANDAM continuum emission which was well-fit by a disk profile \citep{SeguraCox2016}. 
 
\subsubsection{Per-emb-14} \label{vandam:gallery:singles:per14}
The single Class 0 protostar Per-emb-14 lies in the NCG 1333 region of Perseus, first noted by \citet{Wooten1993}.  This source has been referred to as NGC 1333 IRAS4C by some \citep[e.g.,][]{Rodriguez1999,Sandell2001}, though the IRAS4C name has also been used for the source $\sim$40$^{\prime\prime}$ from NGC 1333 IRAS4A \citep[e.g.,][]{Looney2000}.  CARMA CO observations revealed bipolar outflows with a 95$^{\circ}$ position angle \citep{Tobin2015b}.  \citet{Tobin2015b} also found a velocity gradient in C$^{18}$O perpendicular to the outflow orientation.  The VANDAM 8 mm continuum emission was  modeled and consistent with a disk \citep{SeguraCox2016} parallel to the C$^{18}$O velocity gradient, making Per-emb-14 a likely young embedded disk.  Per-emb-14 was also resolved in continuum dust emission at 1.3 mm \citep{Tobin2015b}.  The extent of the dust disk at 1.3 mm is a factor of $\sim$3$\times$ larger than the modeled 8 mm radius \citep{SeguraCox2016}.  
 
\subsubsection{Per-emb-30} \label{vandam:gallery:singles:per30}
 The Barnard 1 region hosts the Class 0 protostar Per-emb-30, first detected by \citet{Hatchell2005}.  This single source drives a monopolar H$_{2}$ jet with a position angle of 109$^{\circ}$ to the northwest of the source \citep{Davis2008}.  An HCO$^{+}$ northwestern monopolar outflow was also detected near Per-emb-30 with a similar position angle \citep{Storm2014}.   The outflow emission could be monopolar because an unseen southeastern component may  be interacting with dense gas near the protostar while the ambient medium is less dense towards the northwest where we see the monopolar flow.  The 8 mm continuum VANDAM data for Per-emb-30 was well-fit by a disk profile \citep{SeguraCox2016} with a mostly face-on ($\sim$30$^{\circ}$) inclination.

\subsubsection{HH211-mms} \label{vandam:gallery:singles:hh211}
Disk candidate HH211-mms is a Class 0 protostar in the IC 348 region of the Perseus molecular cloud with no known companions.  A well-collimated jet was first detected in H$_{2}$ near infrared observations  \citep{McCaughrean1994}, with HH211-mms confirmed as the driving source of the jet with CO observations \citep{Gueth1999}.  The position angle of the jet is $\sim$116$^{\circ}$ \citep{Hull2014}.  The kinematic structure of the envelope was studied in N$_{2}$H$^{+}$ with CARMA \citep{Tobin2011} and NH$_{3}$ with the VLA \citep{Tanner2011}, revealing velocity gradients perpendicular to the jet.  HH211-mms was well-modeled by a disk profile in 8 mm continuum emission \citep{SeguraCox2016}, with the disk parallel to the velocity gradient and perpendicular to the outflows.
   
A SCUPOL map with $\sim$4500 AU resolution toward HH211-mms showed inferred magnetic field morphology from polarization that is misaligned with the both bipolar outflow and VANDAM modeled disk orientation \citep{Matthews2009}.  The orientation of the TADPOL survey CARMA $\sim$600 AU resolution polarized emission is consistent with the misaligned inferred magnetic field morphology on larger scales \citep{Hull2014} and also traces a slight hourglass morphology \citep{Girart2006}.  Polarization towards HH211-mms was also detected with the SMA with $\sim$150 AU resolution \citep{Lee2014}.  The SMA polarized emission is present across only some parts of the protostar and has an unclear inferred magnetic field morphology (see Section \ref{vandam:discussion:polarized}). 

\subsubsection{Per-emb-25} \label{vandam:gallery:singles:per25}
Per-emb-25 is a Class 0/I protostar first detected with \textit{IRAS} \citep{Ladd1993}, near the L1452 region in Perseus.  Outflows were detected in CO with JCMT and CSO with position angles near 110$^{\circ}$ \citep{Dunham2014b}.  This protostar has no detected companions near submillimeter wavelengths, though the extension to the north seen in Figure \ref{Per25_2x4} coincides with a free-free peak in the C-band VANDAM data at 4.1 cm, indicating Per-emb-25 may be a binary source \citep{Tychoniec2018submitted}.  

\subsubsection{Per-emb-62} \label{vandam:gallery:singles:per62}
The Class I protostar Per-emb-62 is a single protostar and lies in the IC 348 region.  It was first observed as part of an optical and near-infrared survey of the IC 348 region \citep{Herbig1998} and has since been observed in millimeter/submillimeter wavelengths.  A monopolar outflow emanating from Per-emb-62 was detected in CO with the JCMT \citep{Hatchell2009} with a position angle of $\sim$35$^{\circ}$. 

\subsubsection{Per-emb-63} \label{vandam:gallery:singles:per63}
The  Class I protostar Per-emb-63 is found in the NGC 1333 region.  It was discovered with observations made by the Palomar Observatory \citep{Cohen1980} and later confirmed  with millimeter/submillimeter observations.  This source is not known to have any companions, and it is unclear if it drives any outflows.   
 

\subsection{Modeled Multiple Sources} \label{vandam:gallery:binaries}
Sources with at least 1 companion detected within 10,000 AU, as reported in \citet{Tobin2016a}. 

\subsubsection{SVS13B}   \label{vandam:gallery:binaries:svs13b}
SVS13B is a Class 0 protostar in NGC 1333.  It is a part of the larger protostellar multiple system SVS13 \citep{Strom1976}, which is comprised of SVS13A, SVS13B, and SVS13C \citep{Looney2000}  SVS13B was first detected at $\sim$3 mm \citep{Grossman1987}, and later confirmed with more sensitive observations \citep{Chini1997}.  SVS13B is $\sim$3400 AU and $\sim$4500 AU from SVS13A and SVS13C, respectively \citep{Tobin2016a}. SVS13A is a Class 0/I protostar, while SVS13B and SVS13C are Class 0 sources.  SiO emission revealed an outflow emanating from SVS13B with a position angle of 160$^{\circ}$ \citep{Bachiller1998}, agreeing well with the outflow position angle of $\sim$170$^{\circ}$ from CO observations \citep{Lee2016}.  The 8 mm VANDAM continuum data for SVS13B was previously modeled and consistent with a protostellar disk profile \citep{SeguraCox2016}.
  
Polarized emission was detected towards SVS13B at 1.3 mm with CARMA \citep{Hull2014}.  The inferred magnetic field morphology from polarization orientation reflects an hourglass morphology \citep{Girart2006}, with the average orientation misaligned with both the outflows and modeled candidate disk orientation of the system.

\subsubsection{IC348 MMS}   \label{vandam:gallery:binaries:ic348mms}
IC348 MMS is a Class 0 source in the IC 348 region, detected first via its molecular H$_{2}$ outflows \citep{Eisloffel2003}. IC348 MMS is a multiple with $\sim$1500 AU separation \citep{Chen2013}, with a third closer component identified more recently \citep{Rodriguez2014}.  \citet{Tobin2016a} denotes our candidate disk modeled here as Per-emb-11-A.  The closer companion, Per-emb-11-B appears to be directly between outflow cavities detected in CO with a position angle of $\sim$150$^{\circ}$ \citep{Pech2012} while Per-emb-11-A lies to the side of the outflow cavity \citep{Tobin2016a}, indicating that our candidate disk may not be the central driving source of the outflows in the system. In \citet{SeguraCox2016}, we found the 8 mm VANDAM continuum data was well-fit by a disk profile extended perpendicular to the outflows.  
    
\subsubsection{Per-emb-8}  \label{vandam:gallery:binaries:per8}
The Class 0 protostar Per-emb-8 is a part of a wide binary system ($\sim$2200 AU separation) along with the Class I protostar Per-emb-55 \citep{Tobin2016a} in the IC 348 region of Perseus.  Per-emb-8 was first detected in the \textit{Spitzer} c2d survey \citep{Rebull2007}.  Per-emb-55 contains two protostars separated by less than 150 AU \citep{Tobin2016a}, bringing the total protostellar count of the system to three.   We modeled the 8 mm VANDAM continuum in \citet{SeguraCox2016} with a disk-shaped profile, and the simple profile of our disk model reasonably fit the data with $\chi^{2}_{reduced}$ near 1.  Recently, ALMA $^{12}$CO observations revealed bipolar outflows emanating from Per-emb-8 with a position angle of $\sim$135$^{\circ}$ (Tobin et al.~2018, submitted), only a 19$^{\circ}$ separation in position angle from the major axis of the 8 mm candidate disk. Tobin et al.~(2018, submitted) also revealed the 1.3 mm ALMA continuum data to have a disk position angle of $\sim$45 degrees, a 71$^{\circ}$ discrepancy from the \citet{SeguraCox2016} 8 mm modeling result.  The Per-emb-8 VANDAM 8 mm data may be contaminated by free-free emission from the jet \citep{Tychoniec2018}, or the large grains traced by the 8 mm emission may have a different morphology than the smaller grains traced by the 3.1 mm ALMA data.  The complications found in this source highlight the importance of considering data at multiple wavelengths when studying disk structures.

\subsubsection{NGC 1333 IRAS1 A}   \label{vandam:gallery:binaries:iras1a}
The Class I protostar NGC 1333 IRAS1 A is located on the edge of the NGC 1333 region, and has a binary source $\sim$440 AU away \citep{Tobin2016a}.  This source was first detected by \textit{IRAS} observations of the NGC 1333 region \citep{Jennings1987}. The outflow which originates from the system has an S-shape \citep{Gutermuth2008}, likely due to binary interactions altering outflow morphology \citep{Tobin2016a}.  The outflow has a position angle of $\sim$123$^{\circ}$ \citep{Stephens2017}.
  
\subsubsection{SVS13C}   \label{vandam:gallery:binaries:svs13c}
The SVS13 system in NGC 1333 also hosts SVS13C, a Class 0 protostar first identified by \citet{Chini1997} and later confirmed at 2.7 mm by  \citet{Looney2000}.  A molecular outflow in the north-south direction was first identified by \citet{Plunkett2013}, and an outflow position angle of 0$^{\circ}$ was later found with more sensitive CO observations \citep{Lee2016}.  The outflow is also detected in the VANDAM C-band data, tracing free-free emission, along the north-south direction \citep{Tychoniec2018}.  Kinematics reveal the outflow to be nearly in the plane of the sky \citep{Plunkett2013,Lee2016}, indicating that the disk launching the outflow should be nearly edge-on.  

\subsubsection{NGC 1333 IRAS4A}   \label{vandam:gallery:binaries:iras4a}
NGC 1333 IRAS4A is the brightest protostar in our sample of disk candidates (Table \ref{diskfulltab2}).  Located in the NGC 1333 region, NGC 1333 IRAS4A has been long known to be a Class 0 protostar based on it's SED \citep{Sandell1991,Andre1993}, and was first observed at 50 and 100 $\mu$m with \textit{IRAS} \citep{Jennings1987}. To date, there has been no confirmation of Keplerian rotation, hard evidence of a disk, in NGC 1333 IRAS4A due to the dense envelope surrounding the protostar, making observations difficult.  Continuum emission observed with the SMA at 335 GHz was modeled with a disk profile by \citet{Persson2016}. NGC 1333 IRAS4A has a close Class 0 companion \citep[separation 420.8 AU,][]{Tobin2016a} called NGC 1333 IRAS4A2, first detected at 0.84 mm \citep{Lay1995}, with the binary system sharing a common envelope \citep{Looney2000}.  The binary system has been resolved at multiple wavelengths \citep{Looney2000,Jorgensen2007,Reipurth2002}.  Distinct outflows from each source were resolved in SiO, SO, and CO \citep{Santangelo2015}. \citet{Lee2016} found the outflows of our candidate disk NGC 1333 IRAS4A, as traced by CO, to have a position angle of $\sim$19$^{\circ}$.   The line-of-sight velocities detected in the CO outflows indicate that the system has an inclination  of $<$45$^{\circ}$ \citep{Hull2014}, consistent with our measured inclination (Table \ref{diskfulltab2}). NGC 1333 IRAS 4A revealed low- and high-density molecular lines tracers with inverse  P-Cygni profiles \citep{Jorgensen2007,DiFrancesco2001}.  The inverse P-Cygni profiles are consistent with material infalling onto the binary protostar system.  Inner-envelope scale kinematic observations have revealed a velocity gradient in the east-west direction across the source \citep{Yen2015}, roughly perpendicular to the outflows and consistent with a disk present in the system.

Polarized observations of dust emission towards NGC 1333 IRAS4A have been made at 850 $\mu$m, 1.3 mm, and 8 mm with SCUPOL, SMA and CARMA, and the VLA respectively \citep{Matthews2009,Girart2006,Hull2014,Cox2015}.  The SCUPOL data, on $\sim$4500 AU scales traces a uniform inferred magnetic field oriented in the northeast-southwest direction.  The SMA data, with $\sim$300 AU resolution, shows an inferred magnetic field consistent with the larger scale SCUPOL field and reveals one of the first hourglass-shaped inferred magnetic field morphologies towards a low-mass protostar.  The TADPOL survey CARMA data also have an inferred magnetic field with an hourglass shape.  The 8 mm VLA polarization data ($\sim$65 AU resolution) was observed as a part of the larger VANDAM survey; 10 mm VANDAM polarized emission was also detected with the same morphology as the 8 mm data \citep{Cox2015}.  The inferred magnetic field orientation from the polarized VLA emission reveals a circular morphology circling the face of the modeled candidate disk, a stark contrast to the hourglass morphology on larger scales.

\subsubsection{NGC 1333 IRAS2A}   \label{vandam:gallery:binaries:iras2a}
NGC 1333 IRAS2A is a relatively well-studied Class 0/I protostar in NGC 1333, reported first in \citet{Jennings1987}.  The IRAS2 core hosts at least three embedded young stellar objects  \citep{Sandell2001}, with NGC 1333 IRAS2A emitting the brightest at millimeter wavelengths.  Two bipolar outflows, almost orthogonal to each other, appear to arise from NGC 1333 IRAS2A in both single dish \citep{Sandell1994} and higher-resolution interferometric CO observations \citep{Engargiola1999}. The north-south outflow has a position angle of $\sim$25$^{\circ}$ while the east-west outflow has a position angle of  $\sim$104$^{\circ}$ \citep{Hull2014}. Two orthogonal outflows do not have a clear, theoretical launching mechanism from a single disk, so this is taken as evidence of a close multiple system \citep{Jorgensen2004}.  High-resolution observations of NGC 1333 IRAS2A did not find close companions \citep{Looney2000,Maury2010} until the VANDAM survey resolved a binary source with $\sim$140 AU separation \citep{Tobin2015a}.  The close binary components are referred to as VLA1 and VLA2; our candidate disk is VLA 1. \citet{Tobin2015a} concluded that our candidate disk drives the north-south outflow, with VLA2 likely driving the east-west outflow.  Plateau de Bure Interferometer (PdBI, now NOEMA) 203.4 GHz continuum data was described by a disk profile \citet{Persson2016}.
 
NGC 1333 IRAS2A has polarization emission, with the inferred magnetic field orientation directly associated with the protostellar peak oriented in the east-west direction \citep{Hull2014}.  The inferred magnetic field is misaligned with both the north-south outflow associated with NGC 1333 IRAS2A and the candidate disk orientation and may reflect an hourglass morphology \citep{Girart2006}.


\subsection{Complicated Sources}   \label{vandam:gallery:weird}
These are extended multiple sources which cannot be modeled by our axisymmetric disk model due to irregular structures, asymmetry, or close-separation binaries.  Three of these sources have emission surrounding close-separation binaries, which are resolved only with the A-array \citep[Figures \ref{Per2_2x4}-\ref{Per18_2x4}; see also][]{Tobin2016a}.  All four sources are extremely extended and have indirect evidence of circumbinary disks; we consider these sources VANDAM candidate disks even without disk modeling.

\subsubsection{IRAS 03292+3039}    \label{vandam:gallery:weird:per2}
IRAS 03292+3039 is a Class 0 protostar first detected by \textit{IRAS} \citep{Bachiller1990} in the Barnard 1 region, and it is the most spatially extended protostellar system out of all protostars detected by the VANDAM survey.  This system is comprised of two close components separated by $\sim$18 AU, resolved with the A array VANDAM data \citep{Tobin2016a}.   The extended resolved structure is non-axisymmetric and includes a southern extension, hence we are unable to model this source according to our prescription in Section \ref{vandam:modeling}.  Indications of rotation in IRAS 03292+3039 were seen on 1000 AU scales in C$^{18}$O \citep{Yen2015}, which appear to be roughly perpendicular to a known bipolar outflow with an orientation of 127$^{\circ}$ \citep{Schnee2012}. 

\subsubsection{IRAS 03282+3035}    \label{vandam:gallery:weird:per5}
The Class 0 protostar IRAS 03282+3035, first observed as a low-luminosity \textit{IRAS} source \citep{Bachiller1991},  is an isolated protostellar system near the Barnard 1 region \citep{Jorgensen2006}.  It was identified as a binary in millimeter emission \citep{Chen2007}, though this feature was not detected in later millimeter studies with better sensitivity and is likely emission from within the outflow cavity \citep{Tobin2015b}. IRAS 03282+3035 was later identified as a close multiple system with $\sim$22 AU separation between components resolved with A array VANDAM data \citep{Tobin2016a}.   While in the combined A+B array VANDAM data the system appears as a single core, the close binary introduces enough deviations from symmetry that we cannot model this candidate disk.  An outflow was detected in CO with a 125$^{\circ}$ position angle \citep{Lee2015}.  CARMA N$_{2}$H$^{+}$ observations \citep{Tobin2011} revealed a velocity gradient both along the outflow, and a component perpendicular to the outflow on the southeast side of the envelope, which could be consistent with interaction between the envelope and central protostar.
      
\subsubsection{Per-emb-18}    \label{vandam:gallery:weird:per18}
Per-emb-18 is a Class 0 protostar, which is a part of the larger NGC 1333 IRAS7 system detected first in the near infrared \citep{Aspin1994}.  Per-emb-21 and Per-emb-49 were also previously known protostars in NGC 1333 IRAS7.  Per-emb-18 and Per-emb-49 were found to have close companions ($<$100 AU separations), meaning NGC 1333 IRAS7 is a quintuple system \citep{Tobin2016a}.  An H$_{2}$ outflow was found in the system with a position angle of 159$^{\circ}$ \citep{Davis2008}.
 Per-emb-18 shows lopsided dust structures in the VANDAM data extended roughly perpendicular to the outflows, with the bright western structure harboring a double source with $\sim$20 AU separation in A array data \citep{Tobin2016a}.   
     
\subsubsection{L1448 IRS3B}    \label{vandam:gallery:weird:irs3b}
L1448 IRS3B is a Class 0 protostar first detected in NH$_{3}$ \citep{Bachiller1986} with two long-known companions, L1448 IRS3A and L1448NW \citep{Terebey1997,Looney2000}, with $\sim$2750 AU and $\sim$4950 AU separations from our survey respectively \citep{Tobin2016a}.    L1448 IRS3B is itself a close triple system and L1448NW is a double \citep{Tobin2016a}, making the protostars part of a sextuple system.  \citet{Tobin2015b} first reported two compact sources in L1448 IRS3B with CARMA data, before the third component was resolved later with the VLA.  L1448 IRS3B and L1448NW are Class 0 protostars while L1448 IRS3A is a Class I source--older and less embedded than its companions \citep{OLinger2006}.  CO observations reveal an outflow with a position angle of 122$^{\circ}$ \citep{Lee2016}, though the outflow was also detected on larger scales with CO previously \citep{Kwon2006}.  The blueshifted lobe from L1448 IRS3B overlaps and may interact with the outflow from L1448NW \citep{Hull2014}.  2.7 mm continuum emission from L1448 IRS3B revealed a protostellar envelope elongated almost perpendicular to the outflow \citep{Looney2000}. A velocity gradient along the extended envelope was found with C$^{18}$O observations \citep{Yen2015}.
  
ALMA Band 6 (1.3 mm) observations of the L1448 IRS3B triple source revealed a disk with spiral arms surrounding the three protostars \citep{Tobin2016b}.  The spiral arm structure and the placement of the protostars with in them demonstrate that protostellar disks can undergo gravitational instability at early times, and may form hierarchical multiples.  ALMA C$^{18}$O maps from \citet{Tobin2016b} also revealed a velocity gradient perpendicular to the outflows and centered around two of the three protostars, with the third---and brightest---protostar of the system lying closer to the edge of the disk.

Polarization towards L1448 IRS3B was detected on $\sim$2500 AU, $\sim$1000 AU, and $\sim$600 AU size scales with SCUPOL, BIMA, and CARMA respectively \citep{Matthews2009,Kwon2006,Hull2014}.  The inferred magnetic field orientation on all scales is nearly perpendicular to the outflow and parallel to the velocity gradient in C$^{18}$O \citep{Yen2015,Tobin2016b}.

\clearpage

\section{BEST-FIT CANDIDATE DISK MODELING RESULTS IN IMAGE AND {\it U,V}-SPACES}    \label{vandam:bestfits}

 \begin{figure}[h]
        \centering
                \includegraphics[width=0.8\textwidth]{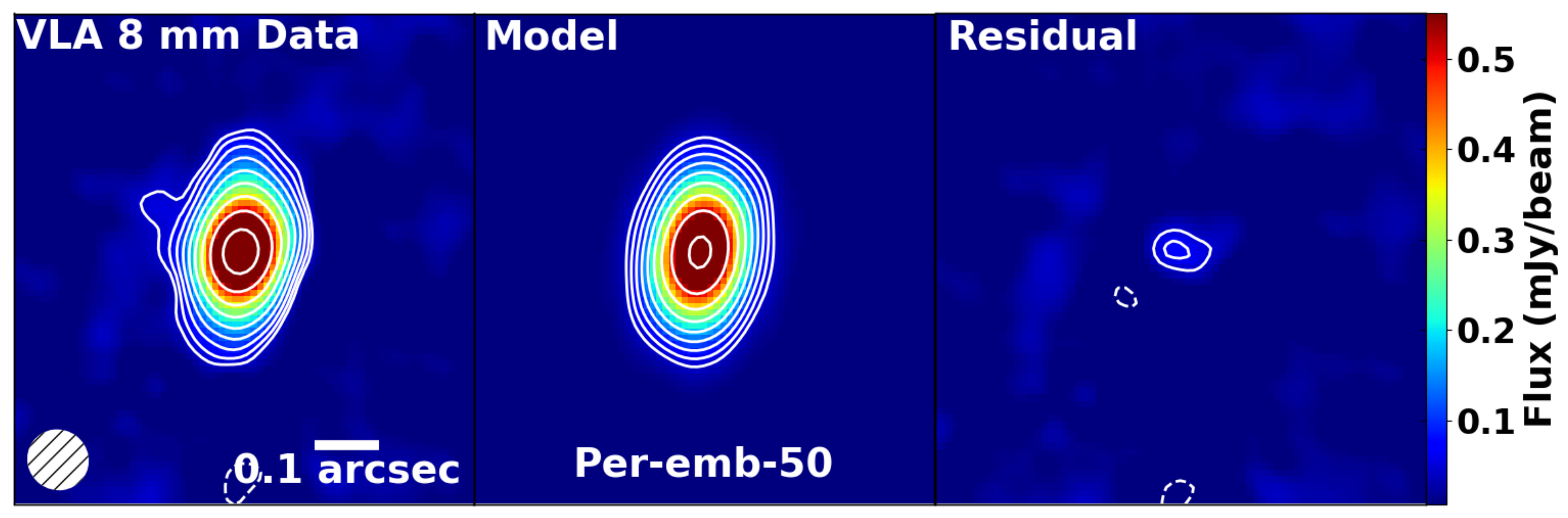}
        \caption{VLA A+B array data (left), $q=0.25$ model from {\it u,v}-plane best-fit (center), and
residual (right) of Per-emb-50. Images were produced with robust = 0.25 weighting. Contours start at 3$\sigma$ ($\sigma$ $\sim$15$\mu$Jy) with a factor of $\sqrt{2}$ spacing. The synthesized beam is in the lower left.}
        \label{Per50dmr}
\end{figure} 

 \begin{figure}[h]
        \centering
                \includegraphics[width=0.7\textwidth]{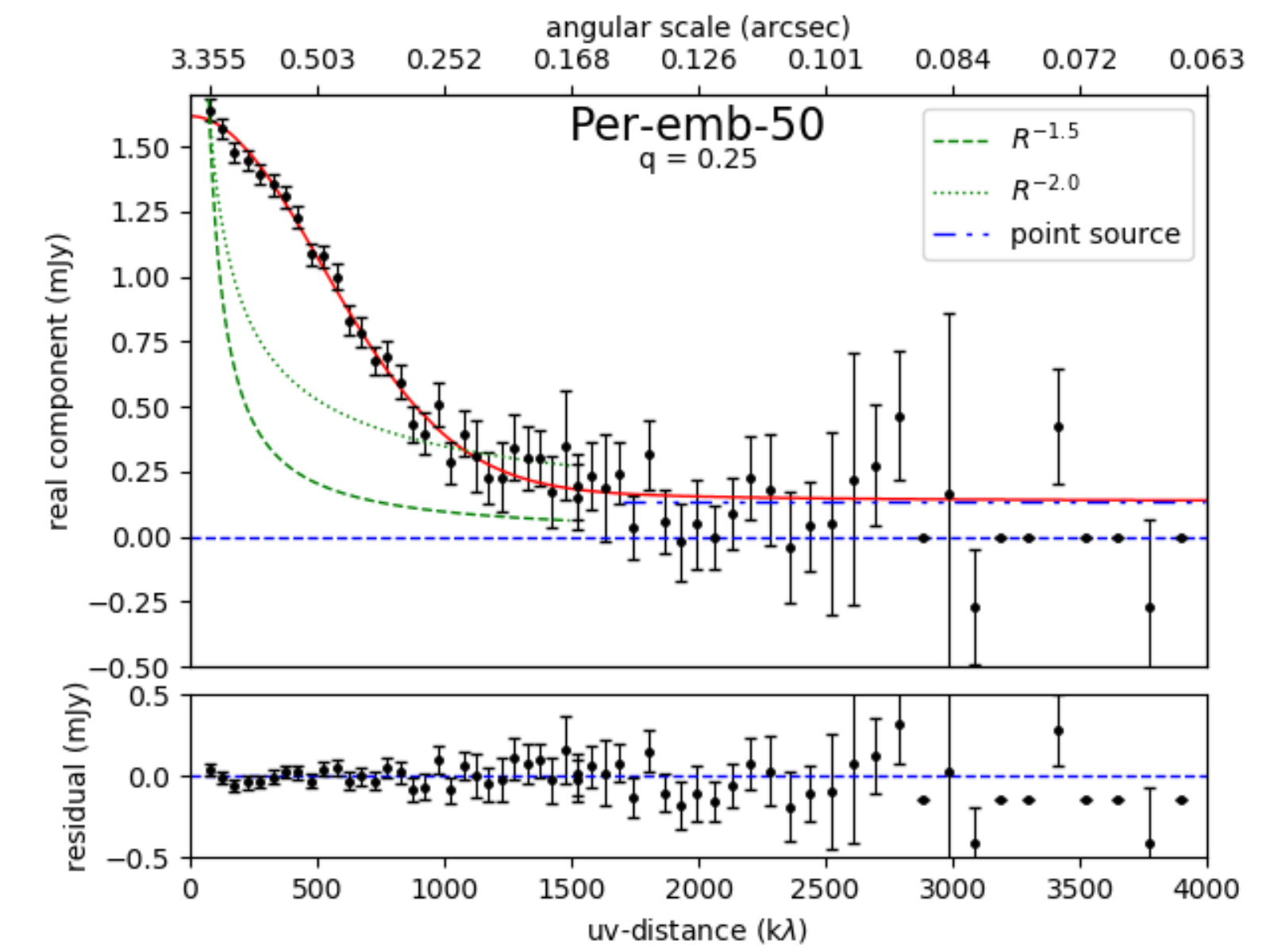}
                \caption{Real vs {\it u,v}-distance plot of 8 mm data for Per-emb-50. Top:  real component of data. The blue dashed line indicates real component of zero.  The red solid line is the best-fit
        model.   The green dashed and dotted lines correspond to R$^{-2.0}$ and R$^{-1.5}$ envelope visibility profiles.   Bottom: residual of real component minus model. }
                \label{Per50uvdist}
\end{figure} 

\clearpage

 \begin{figure}[t]
        \centering
                \includegraphics[width=0.8\textwidth]{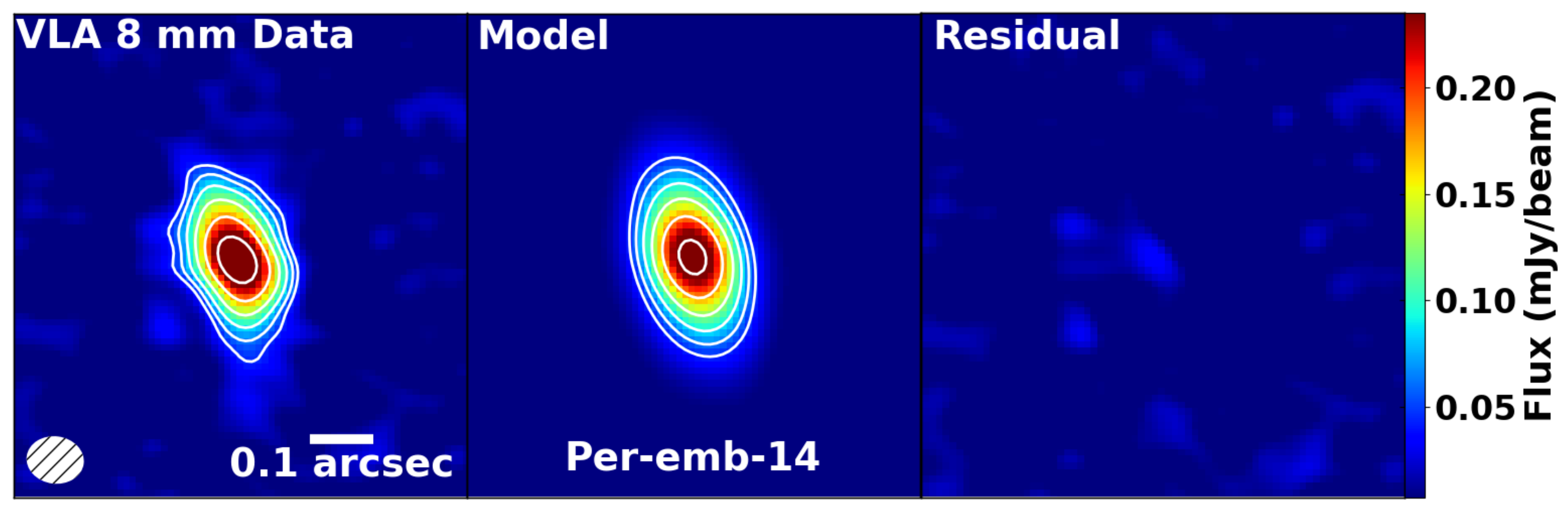}
        \caption{Same as Figure \ref{Per50dmr}, for Per-emb-14, with $q=0.25$.}
        \label{Per14dmr}
\end{figure} 

 \begin{figure}[b]
        \centering
                \includegraphics[width=0.7\textwidth]{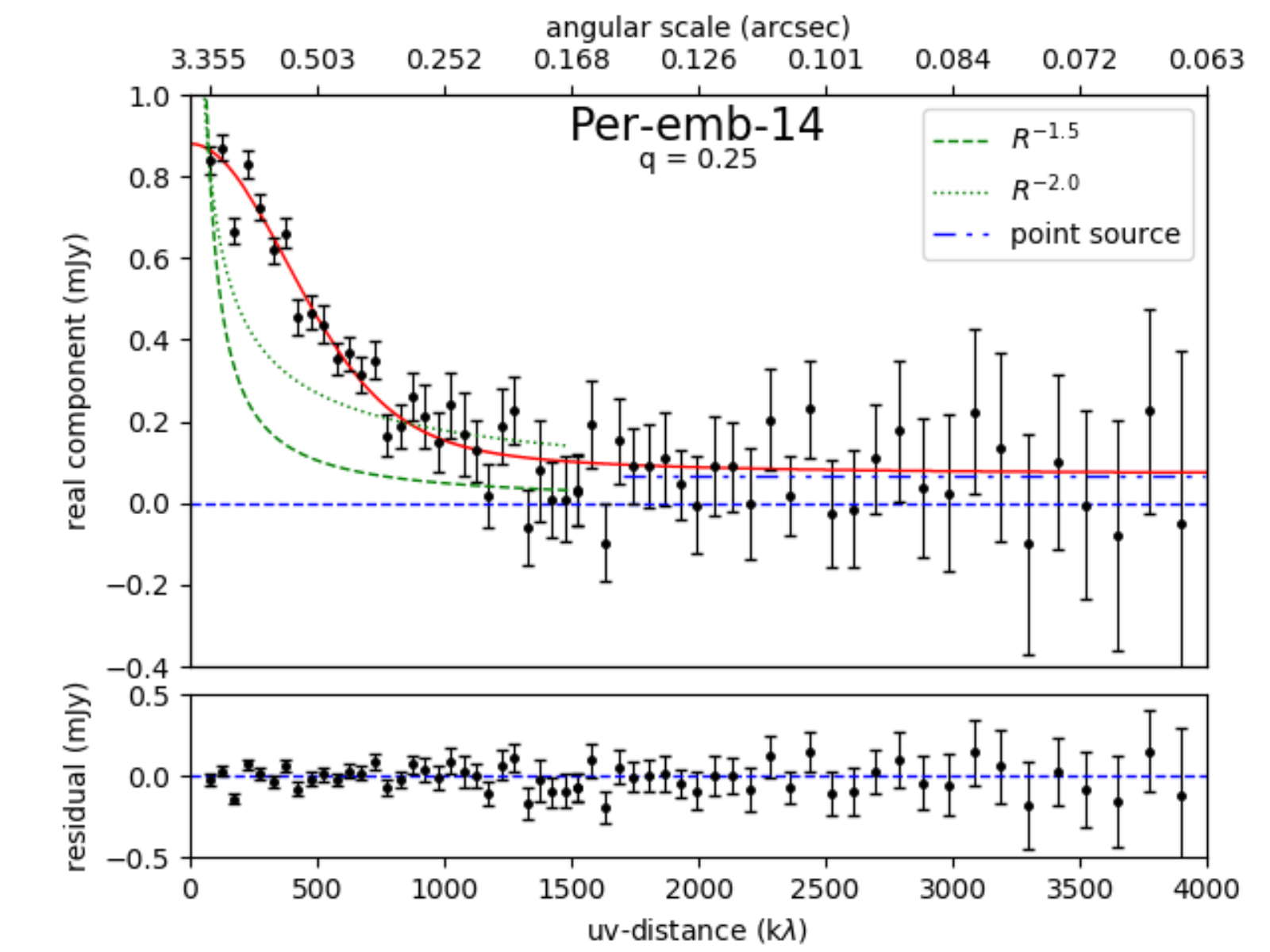}
                 \caption{Same as Figure \ref{SVS13Buvdist}, for Per-emb-14.}
        \label{Per14uvdist}
\end{figure} 
\clearpage

 \begin{figure}[t]
        \centering
                \includegraphics[width=0.8\textwidth]{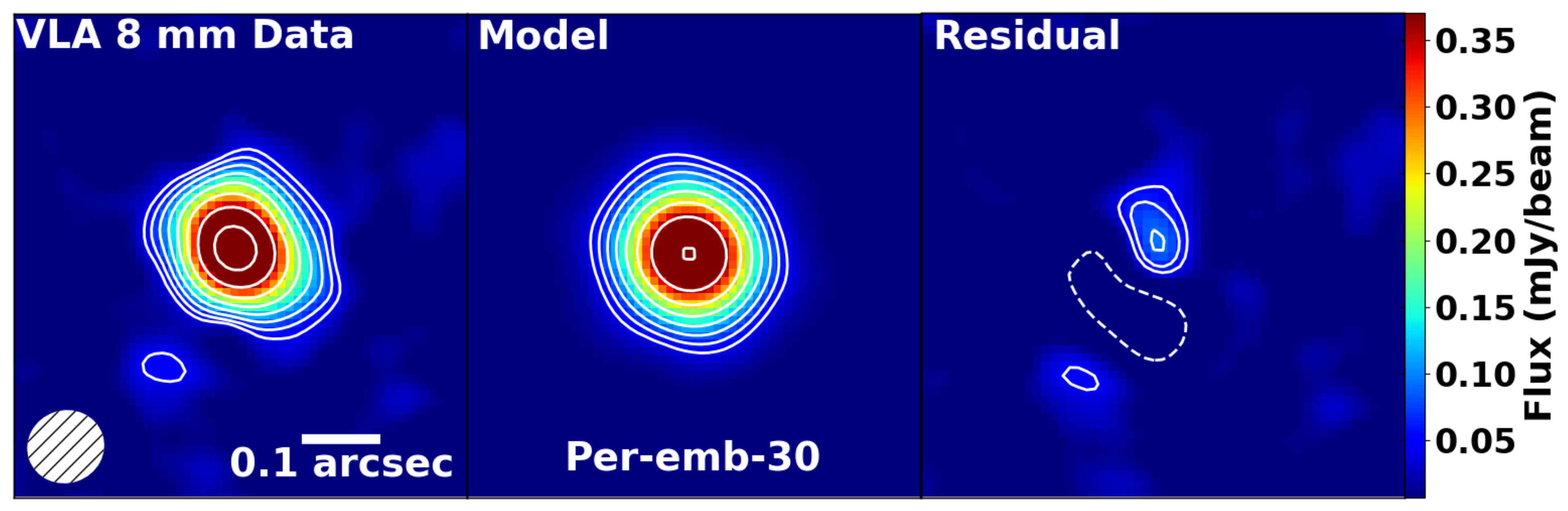}
        \caption{Same as Figure \ref{Per50dmr}, for Per-emb-30, with $q=0.25$.}
        \label{Per30dmr}
\end{figure} 

 \begin{figure}[b]
        \centering
                \includegraphics[width=0.7\textwidth]{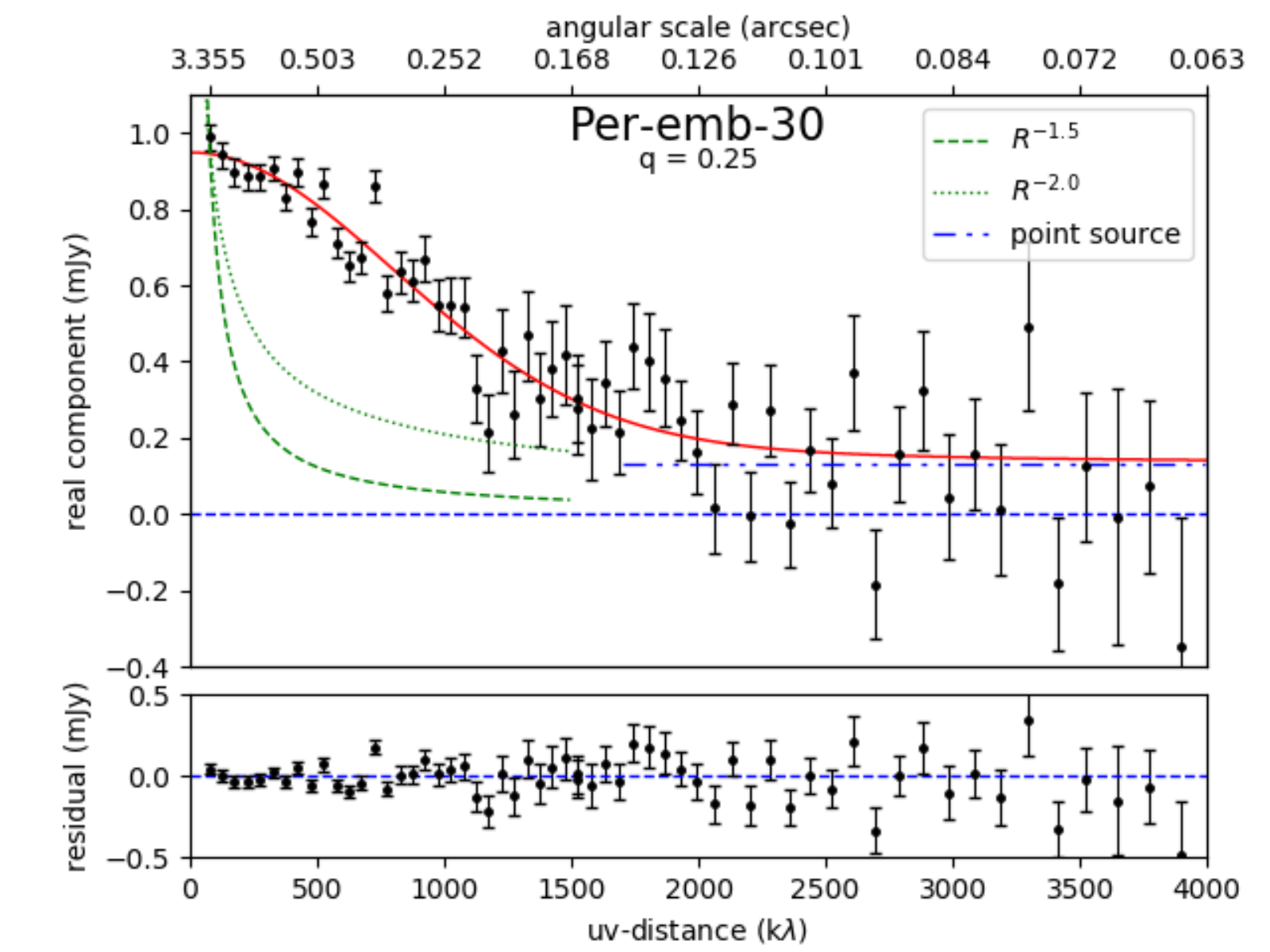}
                \caption{Same as Figure \ref{SVS13Buvdist}, for Per-emb-30.}
                \label{Per30uvdist}
\end{figure} 
\clearpage

 \begin{figure}[t]
        \centering
                \includegraphics[width=0.8\textwidth]{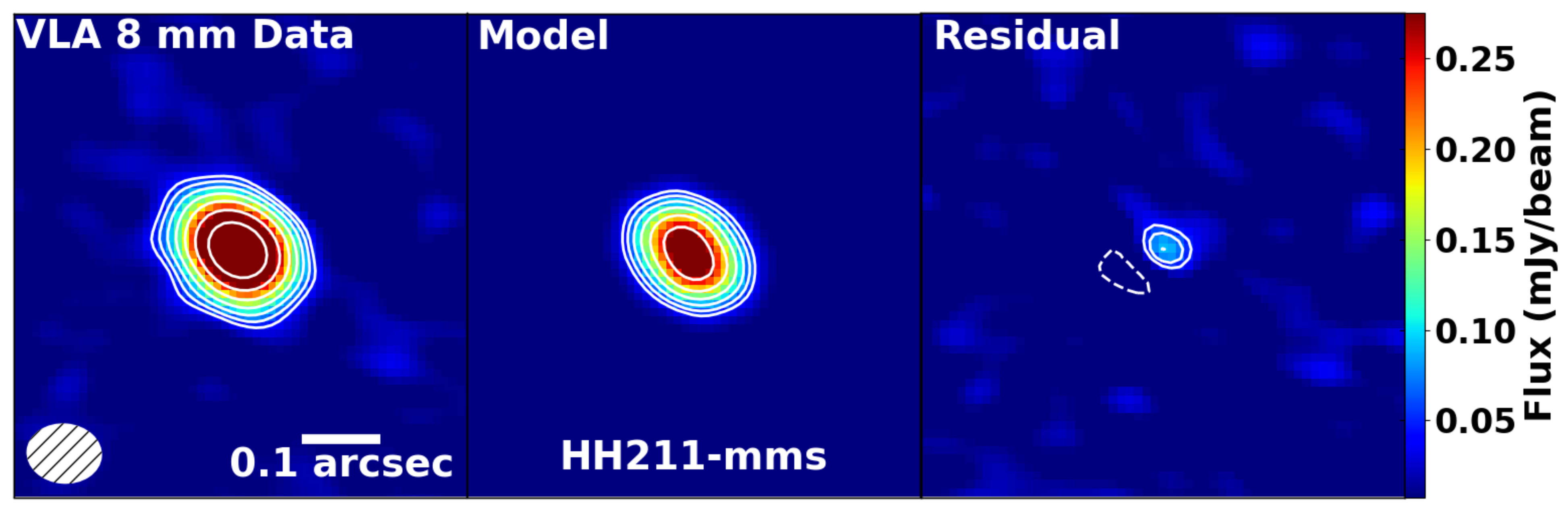}
        \caption{Same as Figure \ref{Per50dmr}, for HH211-mms, with $q=0.25$.}
        \label{HH211dmr}
\end{figure} 

 \begin{figure}[b]
        \centering
                \includegraphics[width=0.7\textwidth]{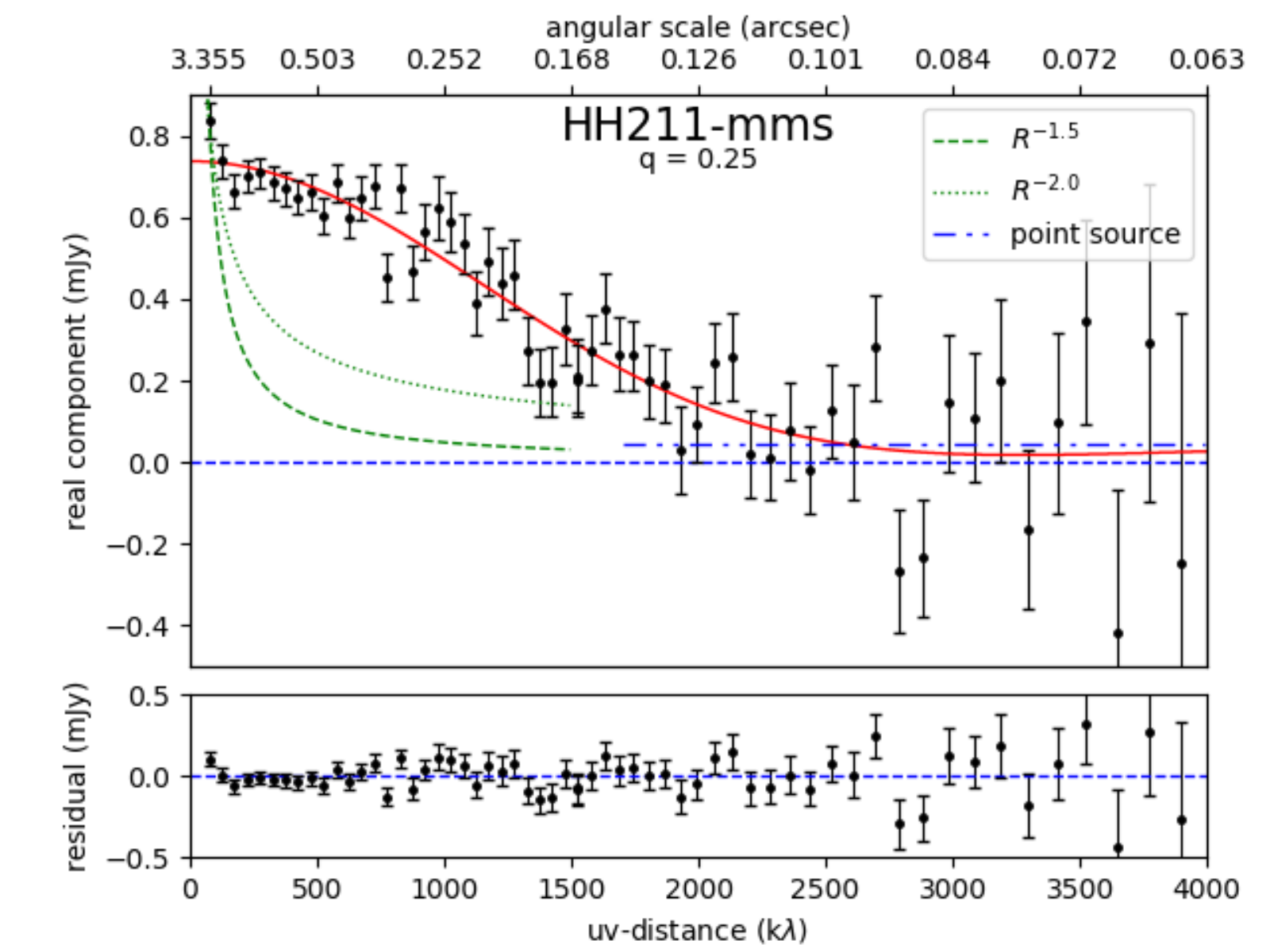}
                \caption{Same as Figure \ref{SVS13Buvdist}, for HH211-mms.}
                \label{HH211uvdist}
\end{figure} 
\clearpage

 \begin{figure}[t]
        \centering
                \includegraphics[width=0.8\textwidth]{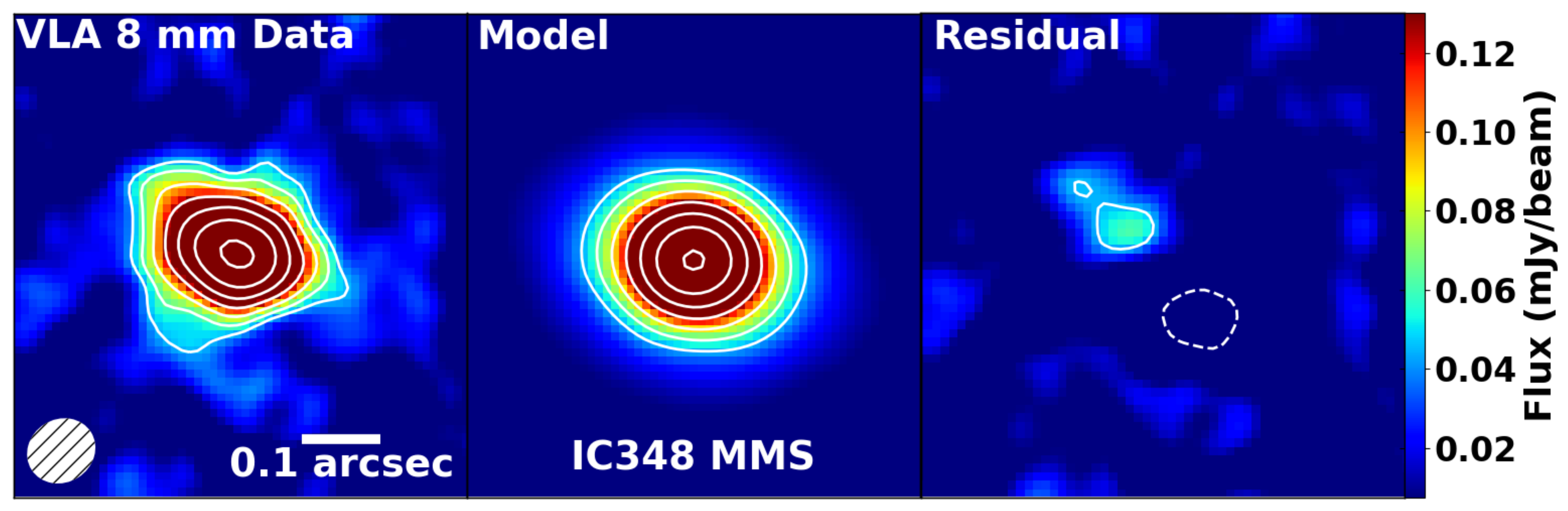}
        \caption{Same as Figure \ref{Per50dmr}, for IC348 MMS, with $q=0.25$.}
        \label{IC348dmr}
\end{figure} 

 \begin{figure}[b]
        \centering
                \includegraphics[width=0.7\textwidth]{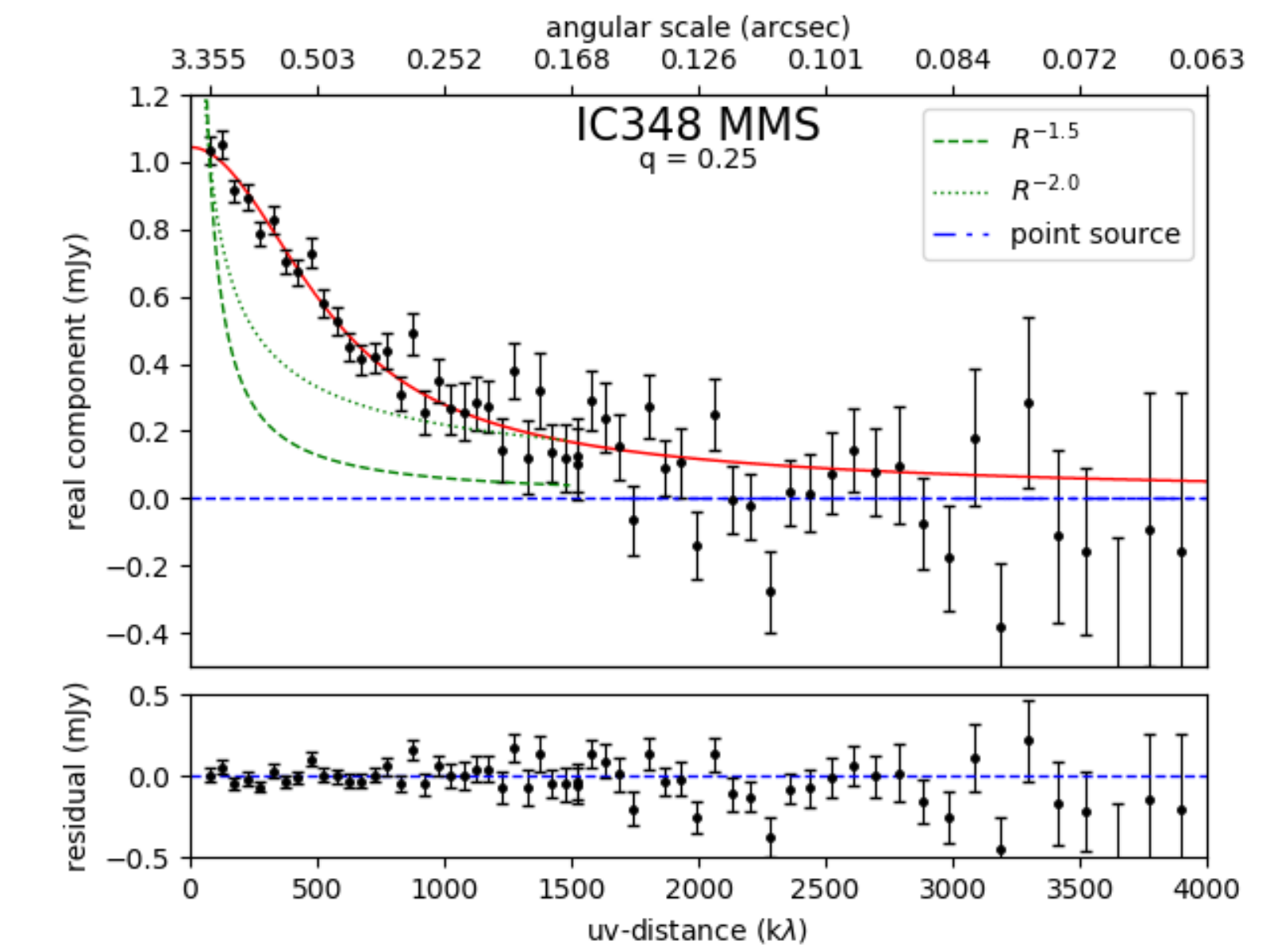}
                \caption{Same as Figure \ref{SVS13Buvdist}, for IC348 MMS.}
                \label{IC348uvdist}
\end{figure} 
\clearpage

 \begin{figure}[t]
        \centering
                \includegraphics[width=0.8\textwidth]{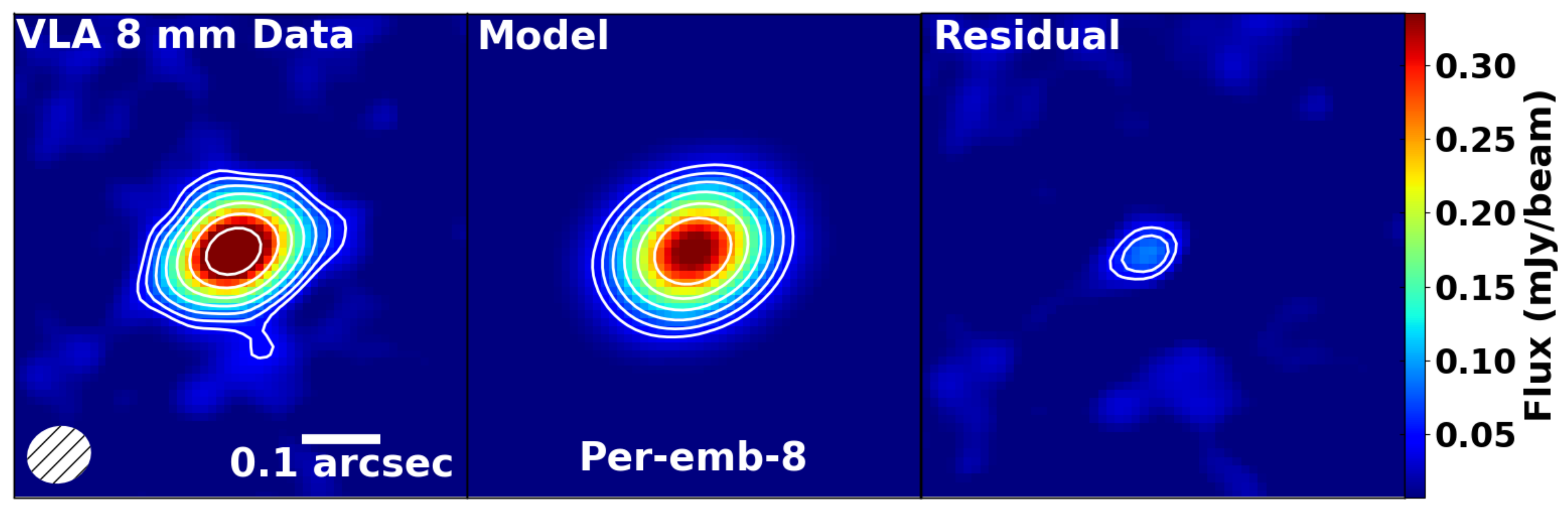}
        \caption{Same as Figure \ref{Per50dmr}, for Per-emb-8, with $q=0.25$.}
        \label{Per8dmr}
\end{figure} 

 \begin{figure}[b]
        \centering
                \includegraphics[width=0.7\textwidth]{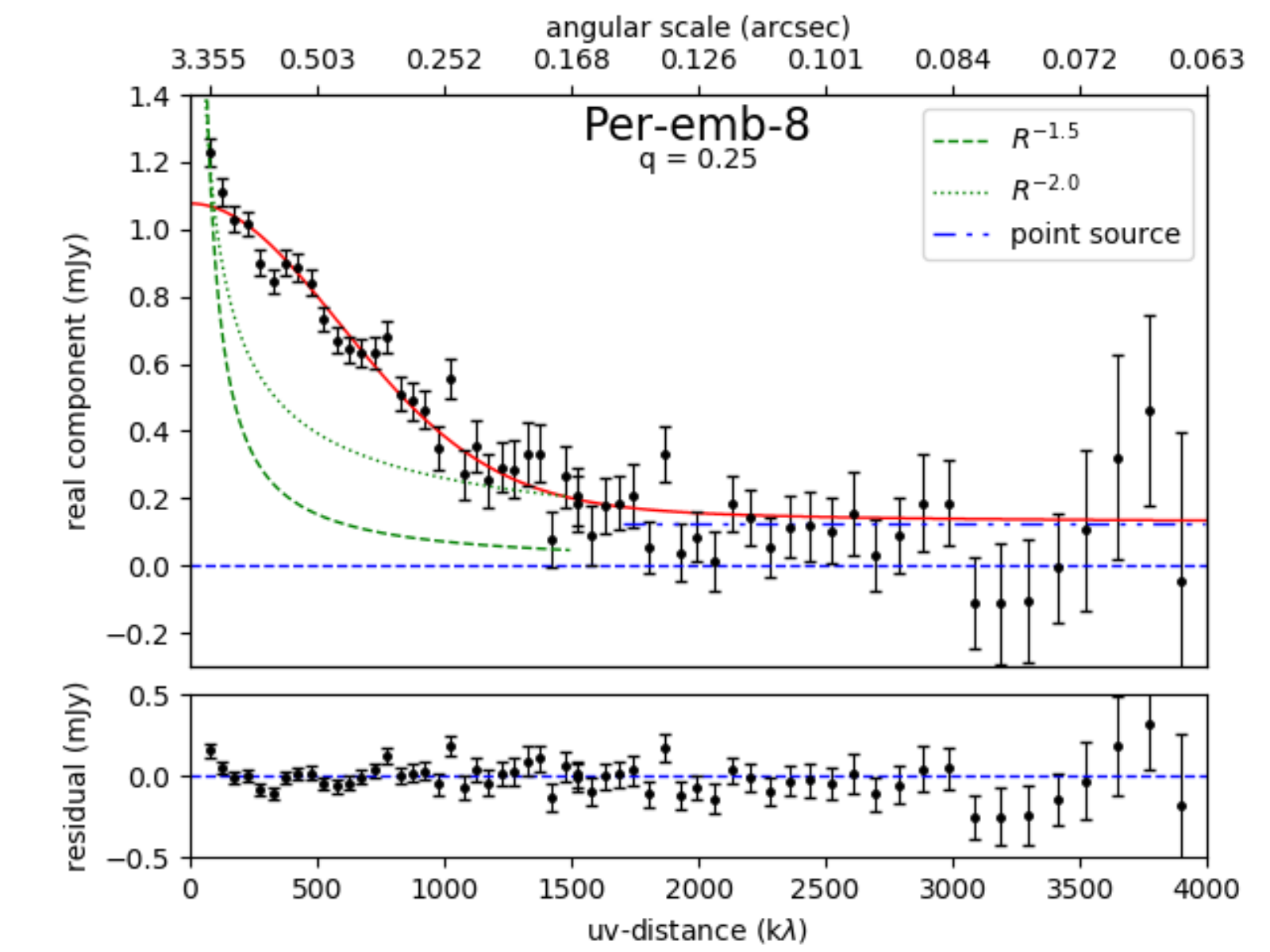}
                \caption{Same as Figure \ref{SVS13Buvdist}, for Per-emb-8.}
                \label{Per8uvdist}
\end{figure} 
\clearpage

 \begin{figure}[t]
        \centering
                \includegraphics[width=0.8\textwidth]{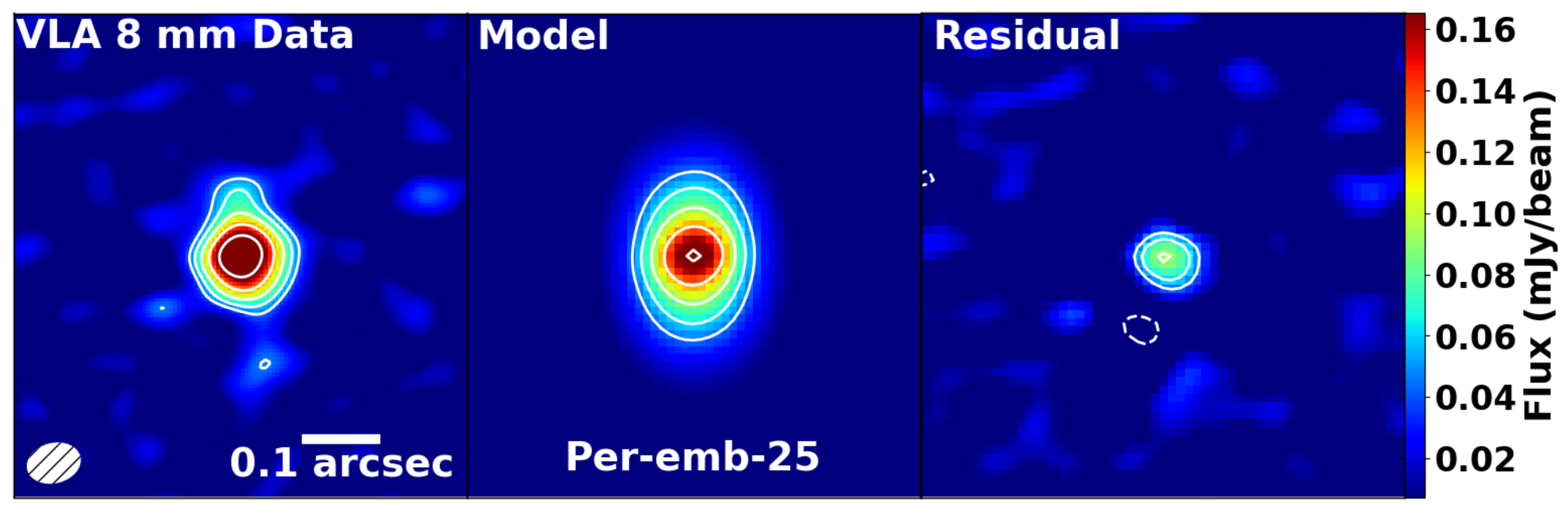}
        \caption{Same as Figure \ref{Per50dmr}, for Per-emb-25, with $q=0.25$.}
        \label{Per25dmr}
\end{figure} 

 \begin{figure}[b]
        \centering
                \includegraphics[width=0.7\textwidth]{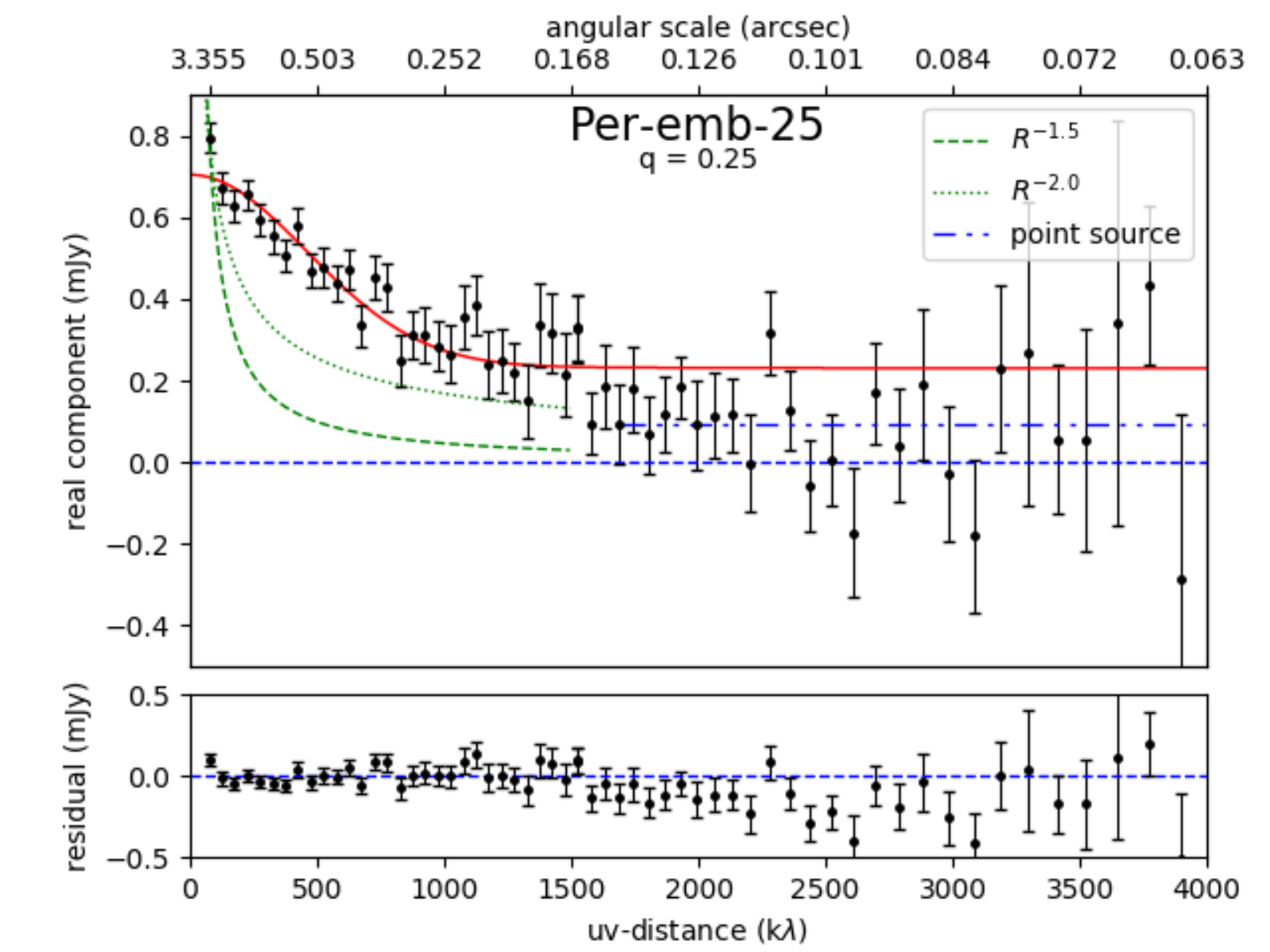}
                \caption{Same as Figure \ref{SVS13Buvdist}, for Per-emb-25.}
                \label{Per25uvdist}
\end{figure} 
\clearpage

\begin{figure}[t]
        \centering
                \includegraphics[width=0.8\textwidth]{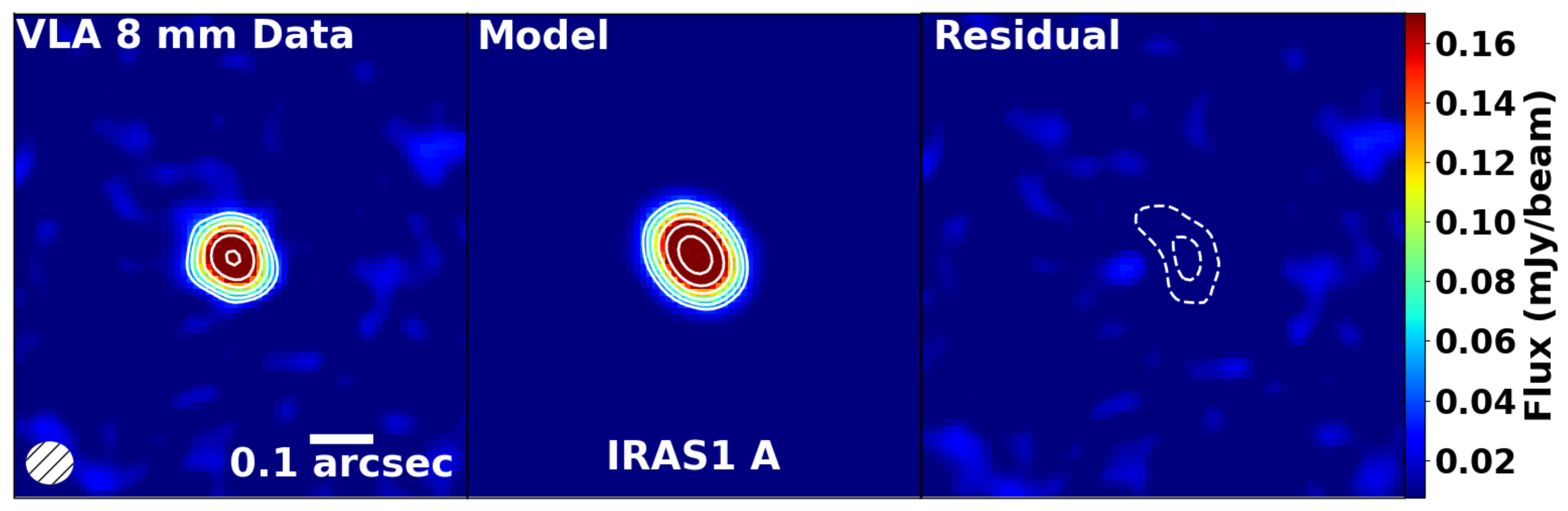}
        \caption{Same as Figure \ref{Per50dmr}, for NGC 1333 IRAS1 A, with $q=0.25$.}
        \label{Per35Admr}
\end{figure} 

 \begin{figure}[b]
        \centering
                \includegraphics[width=0.7\textwidth]{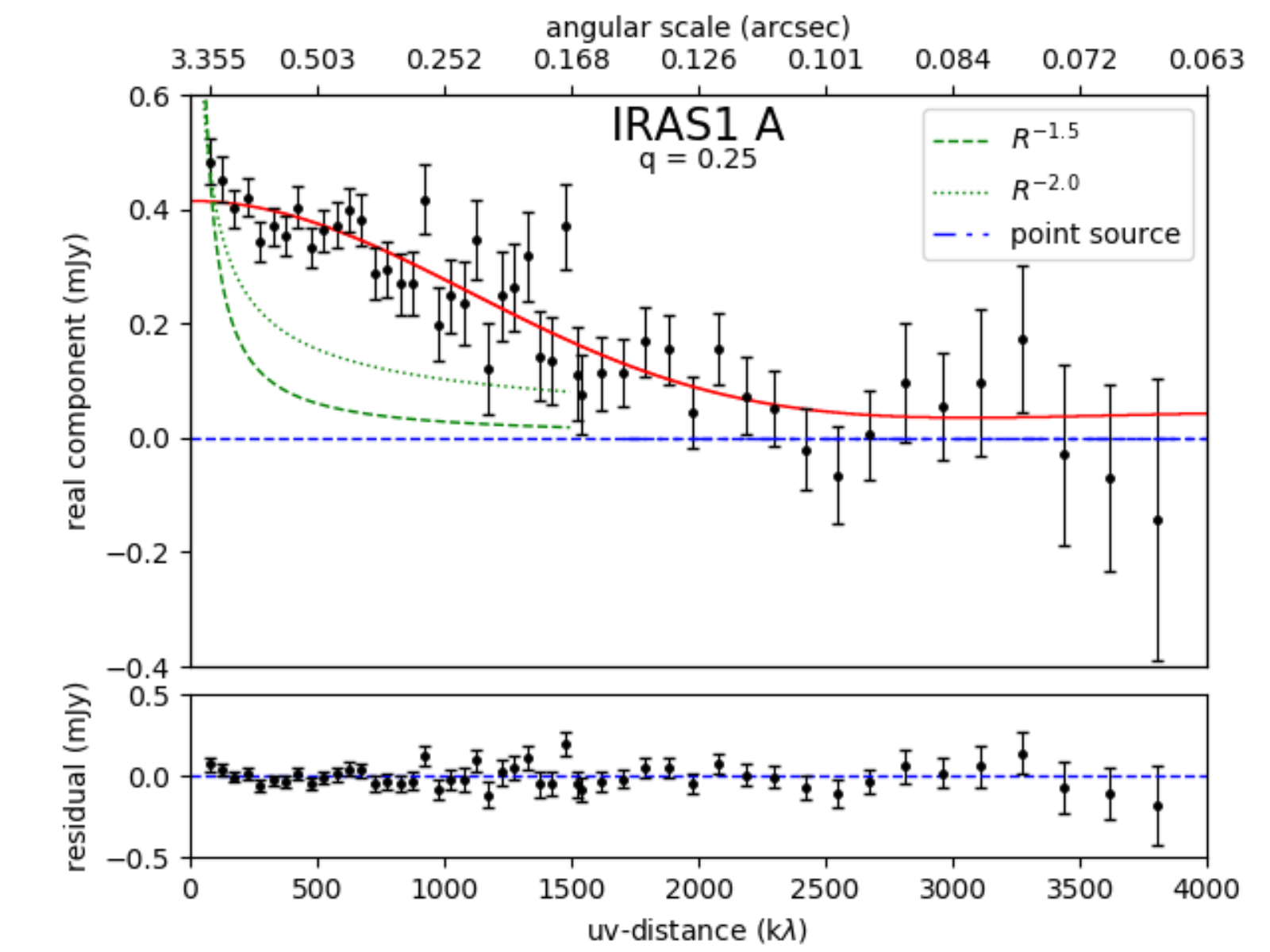}
                \caption{Same as Figure \ref{SVS13Buvdist}, for NGC 1333 IRAS1 A.}
                \label{Per35Auvdist}
\end{figure} 
\clearpage

 \begin{figure}[t]
        \centering
                \includegraphics[width=0.8\textwidth]{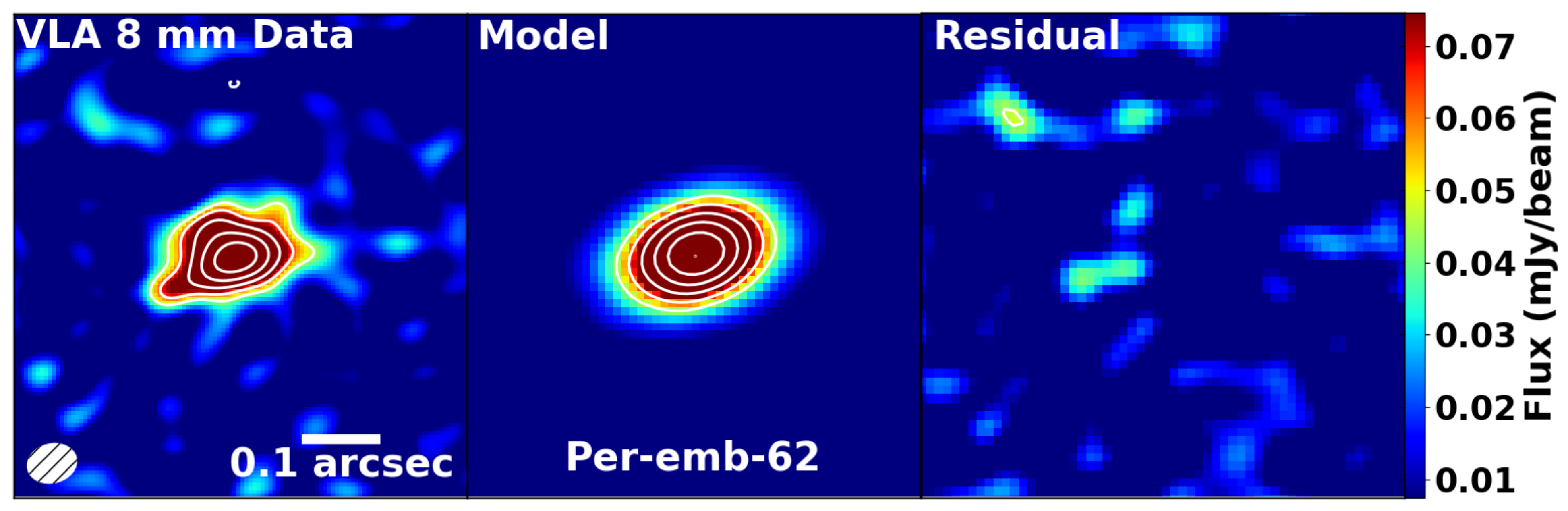}
        \caption{Same as Figure \ref{Per50dmr}, for Per-emb-62, with $q=1.0$.}
        \label{Per62dmr}
\end{figure} 

 \begin{figure}[b]
        \centering
                \includegraphics[width=0.7\textwidth]{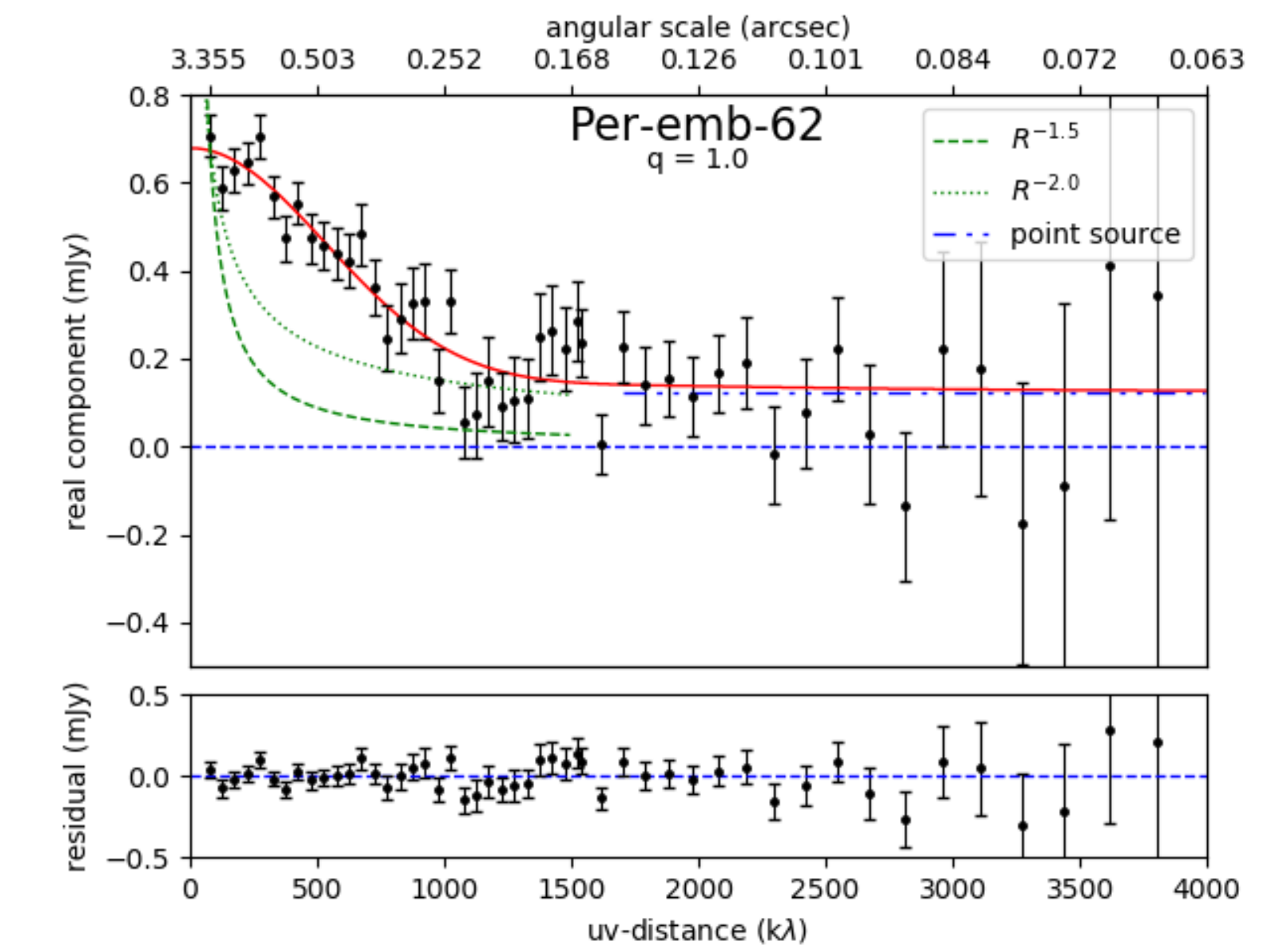}
                \caption{Same as Figure \ref{SVS13Buvdist}, for Per-emb-62.}
                \label{Per62uvdist}
\end{figure} 
\clearpage

 \begin{figure}[t]
        \centering
                \includegraphics[width=0.8\textwidth]{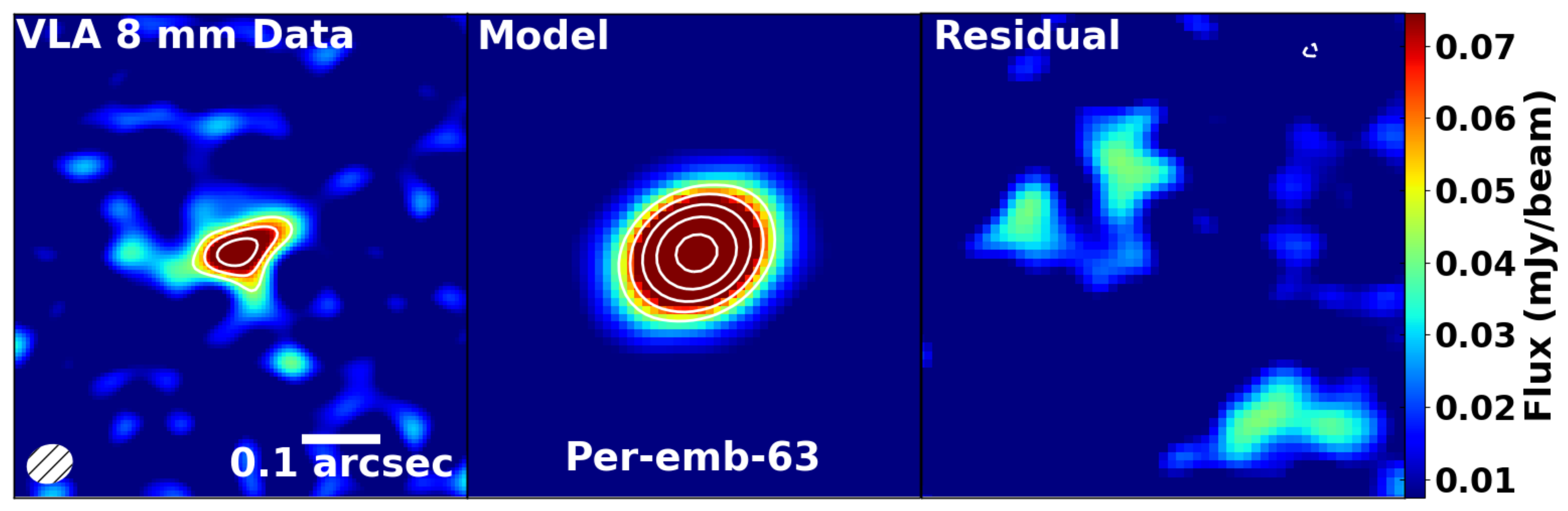}
        \caption{Same as Figure \ref{Per50dmr}, for Per-emb-63, with $q=0.25$.}
        \label{Per63dmr}
\end{figure} 

 \begin{figure}[b]
        \centering
                \includegraphics[width=0.7\textwidth]{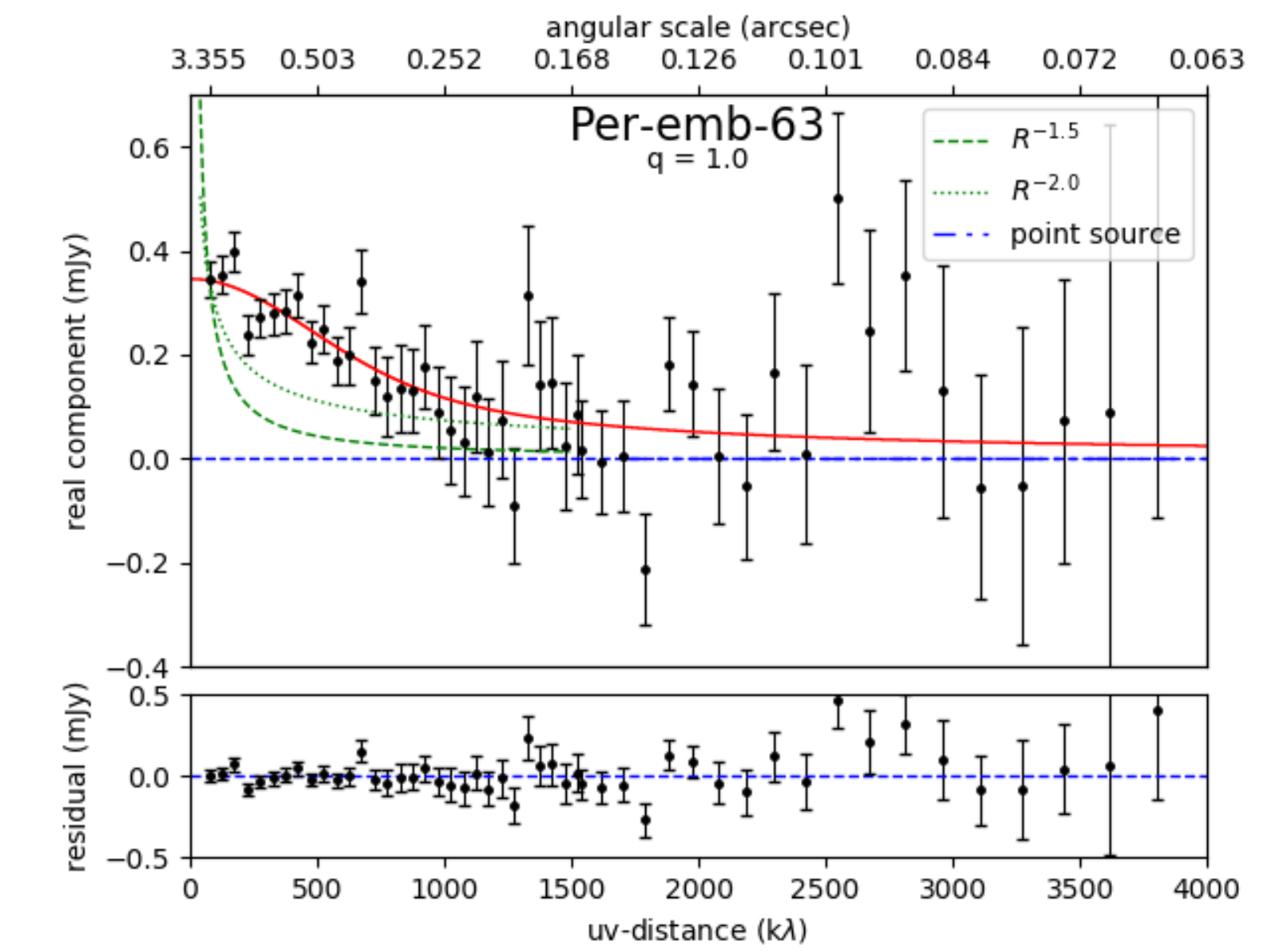}
                \caption{Same as Figure \ref{SVS13Buvdist}, for Per-emb-63.}
                \label{Per63uvdist}
\end{figure} 
\clearpage

 \begin{figure}[t]
        \centering
                \includegraphics[width=0.8\textwidth]{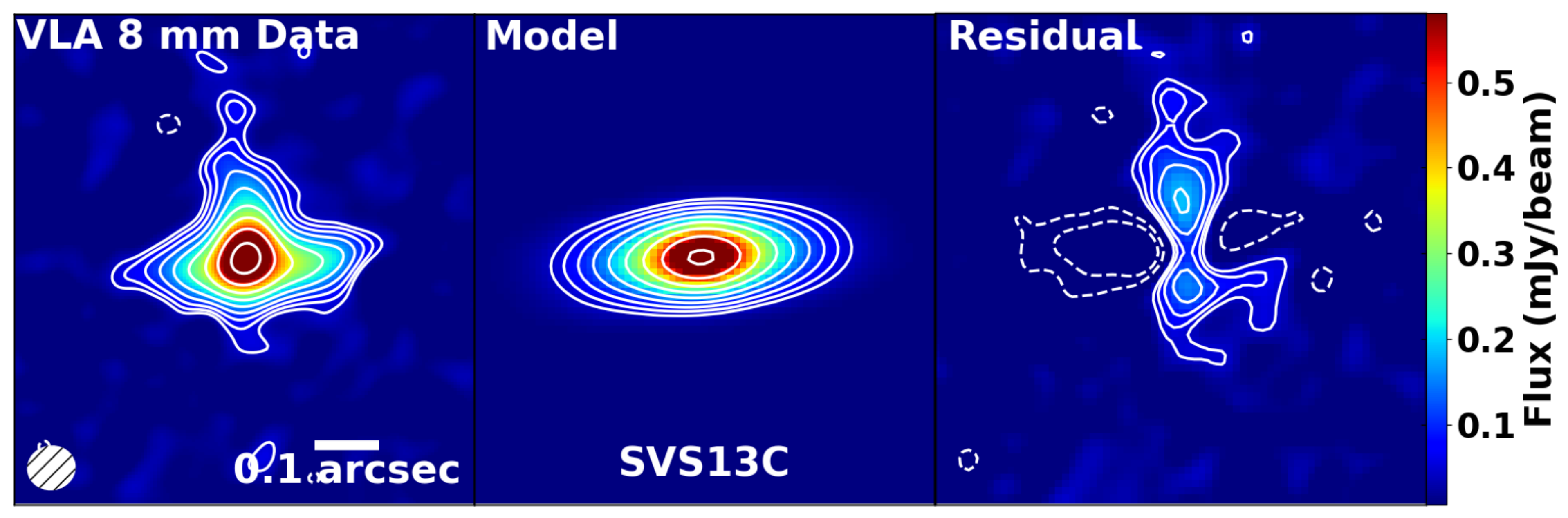}
        \caption{Same as Figure \ref{Per50dmr}, for SVS13C, with $q=0.25$.}
        \label{SVS13Cdmr}
\end{figure} 

 \begin{figure}[b]
        \centering
                \includegraphics[width=0.7\textwidth]{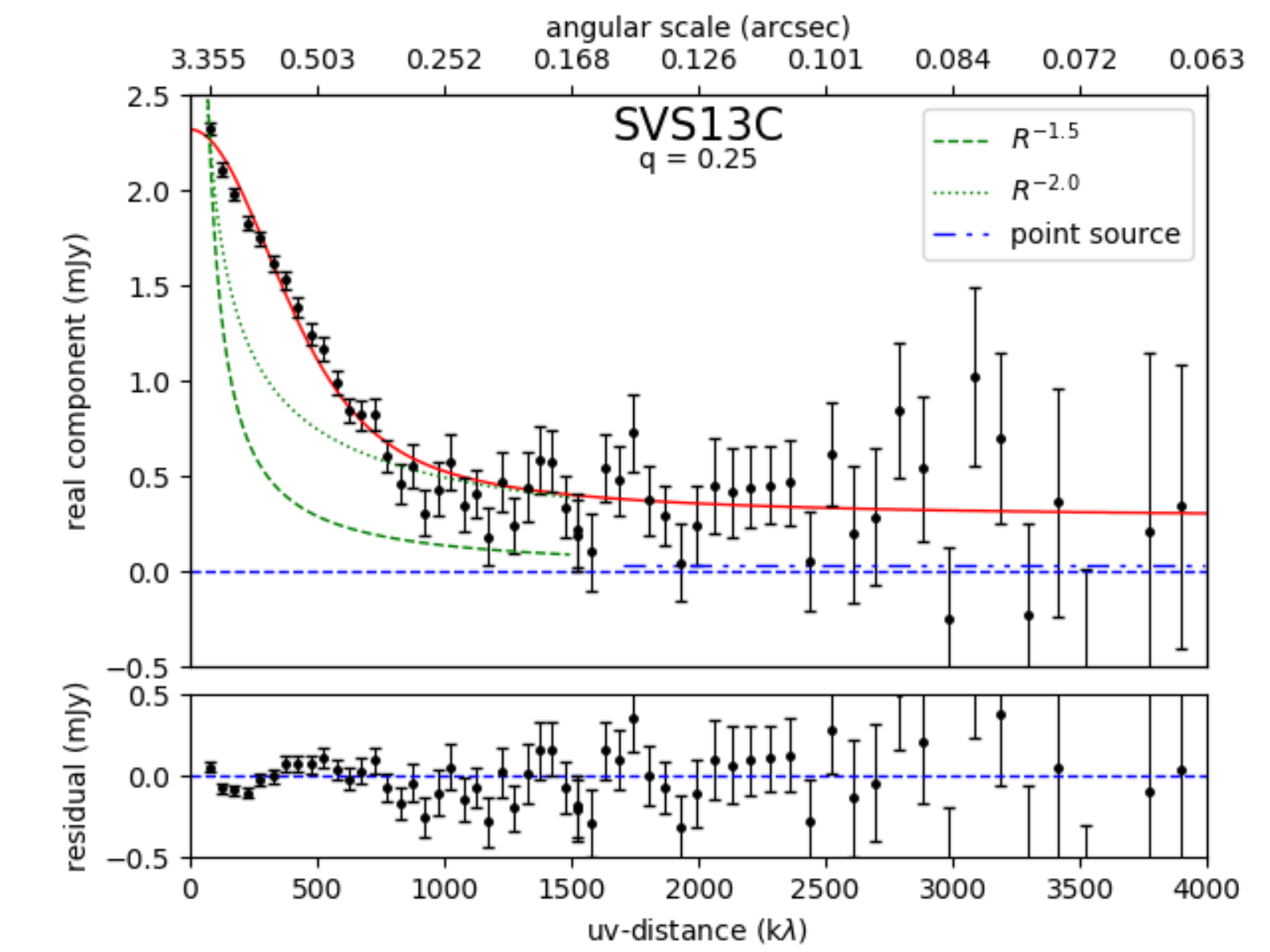}
                \caption{Same as Figure \ref{SVS13Buvdist}, for SVS13C.}
                \label{SVS13Cuvdist}
\end{figure} 
\clearpage

 \begin{figure}[t]
        \centering
                \includegraphics[width=0.8\textwidth]{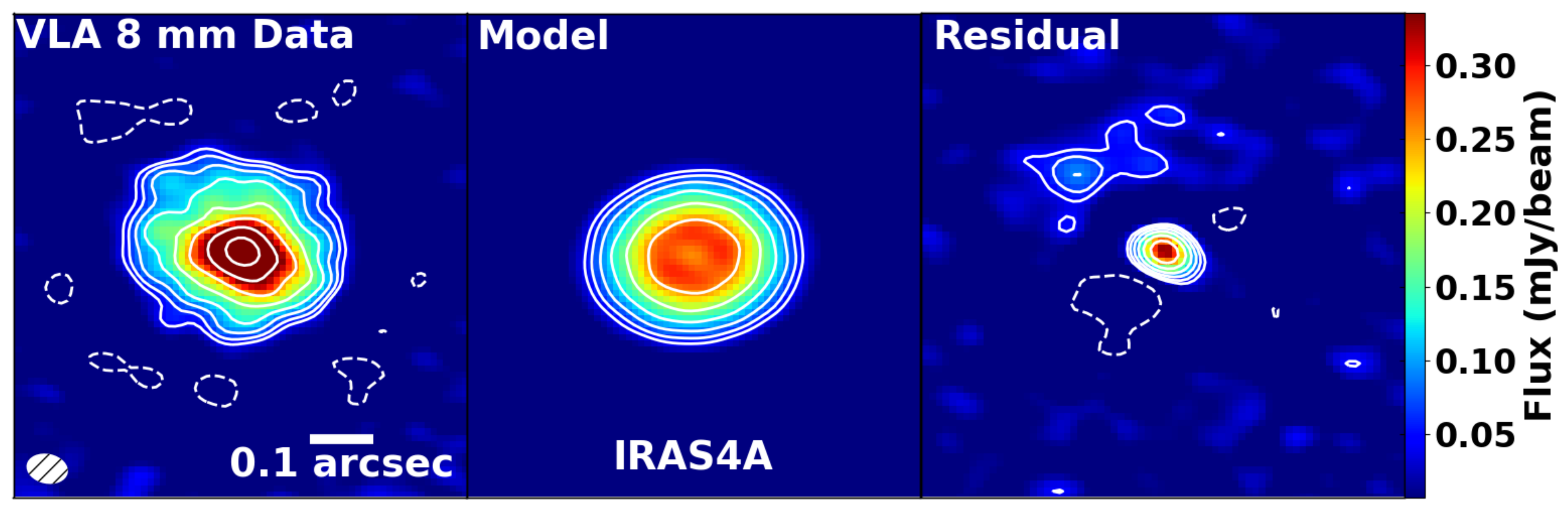}
        \caption{Same as Figure \ref{Per50dmr}, for NGC 1333 IRAS4A, with $q=0.25$ and baselines $<$350 k$\lambda$ removed.}
        \label{IRAS4Admr}
\end{figure} 
 \begin{figure}[b]
        \centering
                \includegraphics[width=0.7\textwidth]{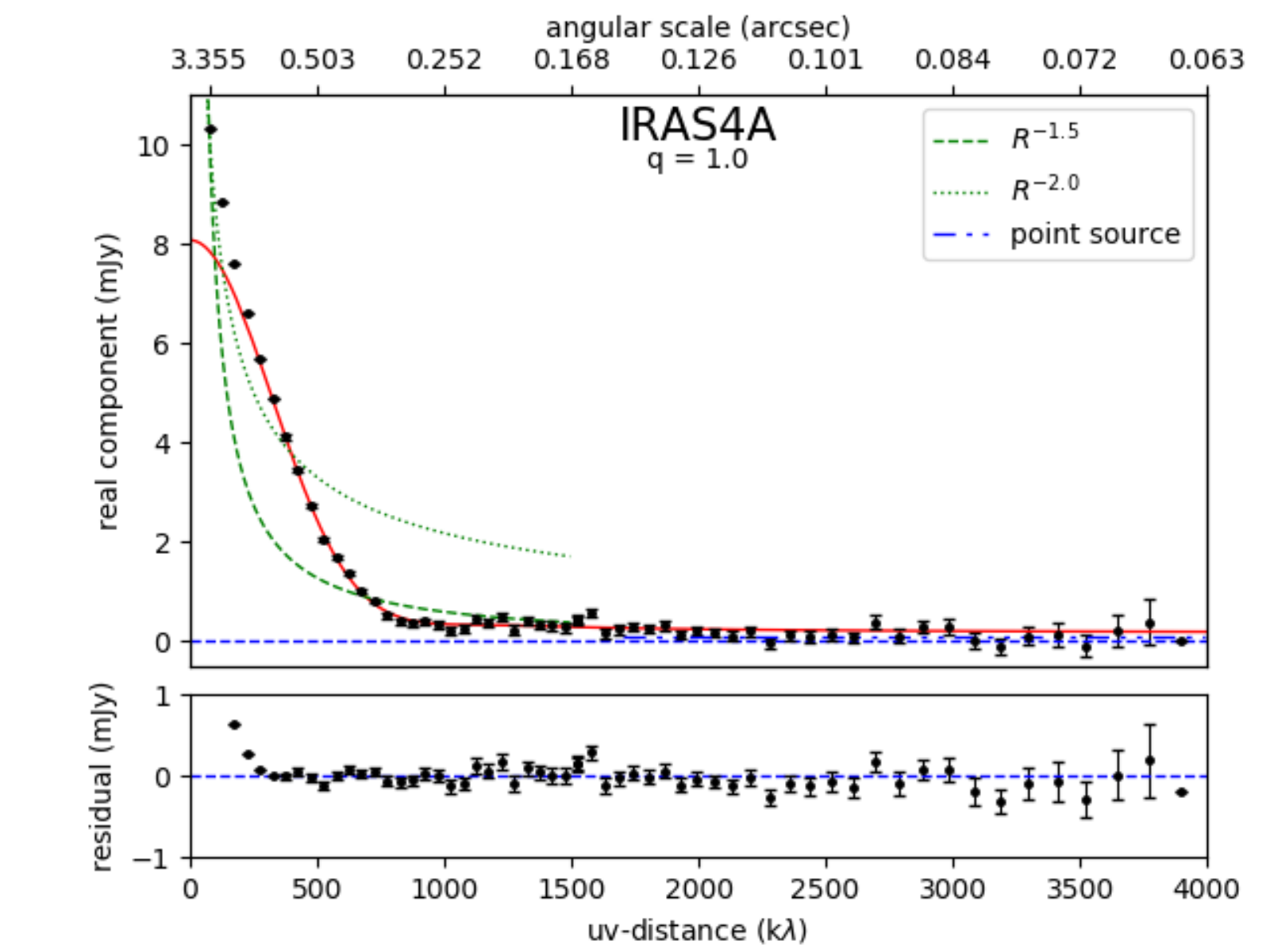}
                \caption{Same as Figure \ref{SVS13Buvdist}, for NGC 1333 IRAS4A.  The baselines $<$350 k$\lambda$ are not fit to the disk component.}
                \label{IRAS4Auvdist}
\end{figure}

\clearpage
 \begin{figure}[t]
        \centering
                \includegraphics[width=0.8\textwidth]{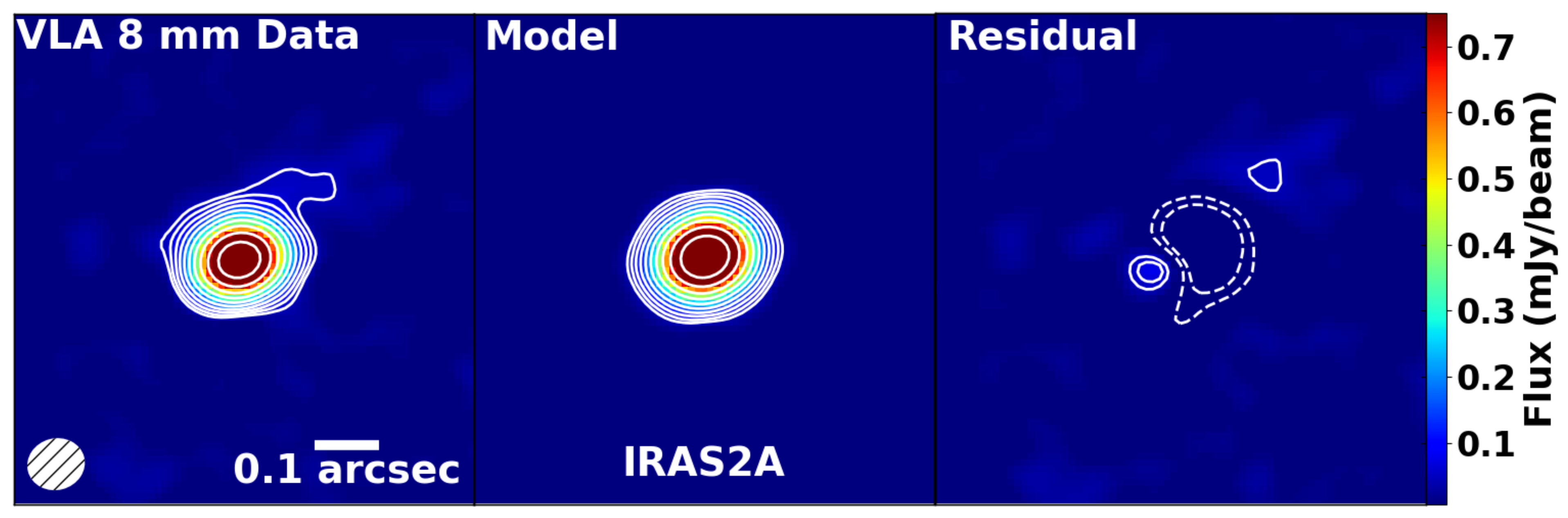}
        \caption{Same as Figure \ref{Per50dmr}, for NGC 1333 IRAS2A, with $q=0.25$.}
        \label{IRAS2Admr}
\end{figure} 
 \begin{figure}[b]
        \centering
                \includegraphics[width=0.7\textwidth]{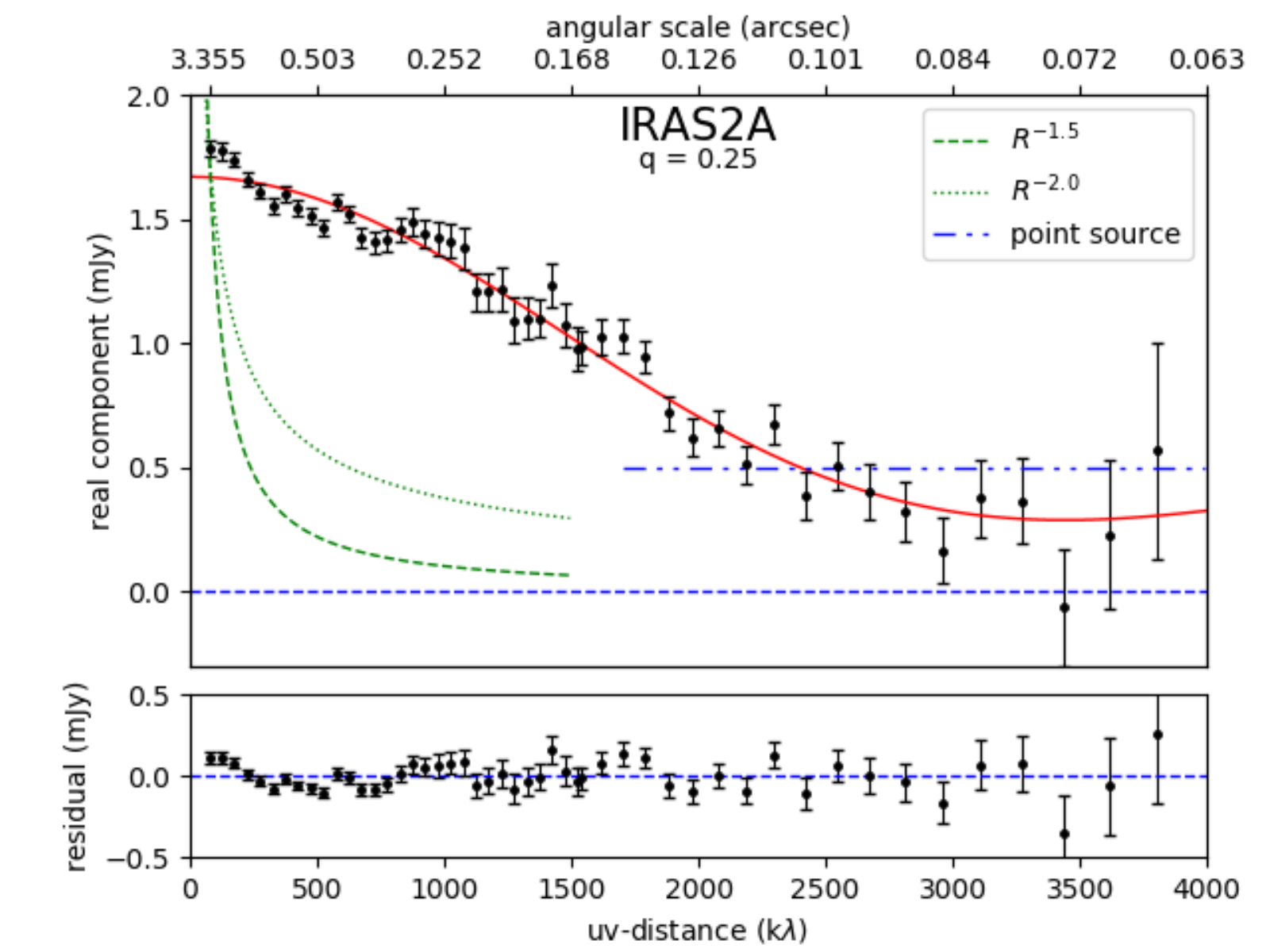}
                \caption{Same as Figure \ref{SVS13Buvdist}, for NGC 1333 IRAS2A.}
                \label{IRAS2Auvdist}
\end{figure} 

\clearpage

 \begin{figure}[t]
        \centering
                \includegraphics[width=0.8\textwidth]{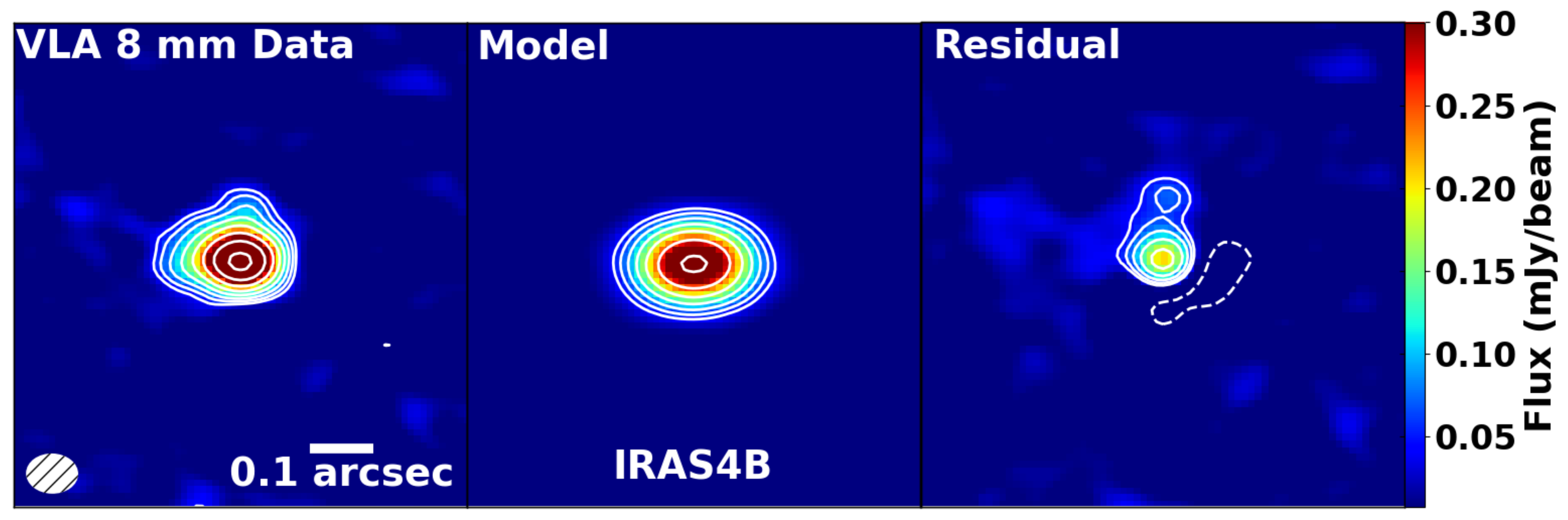}
        \caption{Same as Figure \ref{Per50dmr}, for NGC 1333 IRAS4B, with $q=0.25$ and baselines $<$350 k$\lambda$ removed. We do not consider this source a candidate disk.}
        \label{Per13dmr}
\end{figure} 

 \begin{figure}[b]
        \centering
                \includegraphics[width=0.7\textwidth]{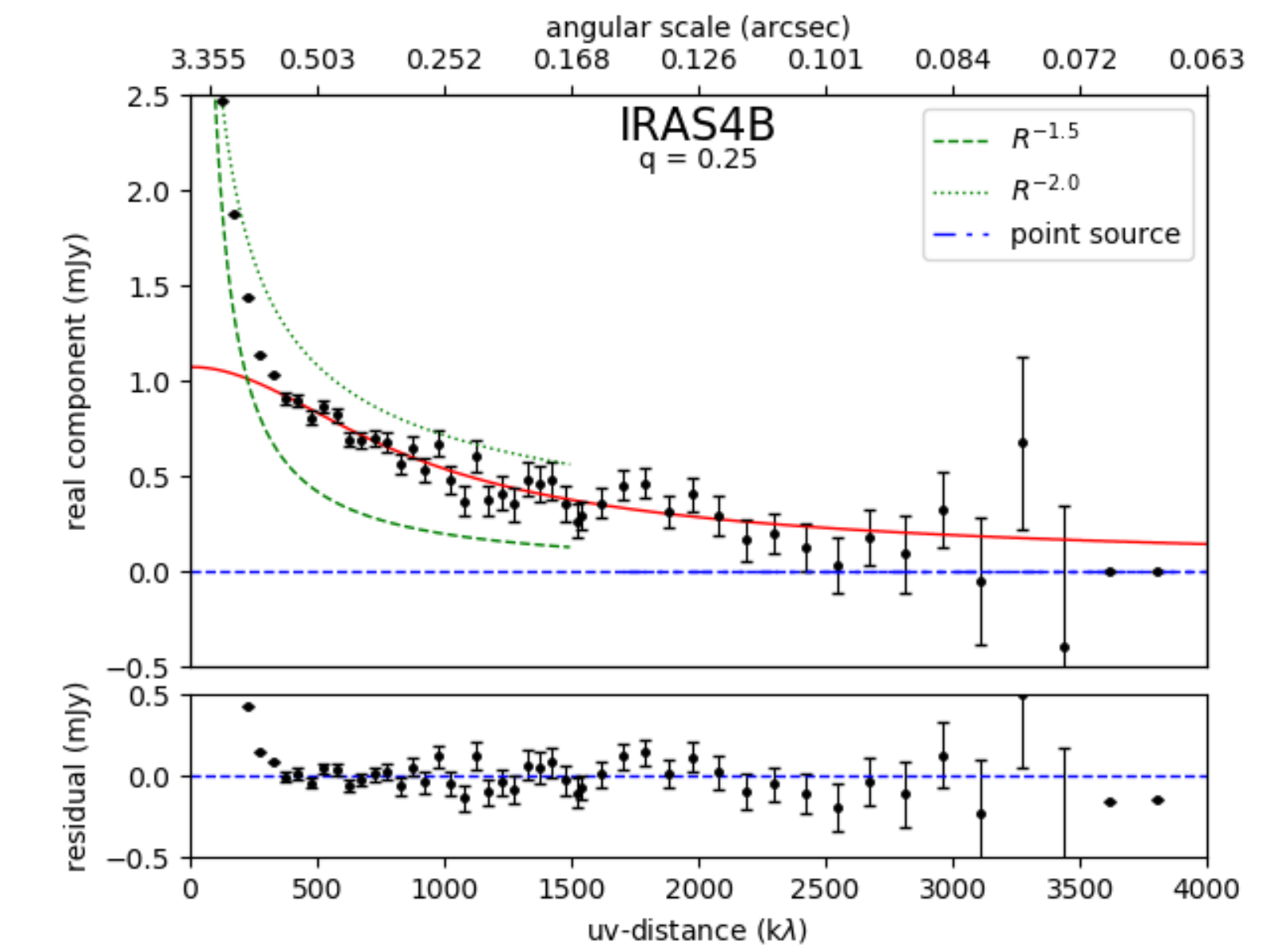}
                \caption{Same as Figure \ref{SVS13Buvdist}, for NGC 1333 IRAS4B.  The baselines $<$350 k$\lambda$ are not fit to the disk component. We do not consider this source a candidate disk.}
                \label{Per13uvdist}
\end{figure} 
\clearpage

 \begin{figure}[t]
        \centering
                \includegraphics[width=0.8\textwidth]{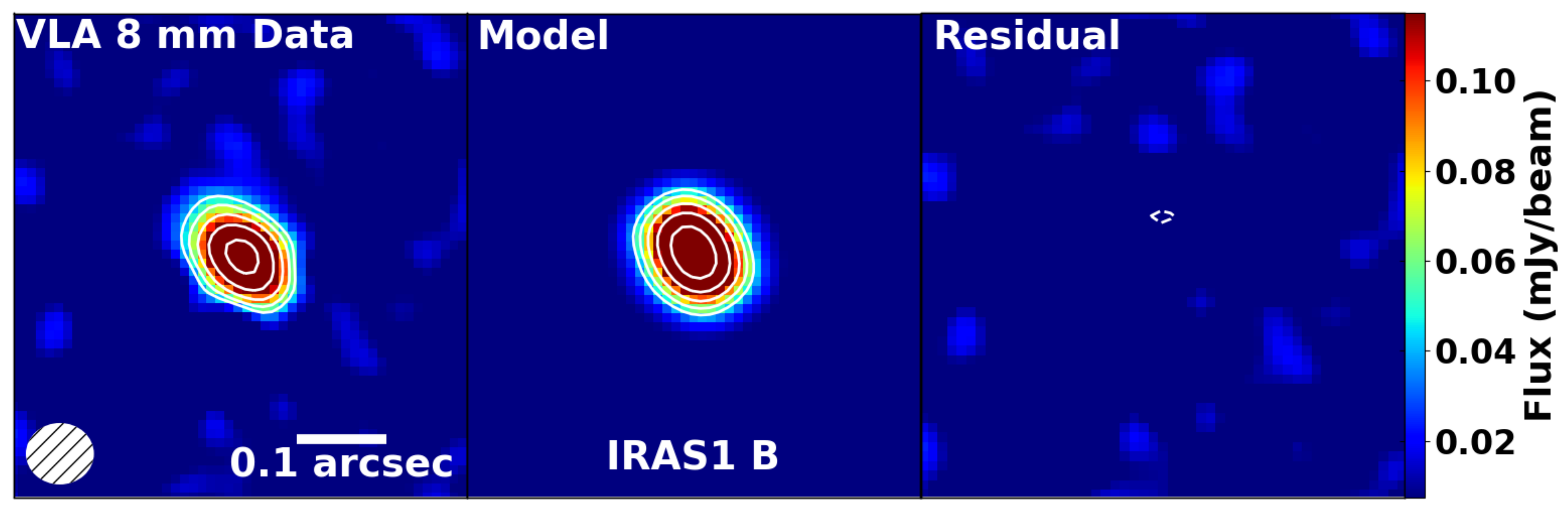}
        \caption{Same as Figure \ref{Per50dmr}, for NGC 1333 IRAS1 B, with $q=0.5$.  We do not consider this source a candidate disk.}
        \label{Per35Bdmr}
\end{figure} 
 \begin{figure}[b]
        \centering
                \includegraphics[width=0.7\textwidth]{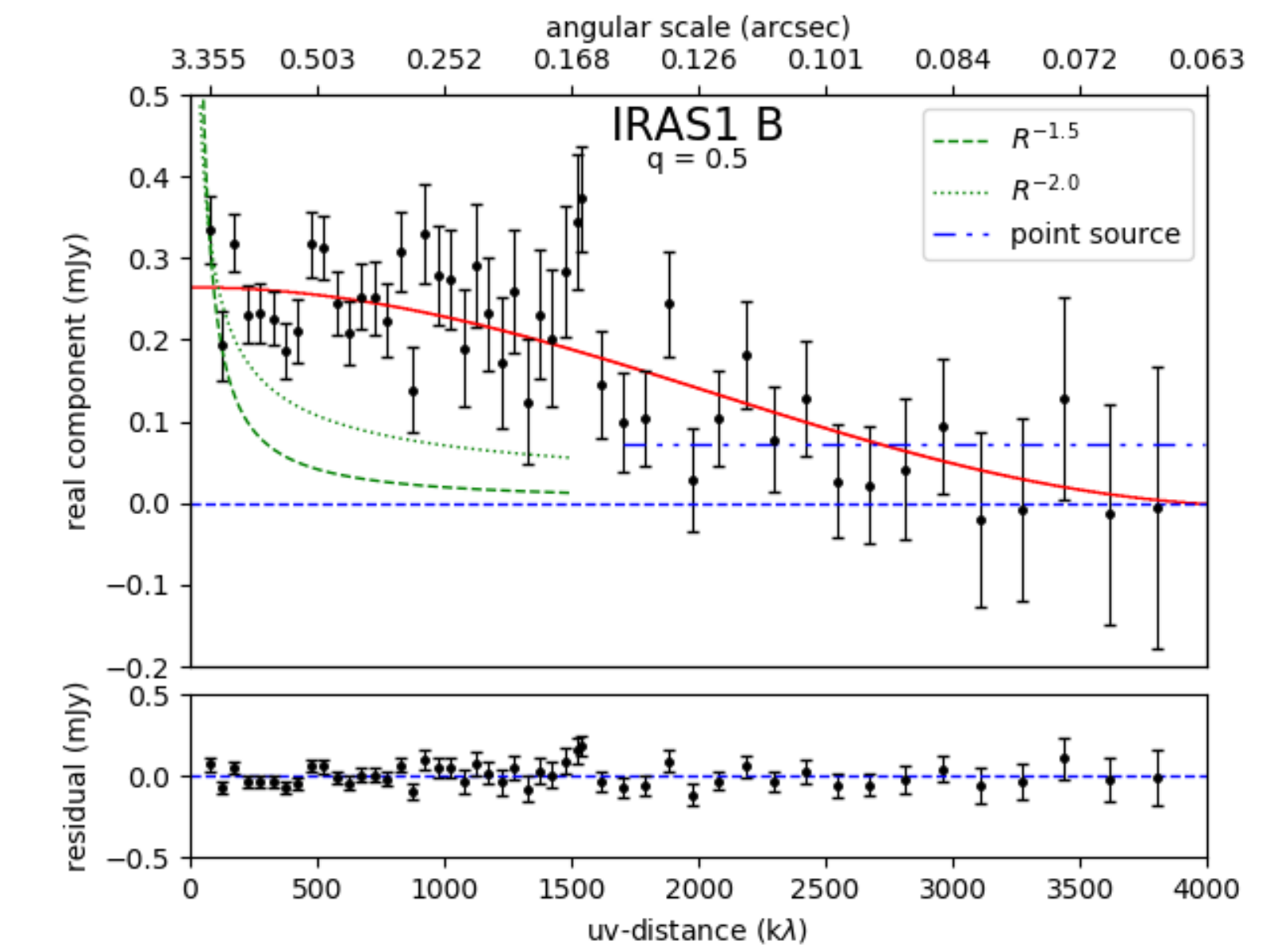}
                \caption{Same as Figure \ref{SVS13Buvdist}, for NGC 1333 IRAS1 B. We do not consider this source a candidate disk.}
                \label{Per35Buvdist}
\end{figure} 
\clearpage

 \begin{figure}[t]
        \centering
                \includegraphics[width=0.8\textwidth]{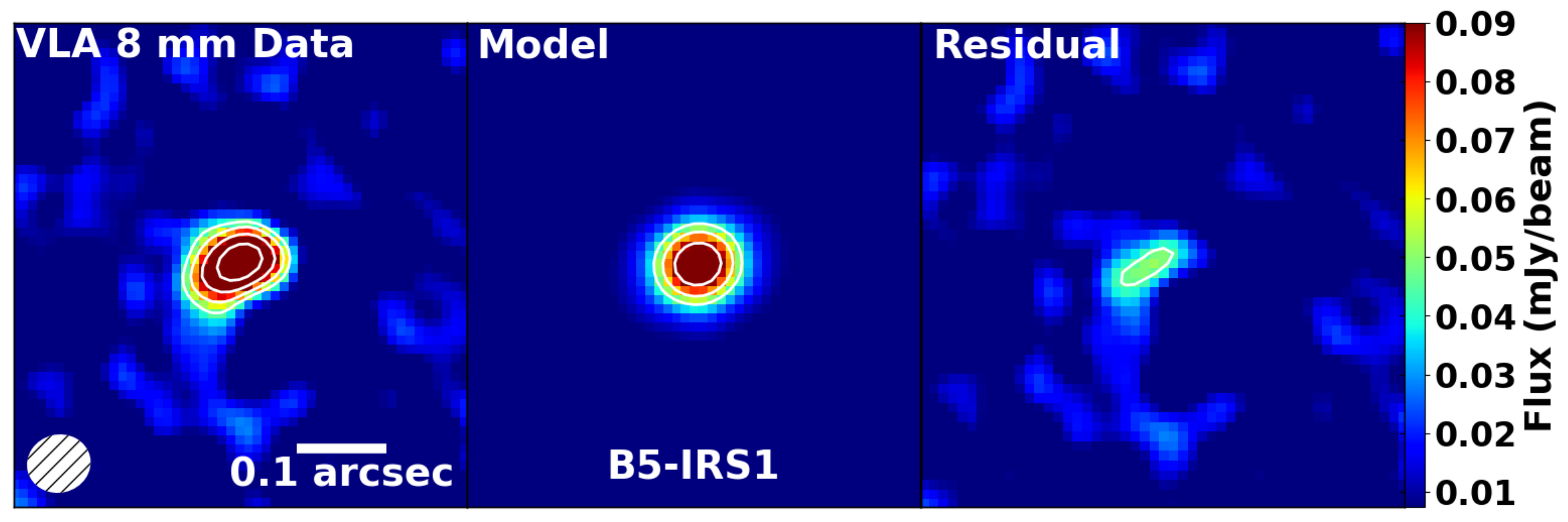}
        \caption{Same as Figure \ref{Per50dmr}, for B5-IRS1, with $q=1.0$.  We do not consider this source a candidate disk.}
        \label{Per53dmr}
\end{figure} 
 \begin{figure}[b]
        \centering
                \includegraphics[width=0.7\textwidth]{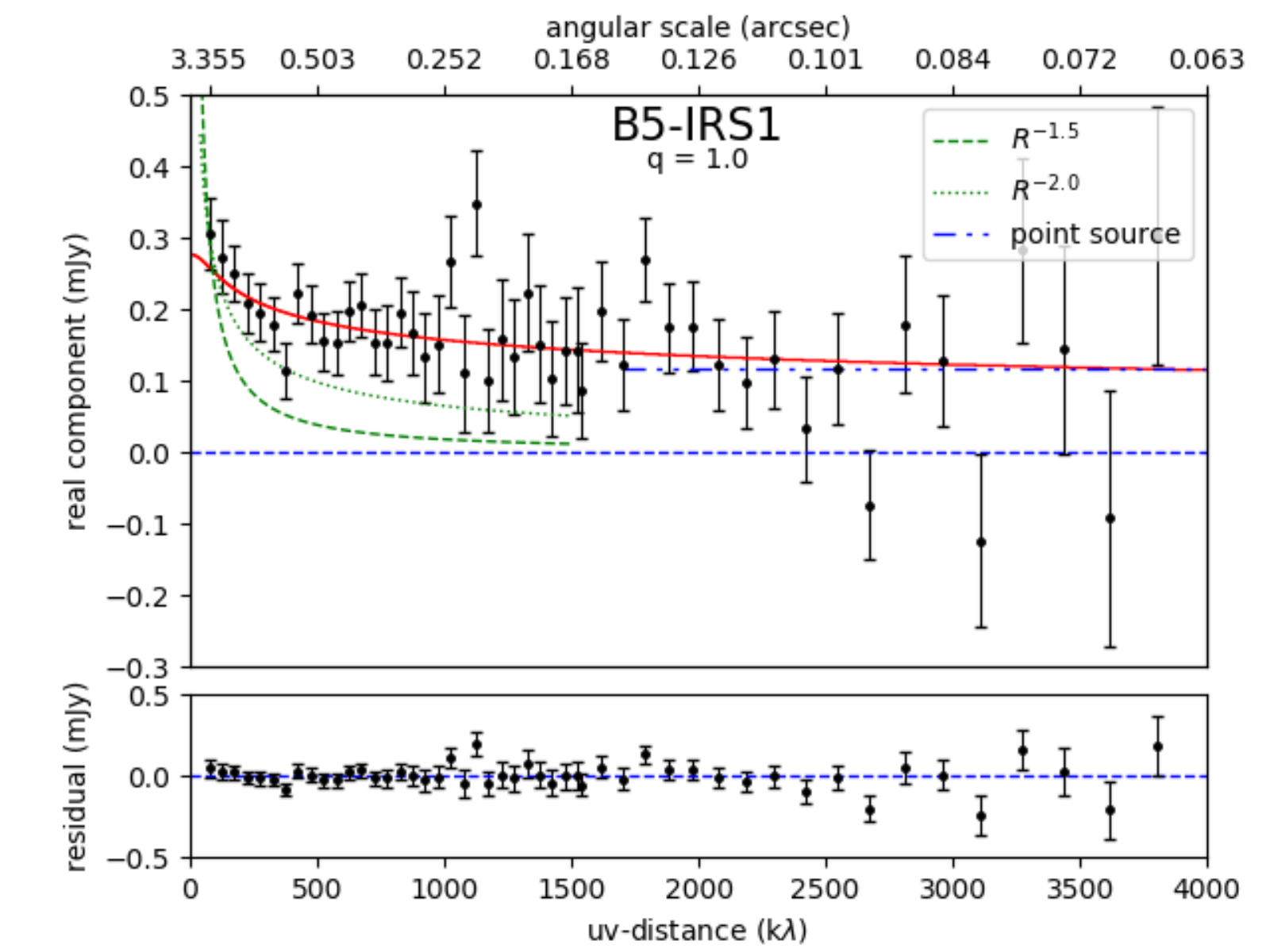}
                \caption{Same as Figure \ref{SVS13Buvdist}, for B5-IRS1.  We do not consider this source a candidate disk.}
                \label{Per53uvdist}
\end{figure} 

\end{document}